\documentclass[]{elsarticle}

\usepackage{floatrow}
\usepackage{algorithm}
\usepackage{amsmath}
\usepackage{amsfonts}
\usepackage{bm}
\usepackage{color}
\usepackage{colortbl}
\usepackage{dlfltxbcodetips}
\usepackage{floatrow}
\usepackage{multirow}
\usepackage{multicol}
\usepackage{psfrag}
\usepackage{url}
\usepackage{lscape}
\usepackage{amsmath}
\usepackage{txfonts}
\usepackage[caption=false,font=footnotesize]{subfig}

\newcommand \D [2]{\frac{\partial #1}{\partial #2}}

\renewcommand{\vec}[1]{\bm{\mathrm{#1}}}

\def \grad{\nabla}

\def \CC{\mathbb{C}}

\def \FF{\mathbb{F}}

\def \PP{\mathbb{P}}
\def \PPs{\PP^{\text{s}}}

\def \RR{\mathbb{R}}

\def \I4fibar{\bar{I}_{4\text{f},i}}

\def \F{\vec{F}}

\def \N{\vec{N}}

\def \X{\vec{\chi}}

\def \e{\vec{e}}
\def \f{\vec{f}}

\def \s{\vec{X}}
\def \t{\vec{t}}
\def \u{\vec{u}}
\def \v{\vec{v}}
\def \x{\vec{x}}
\def \grad{\nabla}

\def \p{\partial}

\def \Dx{\mbox{d} \x}
\def \Ds{\mbox{d} \s}
\def \DA{\mbox{d} A}

\def \Omegaf{\Omega^{\text{f}}}
\def \Omegas{\Omega^{\text{s}}}

\def \sigmas{\vec{\sigma}^{\text{s}}}

\def \dt{\Delta t}

\begin{document}

\begin{frontmatter}

\title{Image-based immersed boundary model of the aortic root}

\author{Ali Hasan}
\author{Ebrahim M.~Kolahdouz}
\address{Department of Mathematics, University of North Carolina, Chapel Hill, North Carolina, USA}

\author{Andinet Enquobahrie}
\address{Medical Computing Group, Kitware, Inc., Carrboro, North Carolina, USA}

\author{Thomas G.~Caranasos}
\address{Division of Cardiothoracic Surgery, Department of Surgery, University of North Carolina School of Medicine, Chapel Hill, North Carolina, USA}

\author{John P.~Vavalle}
\address{Division of Cardiology, Department of Medicine, University of North Carolina School of Medicine, Chapel Hill, North Carolina, USA}

\author{Boyce E.~Griffith\corref{mycorrespondingauthor}}
\address{Department of Mathematics and McAllister Heart Institute, University of North Carolina, Chapel Hill, North Carolina, USA}
\cortext[mycorrespondingauthor]{Corresponding author, Department of Mathematics, Phillips Hall, Campus Box 3250, University of North Carolina, Chapel Hill, NC 27599-3250 USA.}
\ead{boyceg@unc.edu}

\begin{abstract}
Each year, approximately 300,000 heart valve repair or replacement procedures are performed worldwide, including approximately 70,000 aortic valve replacement surgeries in the United States alone.
Computational platforms for simulating cardiovascular devices such as prosthetic heart valves promise to improve device design and assist in treatment planning, including patient-specific device selection.
This paper describes progress in constructing anatomically and physiologically realistic immersed boundary (IB) models of the dynamics of the aortic root and ascending aorta.
This work builds on earlier IB models of fluid-structure interaction (FSI) in the aortic root, which previously achieved realistic hemodynamics over multiple cardiac cycles, but which also were limited to simplified aortic geometries and idealized descriptions of the biomechanics of the aortic valve cusps.
By contrast, the model described herein uses an anatomical geometry reconstructed from patient-specific computed tomography angiography (CTA) data, and employs a description of the elasticity of the aortic valve leaflets based on a fiber-reinforced constitutive model fit to experimental tensile test data.
The resulting model generates physiological pressures in both systole and diastole, and yields realistic cardiac output and stroke volume at physiological Reynolds numbers.
Contact between the valve leaflets during diastole is handled automatically by the IB method, yielding a fully competent valve model that supports a physiological diastolic pressure load without regurgitation.
Numerical tests show that the model is able to resolve the leaflet biomechanics in diastole and early systole at practical grid spacings.
The model is also used to examine differences in the mechanics and fluid dynamics yielded by fresh valve leaflets and glutaraldehyde-fixed leaflets similar to those used in bioprosthetic heart valves.
Although there are large differences in the leaflet deformations during diastole, the differences in the open configurations of the valve models are relatively small, and nearly identical hemodynamics are obtained in all cases considered.
\end{abstract}


\end{frontmatter}

\section{Introduction}

Worldwide, 300,000 heart valve repair or replacement procedures are performed each year \cite{Yoganathan04, LPDasi09, PPiabarot2009}, and this rate is projected to increase to 850,000/year by 2050 \cite{LPDasi09}.
Treatment for severe stenosis of the aortic heart valve is generally to replace the native valve with either a mechanical or a bioprosthetic valve \cite{CarrSavage04}, and in the United States alone, approximately 70,000 aortic valve replacements are performed every year \cite{MAClark2012}.

Surgical valve replacement via open-heart surgery has been performed since 1960, but over the past decade, transcatheter aortic valve replacement (TAVR) has emerged as a less invasive alternative to conventional valve replacement surgery \cite{MBLeon10, CRSmith11}.
In TAVR, a stent-mounted bioprosthetic heart valve is percutaneously implanted via a catheter within the diseased valve.
First-in-man TAVR implantation was performed in 2002 \cite{ACribier02}.
In 2011, the Edwards \emph{SAPIEN} valve became the first TAVR device to be approved by the U.S.~Food and Drug Administration, and approval of a second TAVR device, the Medtronic \emph{CoreValve}, followed in 2014.
TAVR is now approved in the U.S.~for use in both high-~and intermediate-risk patients, and is poised to become an increasingly common alternative to conventional surgical valve replacement.

TAVR device selection, including device sizing, can be challenging.
For instance, leakage flows between the stented valve and the aortic root, which are referred to as paravalvular leaks, are clearly linked to increased long-term mortality \cite{CTamburino11, MHaensig12, SLerakis13}, and interactions between the device and calcification lesions within the aortic root can determine the degree of residual paravalvular leak following TAVR \cite{MHaensig12}.
Improved methods for device selection could also reduce the occurrence of severe complications such as heart block \cite{JBMasson09, SBleiziffer10}.
Computer simulation promises to facilitate device selection and treatment planning for patients who require these devices.

Models also can provide insight into physiological mechanisms, ultimately facilitating improved device design, and can accelerate the testing of implantable medical devices such as prosthetic heart valves.
It is known, for instance, that many of the difficulties of prosthetic heart valves are caused by the fluid dynamics of the replacement valve \cite{Yoganathan04, LPDasi09}.
Computational models of cardiac dynamics can predict the flow patterns of native valves as well as mechanical and tissue valve prostheses.
Simulations also can predict the kinematics and loads experienced by the valve leaflets, which could lead to novel designs that improve device durability.
Indeed, the major limitation of bioprosthetic heart valves is their limited durable lifetime, which is typically 10--15 years.

Fluid-structure interaction (FSI) models can predict the dynamics of the aortic valve leaflets \cite{DeHart03a, DeHart03b, DeHart04, CJCarmody06-aortic_valve_fsi, GMarom13-valve_fsi_collagen, AGilmanov16-comparative_heomdynamics, ALaadhari16-resistive_surfaces, MMega16-native_porcine_fsi}, and recent work has also provide more complete descriptions of the dynamic FSI-mediated coupling between the valve leaflets and the sinuses \cite{MCHsu14-wall_deformation, MCHsu15-bhv_fsi, MGNestola17-aortic_root_stresses}.
To account for patient-specific geometry, a finite element-based fluid-structure interaction model of the aortic root was constructed using magnetic resonance imaging that accounts for geometric variability among a population of ten patients \cite{conti2010dynamic}.
An emphasis is placed on correctly modeling the nonlinearities associated with the mechanical response of the valve leaflets. 
In another study, data from transesophageal echocardiography was used to construct aorta models, integrating tissue properties derived from the patient's age \cite{labrosse2015subject}. 
The model is accurately constructed by determining anatomical landmarks for leaflet and vessel response are further tailored to the patient through the use of age dependent material properties.
The efficacy of the model is assessed by comparison with patient imaging data.
Recently, a study using an Arbitrary Lagrangian-Eulerian (ALE) approach compared the performance of native valves as well as stentless and stented bioprosthetic valves in realistic patient anatomies \cite{MGNestola17-aortic_root_stresses}.

This paper describes ongoing work to develop a simulation framework based on the immersed boundary (IB) method \cite{Peskin02} to model the dynamics of the aortic root, with the goals of facilitating prosthetic valve design and personalized approaches to treatment planning.
We previously developed three-dimensional FSI models of the aortic root using several different versions of the IB method, including a cell-centered IB method based on an efficient approximate projection method \cite{BEGriffith07-ibamr_paper, BEGriffith09-heart_valves}, a staggered-grid IB method with improved volume conservation \cite{BEGriffith12-aortic_valve, BEGriffith12-ib_volume_conservation}, and an IB method with support for finite element elasticity models \cite{VFlamini16-aortic_root, BEGriffithXX-ibfe}.
These initial models used realistic driving and loading conditions, including realistic diastolic pressure loads on the closed valve, and produced physiological stroke volume and cardiac output at realistic Reynolds numbers.
However, all of these studies employed a highly stylized aortic geometry derived from patient image data \cite{Reul90} that did not account for the asymmetry of the aortic root or the curvature of the ascending aorta.
In addition, these earlier studies all used simplified descriptions of the mechanics of the aortic valve cusps based on fiber-based models that were discretized using systems of springs and beams.

This paper extends these earlier IB models of aortic valve dynamics \cite{BEGriffith09-heart_valves, BEGriffith12-aortic_valve, VFlamini16-aortic_root} towards clinical utility by incorporating a realistic, three-dimensional anatomical model of the aortic root and ascending aorta.
Within this image-based anatomical geometry, we construct a rule-based model of the fiber structure of the aortic valve leaflets using an approach based on Poisson interpolation that is similar to methods used previously to generate models of the fiber architecture of the heart \cite{JDBayer12-rule_based, JWong2014, SRossi14-contraction}.
The biomechanics of the valve leaflets is described by a fiber-reinforced hyperelastic constitutive model for the aortic valve leaflets \cite{NJBDriessen05} with constitutive parameters fit to experimental tensile test data from fresh or glutaraldehyde-fixed porcine aortic valve leaflets \cite{KLBilliar00-I, KLBilliar00-II}.
Three-dimensional FSI simulations are performed by an IB method that supports finite-strain continuum mechanics structural models discretized via a nodal displacement-based finite element scheme \cite{BEGriffithXX-ibfe}.
These methods are well suited for complex anatomical geometries, experimentally based constitutive models, and large-scale simulation.
The model is driven using clinical hemodynamic data \cite{JPMurgo80}, and afterload is modeled by a three-element Windkessel model fit to the same data \cite{Stergiopulos99}.
The FSI model is demonstrated to yield physiological cardiac output and stroke volume at realistic driving and loading pressures and Reynolds numbers.
Further, the model accounts for contact between the valve leaflets when the valve is closed and loaded, yielding a fully competent model valve.
Numerical tests described herein show that the model is able to resolve the leaflet biomechanics in diastole and early systole at practical grid spacings.
The model is also used to examine differences in the mechanics and fluid dynamics yielded by fresh valve leaflets and glutaraldehyde-fixed leaflets similar to those used in bioprosthetic heart valves.

\section{Methods}

\subsection{Model construction}

\begin{figure}[t]
\centering
\includegraphics[width=\textwidth]{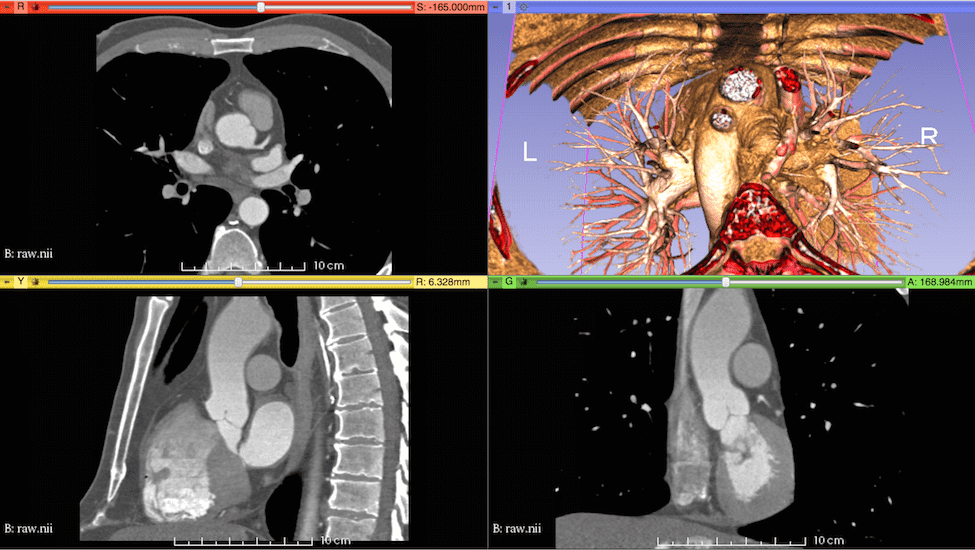}
\caption{The computed tomography (CT) image data used in this study.}
\label{f:CT-data}
\end{figure}

A three-dimensional representation of the aortic root and ascending aorta was generated from publicly available medical image data.
This study uses one of the sample data sets from the OsiriX DICOM Image Library that consists of three-dimensional computed tomography angiography (CTA) of an anonymous patient's chest following administration of a contrast agent.
The use of a contrast agent enables better discrimination of the blood vessels and heart chambers than non-contrast CT images.
The image data were acquired at the Ronald Reagan University of California at Los Angeles Medical Center in Santa Monica, CA using a Siemens SOMATOM Sensation 16 CT scanner.
The image resolution is 512$\times$512$\times$355 with a voxel size of 0.47$\times$0.47$\times$0.5~mm.
The image was processed by an anisotropic diffusion filter to mitigate noise.
Sample renderings of the patient data are shown in Fig.~\ref{f:CT-data}.

\begin{figure}
\centering

\subfloat[][]{\includegraphics[width=0.48\textwidth]{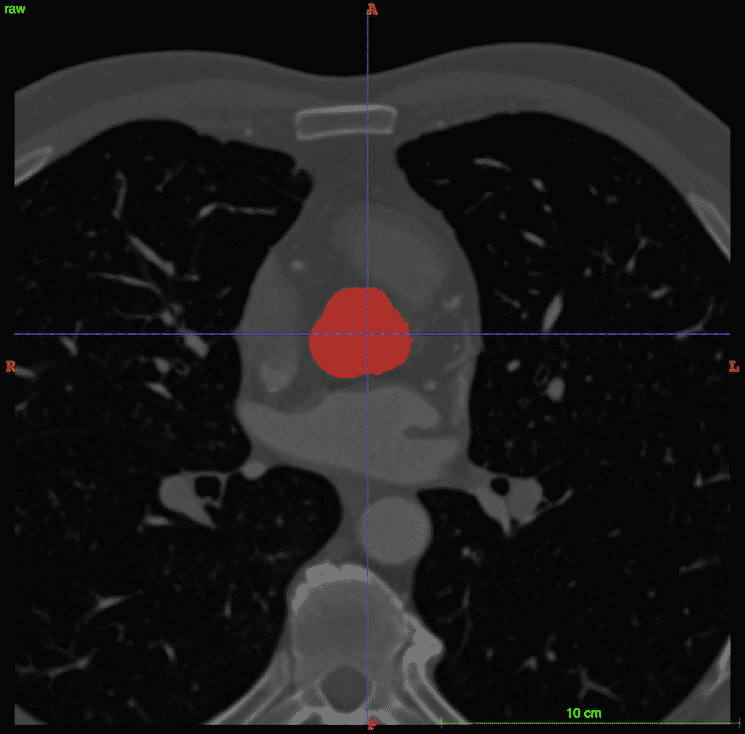}}
\hspace{10pt}
\subfloat[][]{\includegraphics[width=0.48\textwidth]{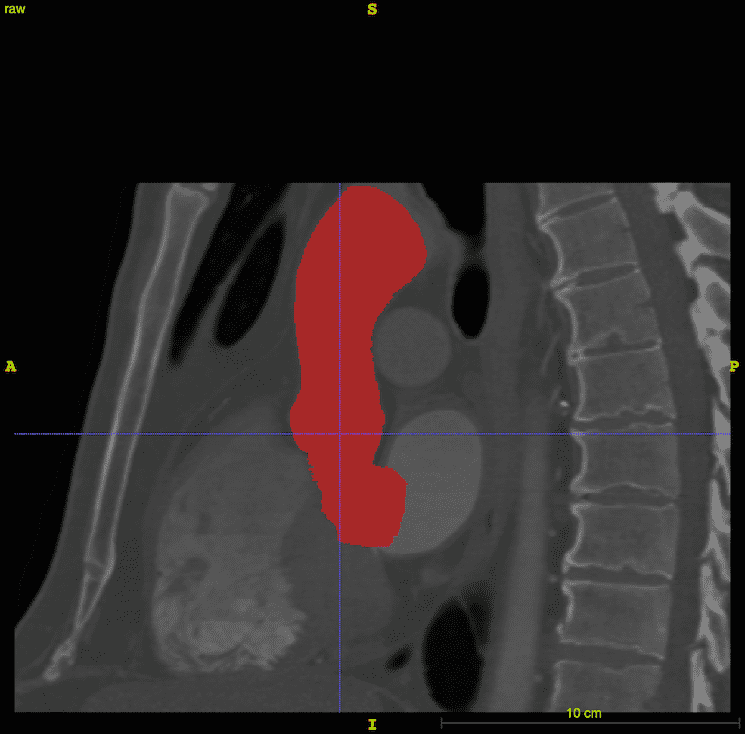}}

\subfloat[][]{\includegraphics[width=0.48\textwidth]{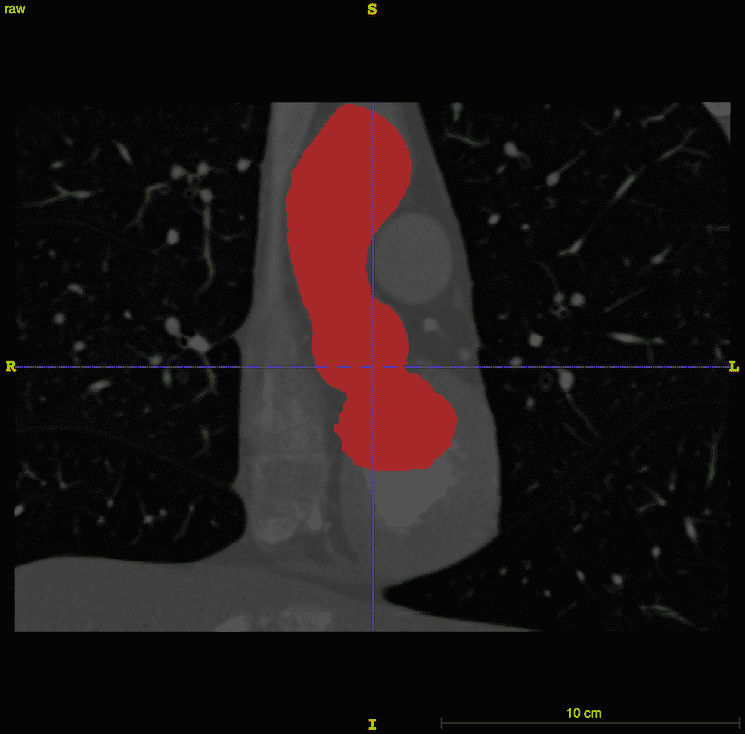}}
\hspace{10pt}
\subfloat[][]{\includegraphics[width=0.48\textwidth]{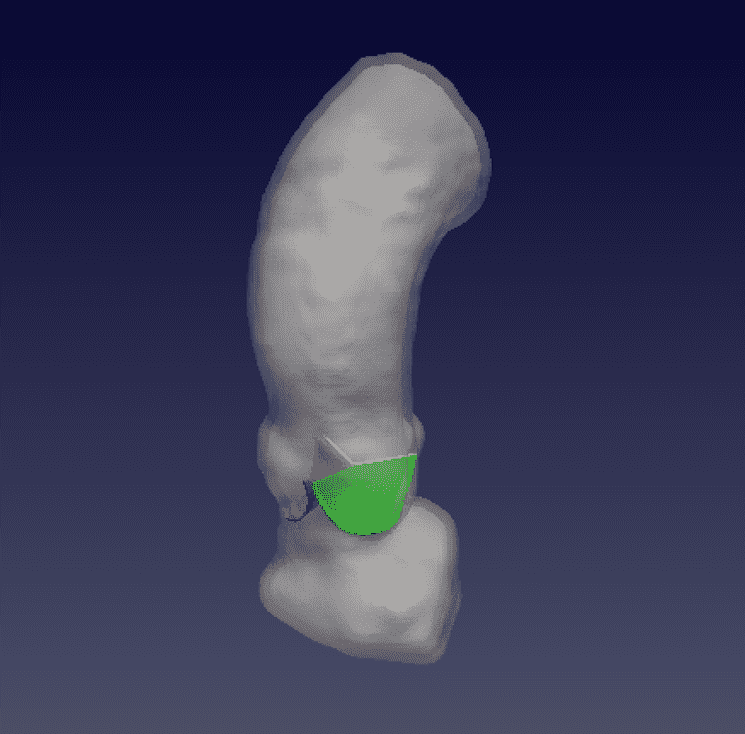}}

\caption{Semi-automatically generated segmentation of the aortic root and ascending aorta, using the CTA dataset shown in Fig.~\ref{f:CT-data}.  Panel (a)~shows the transverse plane, (b)~shows the sagittal plane, (c)~shows the coronal plane of the segmentation, and (d)~shows the three-dimensional reconstruction.}
\label{f:segmentation}
\end{figure}

The aortic root and ascending aorta were segmented by a semi-automated method implemented in the ITK-SNAP software.
ITK-SNAP \cite{py06nimg, ITK-SNAP-web-page}, which is based on the Insight Segmentation and Registration Toolkit (ITK) \cite{ITK, ITK-web-page}, provides a graphical interface for the implementation of the active contour model, also known as Kass snakes \cite{MKass88}, for semi-automatic segmentation. The algorithm works by minimizing an energy functional that is determined by voxel intensities.
The generated segmentation is shown in Fig.~\ref{f:segmentation}.

\begin{figure}[t]
\centering
\subfloat[][]{\includegraphics[height=.34\textheight]{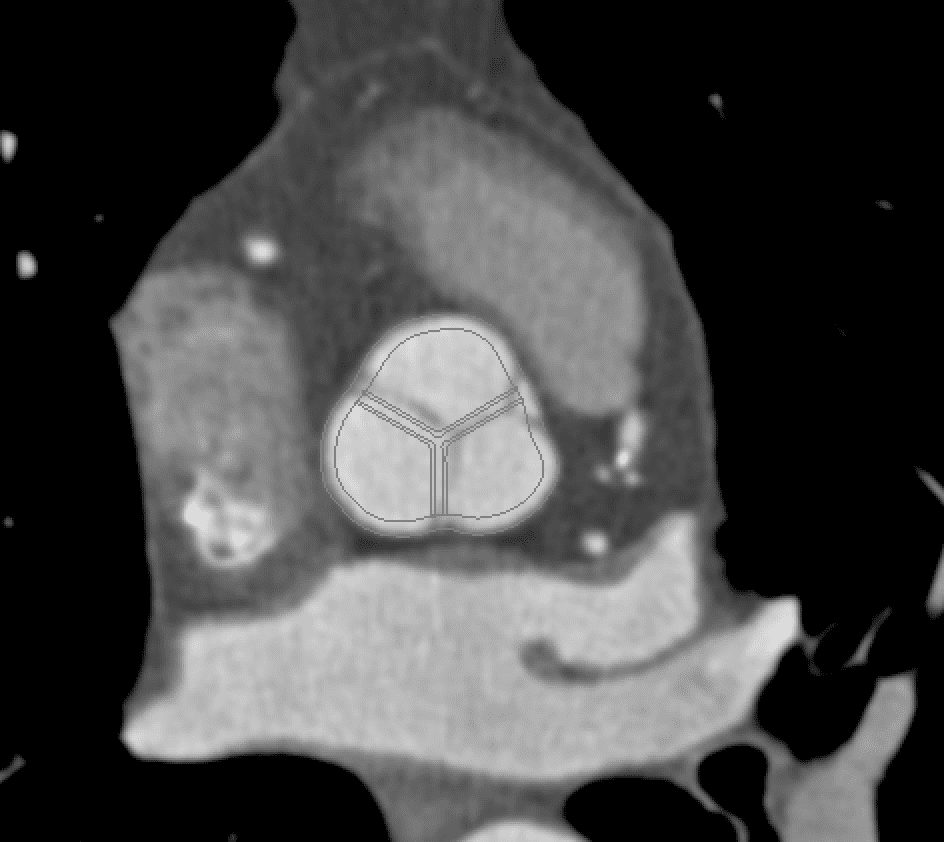}}
\hspace{5pt}
\subfloat[][]{\includegraphics[height=.34\textheight]{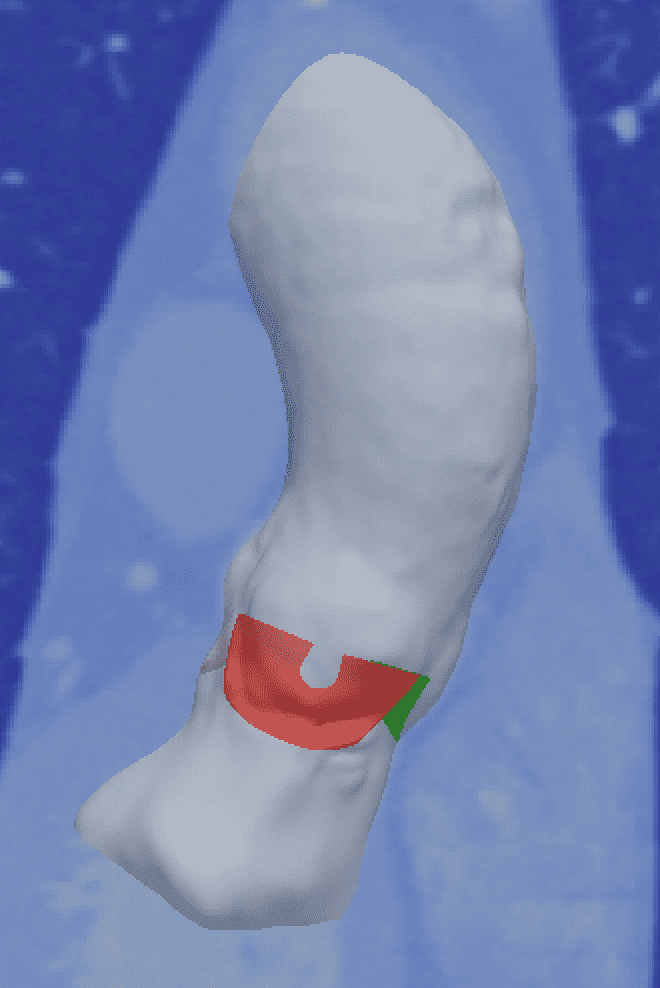}}
\hspace{5pt}
\caption{Alignment of the model and the image data. Panel (a)~shows the leaflet alignment via a two-dimensional axial projection, and panel (b)~shows the three-dimensional along with a sagittal view of the image data.}
\label{f:alignment}	
\end{figure}

Aortic valve leaflets are thin layered structures that are reinforced by collagen fibers running from commissure to commissure \cite{Sauren81, DeHart02, NJBDriessen05}.
To obtain geometrical models of the aortic valves, we first constructed idealized aortic valve leaflets with dimensions based on the study of Driessen et al.~\cite{NJBDriessen05}.
We then manually trimmed and morphed the leaflets to fit within the image data.
Fig.~\ref{f:alignment} shows the alignment between the generate model geometry and the image data.

\begin{figure}[t]
\centering
\subfloat[][]{\includegraphics[width=0.315\textwidth]{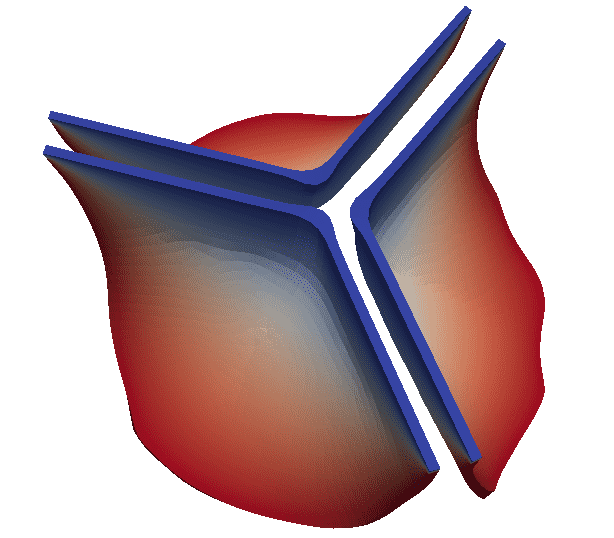}}
\hspace{5pt}
\subfloat[][]{\includegraphics[width=0.315\textwidth]{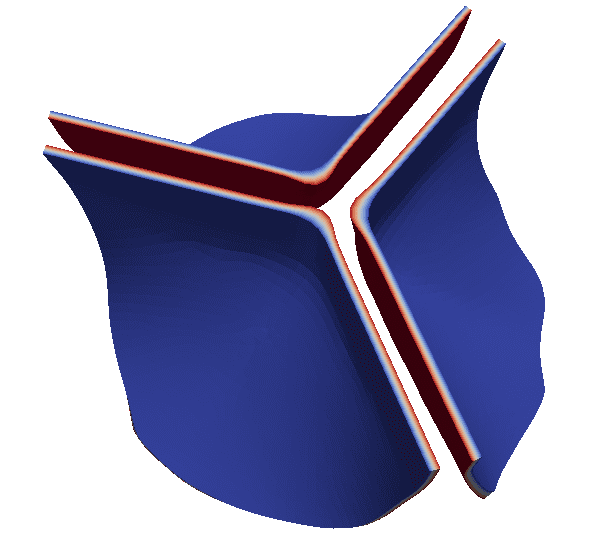}}
\hspace{5pt}
\subfloat[][]{\includegraphics[width=0.315\textwidth]{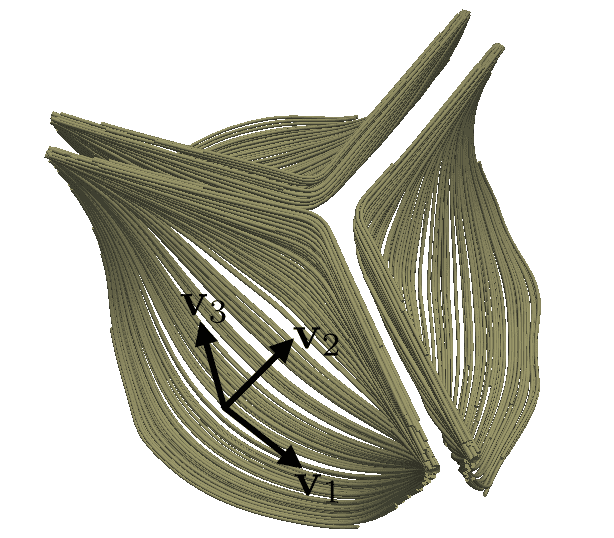}}
\caption{
Idealized aortic valve leaflet geometry along with Poisson interpolation \cite{JDBayer12-rule_based, JWong2014, SRossi14-contraction} based collagen fiber model.
Panel (a)~shows the field $u$ that is used to define the leaflet fiber architecture, and panel (b)~shows $v$.
These fields are two solutions to the Poisson equation on the leaflet geometry, but using different boundary conditions.
Panel (c)~shows the resulting (mean) fiber direction, which is computed as $\v_1 = \grad u \times \grad v$.  Notice that the fibers run from commissure to commissure.
Panel (c)~also shows the local material axes $\v_2$ and $\v_3$.
$\v_3 = \grad v$ points ``through'' the leaflet, and $\v_2 = \v_3 \times \v_1$ lies in the ``plane'' of the leaflet and is oriented to point from the fixed margin to the free margin of the leaflet.
}
\label{f:valve_construction}	
\end{figure}

Modeling the fiber structure of the valve leaflets is important in describing the elasticity of the leaflets, especially when the valve is closed and is bearing a realistic pressure load.
Because CTA data do not provide information about the fiber architecture of the valve leaflets, we use a rule-based approach to describe the fiber reinforcement.
To do so, we employ a method based on Poisson interpolation \cite{JDBayer12-rule_based, JWong2014, SRossi14-contraction}.
Briefly, we construct a local material coordinate system at each position in the leaflets by solving two Poisson problems with different sets of boundary conditions.
One of these fields, $u$, satisfies $\grad^2 u = 0$ with Dirichlet boundary conditions along the free edge and line of attachment to the aortic root and homogeneous Neumann boundary conditions along the remainder of the boundary.
The integral curves of $\grad u$ connect the wall of the aortic root to the free edge of the leaflets, and the iso-surfaces of $u$ connect the commissures and are approximately aligned with the mean fiber direction.
The second field, $v$, satisfies $\grad^2 v = 0$ with Dirichlet boundary conditions only along the top and bottom faces of the leaflets, so that $\vec{v}_3 = \grad v$ is approximately normal to the layers of the leaflets.
The mean fiber direction is modeled as $\vec{v}_1 = \grad u \times \grad v$, and an orthogonal vector to $\vec{v}_1$ confined to lie within a single layer of the leaflet is obtained by computing $\vec{v}_2 = \vec{v}_3 \times \vec{v}_1$.
Notice that by construction, $\vec{v}_1$, $\vec{v}_2$, and $\vec{v}_3$ form a local right-handed orthogonal coordinate system.
The fields $u$ and $v$ and the resulting model fiber structure are shown in Fig.~\ref{f:valve_construction}.

\begin{algorithm}[t]
\caption{Model construction workflow.}	
\label{f:workflow}
\centering
\fbox{ \begin{minipage}{0.95\textwidth}
\small \raggedright
\begin{enumerate}
	\item Pre-process images and establish intensity thresholds in the aortic root region.
	\item Identify and apply initializing points for the semi-automated segmentation algorithm.
	\item Iteratively update segmentation parameters to prevent ``leaking'' and to ensure propagation of the snake.
	\item Export the segmentation as a STereoLithography (STL) triangulated surface mesh.
	\item Import the STL geometry in CAD software to apply Gaussian smoothing and other minor modifications.
	\item Export the geometry as an ACIS geometry and STL file.
	\item Import the aorta geometry into Bolt for hexahedral meshing.
	\item Import the geometry into Trelis for final geometry manipulation, including attachment of model valve leaflets, and valve meshing.
	\item Establish boundary conditions on the valve leaflets needed for Poisson interpolation.
	\item Export the mesh and boundary data as an ExodusII file for simulation.
\end{enumerate}
\end{minipage}}
\end{algorithm}

Several distinct software applications are used to construct the full model.
The construction pipeline starts with three-dimensional NRRD (nearly raw raster data) files containing the medical image data.
Segmentation and initial geometry construction is generated using ITK-SNAP.
Further adjustments are made in SOLIDWORKS (Dassault Syst\`{e}mes SOLIDWORKS Corporation, Waltham, MA, USA) to fix any irregularities.
SOLIDWORKS is also used to generate the geometry of the aortic valve leaflets.
The STL (STereoLithography) geometry constructed in SOLIDWORKS is then used in Bolt (Computational Simulation Software, LLC, American Fork, UT, USA) to construct a hexahedral mesh for the aortic root.
Placement of the model valve leaflets within the aortic root is finalized in Trelis (Computational Simulation Software, LLC, American Fork, UT, USA), which is a mesh generation software application based on CUBIT from Sandia National Laboratory.
Trelis is also used to generate tetrahedral meshes of the aortic valve leaflets.
Because the leaflets and vessel wall are both modeled using the IB method, it is not necessary to use conforming discretizations of the leaflets and wall, which greatly simplifies the mesh generation process.
The overall model construction workflow is detailed in Algorithm~\ref{f:workflow}.
The inflow section of the model is truncated at the left ventricular outflow tract, and the outflow section of the model is truncated in the aortic arch before the first bifurcation.
In the FSI simulation, the inflow and outflow sections are coupled to reduced-order models that provide driving and loading conditions, and that establish realistic pressure differences across the model vessel.

\subsection{The IB method with finite-strain continuum mechanics models}

We use an immersed boundary (IB) formulation to describe interactions between the aortic valve leaflets, the walls of the left ventricular outflow tract, the aortic sinuses, and the ascending aorta, and the blood, which we model as an incompressible Newtonian fluid.
The equations of momentum conservation and incompressibility are posed on a fixed Eulerian computational domain $\Omega \subset \RR^3$, whereas the solid mechanics formulation uses a Lagrangian material coordinate system.
We use fixed physical coordinates $\x = (x_1,x_2,x_3) \in \Omega$ to describe the computational domain.
This domain is subdivided into non-overlapping fluid and solid regions, $\Omega = \Omegaf_t \cup \Omegas_t$, that are indexed by time $t$.
The deformations of the solid body are described using reference coordinates at time $t = 0$, $\s = (X_1,X_2,X_3) \in \Omegas_0$.
The outward unit normal to $\p \Omegas_0$ is $\N(\s)$.
A time-dependent mapping $\X(\s,t)$ relates reference coordinates to current physical coordinates and implicitly determines the solid domain at time $t$ via $\Omegas_t \equiv \X(\Omegas_0,t)$.
We describe the solid stresses in terms of the deformation gradient tensor $\FF(\s,t) = \left(\p \X/\p \s\right) (\s,t)$.
The immersed solid is assumed to be incompressible, so that $J = \det(\FF) \equiv 1$.
The equations of fluid-solid interaction can be written as \cite{DBoffi08, HGao14-iblv_diastole, BEGriffithXX-ibfe}:
\begin{align}
  \rho\frac{\mathrm{D}\u}{\mathrm{D}t}(\x,t) &= - \grad p(\x,t) + \mu \grad^2 \u(\x,t) + \f(\x,t) + \t(\x,t),  \label{e:v2momentum} \\
  \grad \cdot \u(\x,t) &= 0,                                                                                   \label{e:v2incompressibility} \\
  \f(\x,t)             &= \int_{\Omegas_0} \F(\s,t) \, \delta(\x - \X(\s,t)) \, \Ds \nonumber \\
  & \ \ \ \ \ \ \mbox{} + \int_{\Omegas_0} \grad \cdot \PPs(\s,t) \, \delta(\x - \X(\s,t)) \, \Ds,                          \label{e:interior_force_density} \\
  \t(\x,t)             &= -\int_{\p \Omegas_0} \PPs(\s,t) \, \N(\s) \, \delta(\x - \X(\s,t)) \, \DA,                   \label{e:v2transmission_force_density} \\
  \PPs(\s,t)           &= J \, \sigmas(\s,t) \, \FF^{-T},                                                     \label{e:v2solid_stress} \\
  \D{\X}{t}(\s,t)      &= \int_\Omega \u(\x,t) \, \delta(\x - \X(\s,t)) \, \Dx = \u(\X(\s,t),t),                                \label{e:v2interp}
\end{align}
in which $\u(\x,t)$ is the velocity field, $p(\x,t)$ is the pressure, $\rho$ is the mass density, $\mu$ is the viscosity, $\f(\x,t)$ and $\t(\x,t)$ are volumetric and surface force densities, $\F(\s,t)$ is a Lagrangian body force density, $\PPs(\s,t)$ is the first Piola-Kirchhoff elastic stress tensor, $\sigmas$ is the corresponding Cauchy elastic stress, and $\delta(\x) = \delta(x_1) \, \delta(x_2) \, \delta(x_3)$ is the three-dimensional Dirac delta function.
Notice that in this formulation, $\u(\x,t)$ is a common velocity field for both the fluid and the solid.
Because of viscosity, this velocity field is continuous at fluid-solid interfaces (i.e.~there is a no-slip condition at fluid-solid interfaces).
This property of the IB formulation is especially useful for models involving contact (including self contact) because structures that move according to the same continuous velocity field cannot interpenetrate.
Because $\p\X(\s,t)/\p t = \u(\X(\s,t),t)$ and $\grad \cdot \u \equiv 0$, the solid deformation is automatically incompressible, i.e.~$J \equiv 1$.
Nonetheless, we find that it is useful in practice to use a nearly incompressible formulation for the structural mechanics, as detailed below.

\subsection{Solid mechanics models}

Different constitutive models are used for the leaflets and for the wall of the aortic root and ascending aorta.
In both cases, we assume a hyperelastic material response, so that the first Piola-Kirchhoff stress is determined from a strain energy functional $W(\FF)$ via
\begin{equation}
	\PPs = \D{W}{\FF}.
\end{equation}
We further assume that the strain energy functional $W(\FF)$ may be additively split into isochoric and volumetric parts,
\begin{equation}
	W(\FF) = \bar{W}(\bar{\FF}) + U(J),
\end{equation}
in which $\bar{\FF} = J^{-1/3} \FF$, so that $\det(\bar{\FF}) = 1$, and $U(J)$ represents the volumetric energy, which here we use to penalize compressible deformations of the elastic solid.

We describe the valve leaflets using the fiber-reinforced model of Driessen, Bouten, and Baaijens \cite{NJBDriessen05}.
This model uses a neo-Hookean description of the leaflet matrix along with an fiber-aligned stress with an exponential length-tension relationship to describe the collagen fibers distributed within the valve leaflet.
In this model, the isotropic matrix is described by
\begin{equation}
	\bar{W}_\text{matrix} = \frac{c_1}{2} (\bar{I}_1 - 1),
\end{equation}
in which $\bar{\CC} = \bar{\FF}^T \, \bar{\FF}$ is the (isochoric) right Cauchy-Green strain and $\bar{I}_1 = \text{tr}(\bar{\CC})$ is the first invariant of $\bar{\CC}$, and the elastic fibers are described by
\begin{equation}
	\bar{W}_{\text{f},i} = \frac{k_1}{2 k_2} \left(\exp\left(k_2 (\I4fibar^\star - 1)\right) - k_2 \I4fibar^\star\right),
\end{equation}
in which $i$ indexes a discrete collection of fiber directions, $\e_{\text{f},i}^0$ is a unit vector in the $i^\text{th}$ fiber direction in the reference configuration, $\I4fibar$ is the fiber invariant,
\begin{equation}
	\I4fibar = \e_{\text{f},i}^0 \cdot \bar{\CC} \, \e_{\text{f},i}^0,
\end{equation}
and $\I4fibar^\star$ is the modified fiber invariant,
\begin{equation}
	\I4fibar^\star = \begin{cases}
 		\I4fibar & \text{if $\I4fibar \ge 1$,} \\
 		1 & \text{otherwise.}
 \end{cases}
\end{equation}
We use material parameters determined by Driessen et al.~\cite{NJBDriessen05} from the experimental studies of Billiar and Sacks \cite{KLBilliar00-I, KLBilliar00-II}.

In the model of Driessen et al.~\cite{NJBDriessen05}, the total elastic energy associated with the leaflet is a weighted sum of energies associated with the matrix and fibers,
\begin{equation}
	\bar{W} = \bar{W}_\text{matrix} + \sum_{i=1}^{N} \phi_{\text{f},i} \left(\bar{W}_{\text{f},i} - \frac{c_1}{2} (\bar{I}_{4\text{f},i} - 1) \right), \label{e:leaflet_energy}
\end{equation}
in which $\phi_{\text{f},i}$ is the fiber volume fraction of the $i^\text{th}$ fiber direction.
Notice that in Eq.~\eqref{e:leaflet_energy}, the fiber response associated with a given fiber direction replaces the response of the isotropic matrix in that same direction.
See Driessen et al.~\cite{NJBDriessen05} for further discussion.

For a given fiber angle $\gamma_i$, the fiber direction in the reference configuration is determined via
\begin{equation}
\e_{\text{f},i}^0 = \cos(\gamma_i) \v_1 + \sin(\gamma_i) \v_2,
\end{equation}
in which $\v_1$ and $\v_2$ are the local material axes defined previously.
As in the work of Driessen et al.~\cite{NJBDriessen05}, we assume that the fiber angles are normally distributed with mean $0^\circ$, and that the volume fraction of fibers in the leaflet is 0.5, so that $\sum_i \phi_{\text{f},i} = 0.5$.
Different fiber angle standard deviations are used for the fresh and glutaraldehyde-fixed leaflets, as in Driessen et al.~\cite{NJBDriessen05}.
We discretize the normal distribution by considering angles ranging from $-60^\circ$ to $+60^\circ$ in increments of $3^\circ$.
The leaflets also include a volumetric penalty stress with an elastic energy of the form \cite{CHLiu1994}
\begin{equation}
	U(J) = \beta(J \, \log(J) - J + 1).
\end{equation}
In our computations, we use $\beta = 1~\text{MPa}$.

We model the vessel wall as a stiff (nearly rigid) neo-Hookean material, so that $\bar{W} = \frac{c}{2} (\bar{I}_1 - 1)$.
To keep the vessel approximately tethered in place, additional structural forces of the form $\F(\s,t) = \kappa \left(\s - \X(\s,t)\right)$ are included in the vessel.
In the limit $c \rightarrow \infty$, the material becomes perfectly rigid, and as $\kappa \rightarrow \infty$, the structure becomes completely stationary.
In our simulations, we choose $c$ and $\kappa$ to be sufficiently large to prevent the vessel wall from moving more than a fraction of a meshwidth during the course of the simulation.
(Imposing exact rigidity constraints within the IB framework is difficult and necessitates the use of complex solvers for an extended saddle point system \cite{BKallemov16-RigidIBAMR}.
In practice, we find that penalty approaches, as used here, are effective in the flow regime considered in this study \cite{BEGriffithXX-ibfe}.)
Because this model of the vessel wall does not experience substantial deformations, we omit the volumetric penalty from the stress.
We plan to consider realistically flexible models of the aortic sinuses and ascending aorta in future work.

\subsection{Driving and loading conditions}

\begin{figure*}
\centering
\includegraphics[width=0.5\textwidth]{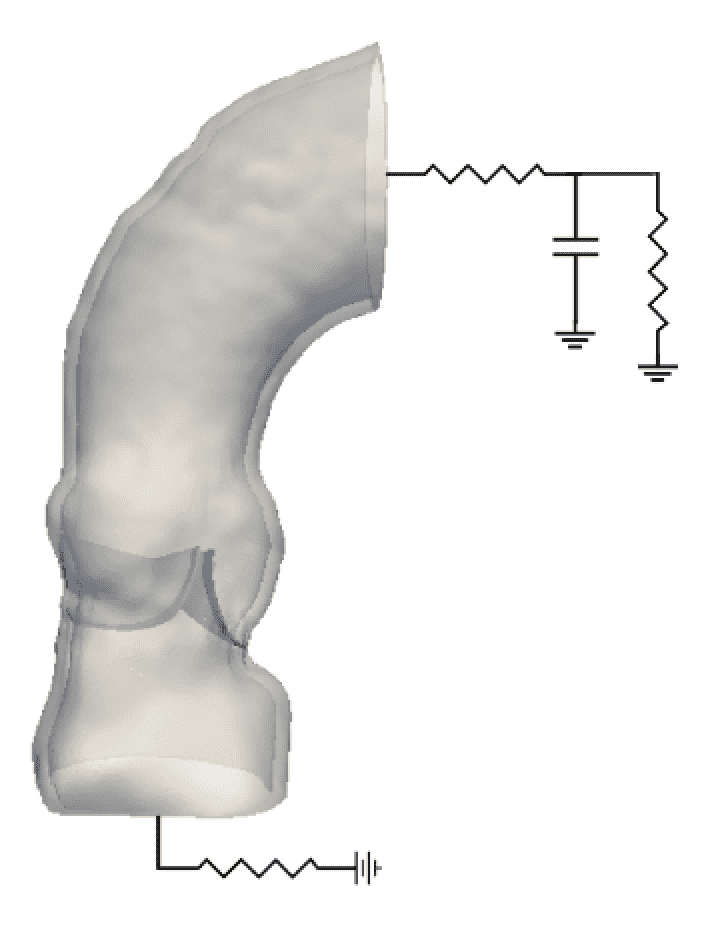}
\caption{Reduced-order models that provide driving and loading conditions.}
\label{f:Windkessel}
\end{figure*}

To drive flow through the model aortic root, we specify driving and loading conditions at the inlet and outlet using reduced-order models, as illustrated in Fig.~\ref{f:Windkessel}.
Let $\partial\Omega_\text{in}$ and $\partial\Omega_\text{out}$ indicate the inlet and outlet boundaries of the domain, and let $P_\text{in}$ and $Q_\text{in}$ (respectively, $P_\text{out}$ and $Q_\text{out}$) indicate the mean pressure and net flow rate along the inlet (respectively, outlet) boundary.
At the outlet, we couple the three-dimensional FSI model to the three-element Windkessel model of Stergiopulos et al.~\cite{Stergiopulos99} using an explicit coupling approach described previously \cite{BEGriffith09-heart_valves}.
Briefly, a first-order (Godunov) time step splitting is used to decouple the three-dimensional FSI model from the Windkessel model.
The mean flow rate through the outlet surface, $Q_\text{out}$, computed by the three-dimensional FSI model is provided as an input to the Windkessel model.
The pressure generated in the Windkessel model is used, along with the flow rate, to determine the pressure along the outlet surface, $P_\text{out}$, which is used as a boundary condition for the three-dimensional model.
See Griffith et al.~\cite{BEGriffith09-heart_valves} for further details.
In our simulations, we use the ``Type A'' parameters from Stergiopulos et al.~\cite{Stergiopulos99} fit to clinical pressure and flow rates collected by Murgo et al.~\cite{JPMurgo80}.
At the inlet, we use a simple linear resistance model to drive flow, so that
\begin{equation}
	(P_\text{LV} - P_\text{in}) = R_\text{src} Q_\text{in},
\end{equation}
with $R_\text{src} = 0.05~\text{mmHg}/(\text{ml}/\text{s})$.
This additional upstream resistance eliminates unrealistic ringing (i.e.~the \emph{waterhammer} effect) upon valve closure that otherwise occurs if $R_\text{src} = 0$.
The upstream pressure waveform $P_\text{LV}(t)$ is determined from the ``Type A'' clinical pressure and flow rates collected by Murgo et al.~\cite{JPMurgo80}.

\subsection{Numerical methods}

The Eulerian equations are discretized on a block-structured locally refined Cartesian grid following a staggered-grid discretization approach \cite{BEGriffith12-aortic_valve}.
The Lagrangian equations are discretized using standard nodal finite element methods \cite{Belytschko00}.
The vessel is discretized using first-order hexahedral ($Q^1$) elements, and the leaflets are discretized using second-order tetrahedral ($P^2$) elements.
The coupling between the Eulerian and Lagrangian discretizations is handled as described by Griffith and Luo \cite{BEGriffithXX-ibfe}, except that here, we employ a standard selective reduced integration approach \cite{Belytschko00} to treat the volumetric penalty terms present in the leaflet model.
In this method, fluid-solid coupling is mediated by replacing the singular Dirac delta function by a regularized version of the delta function, $\delta_h(x)$.
For the vessel wall, we use a three-point delta function introduced by Roma et al.~\cite{RomaPeskinBerger99}, and for the valve leaflets, we use a new ``Gaussian-like'' six-point function with improved regularity and translation invariance that also greatly reduces spurious leakage flows through IB structures \cite{YBao15}.
The integral transforms are discretized using adaptive-order Gaussian quadrature rules that ensure that the structures do not develop ``leaks'', even under very large deformations.
 
\subsection{Software infrastructure}

We use the IBAMR \cite{IBAMR-web-page} software, which is an open-source implementation of the IB method and various extensions, including the methods described herein.
IBAMR uses SAMRAI \cite{samrai-web-page, HornungKohn02} for Cartesian grid adaptive mesh refinement and PETSc \cite{petsc-web-page, petsc-user-ref, petsc-efficient} for linear and nonlinear solvers and unstructured (Lagrangian) data management.
Finite element computations in IBAMR are performed using the libMesh library \cite{libMesh-web-page, libMeshPaper}.

\section{Results}

Using this model, we perform dynamic simulations of the aortic root and ascending aorta.
The model is driven using a left ventricular pressure waveform derived from clinical measurements \cite{JPMurgo80}, and afterload is provided by a three-element Windkessel model fit to those same data \cite{Stergiopulos99}.
In our simulations, we use a fluid viscosity of $4~\text{cP}$ and mass density of $1~\text{g}/\text{cm}^\text{3}$.
We set the computational domain $\Omega$ to be a $5.5~\text{cm} \times 5.5~\text{cm} \times 11~\text{cm}$ box that is discretized using an adaptively refined Cartesian grid with an effective resolution of either $0.86~\text{mm}$ or $0.43~\text{mm}$ along with corresponding spatial discretizations of the solid models of the valve leaflets and vessel wall.
We use a fixed time step size $\dt = 2.5 \times 10^{-5}~\text{s}$, which is required to maintain stability.
Because we use an explicit time stepping scheme to couple the Eulerian and Lagrangian variables, this time step size is primarily determined by the high stiffness of the immersed elastic structures.
This time step size restriction can be relaxed by using an implicit time stepping scheme, and at least for simple material models, it is possible to develop unconditionally stable implicit IB time stepping schemes \cite{Newren07}.
The development of efficient solvers for such formulations is challenging \cite{RDGuy15-gmgiib, APSBhallaXX-scalable_implicit}, however, and in practice, we typically use explicit solvers.

\subsection{Representative simulation results}

\begin{figure}
\centering
\includegraphics[width=0.235\textwidth]{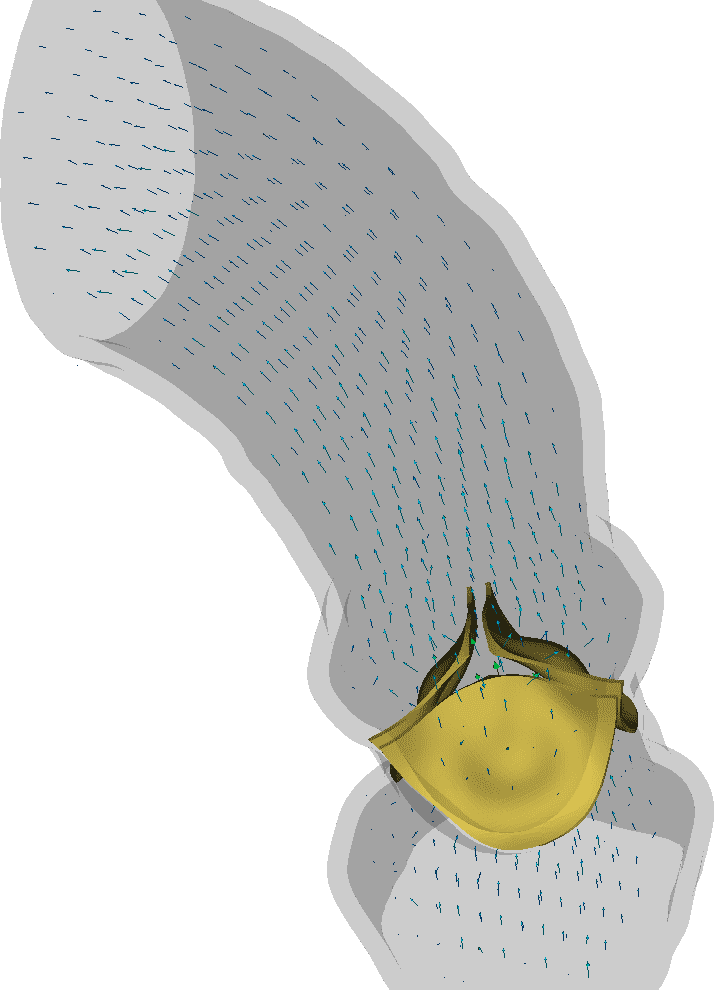} \ \
\includegraphics[width=0.235\textwidth]{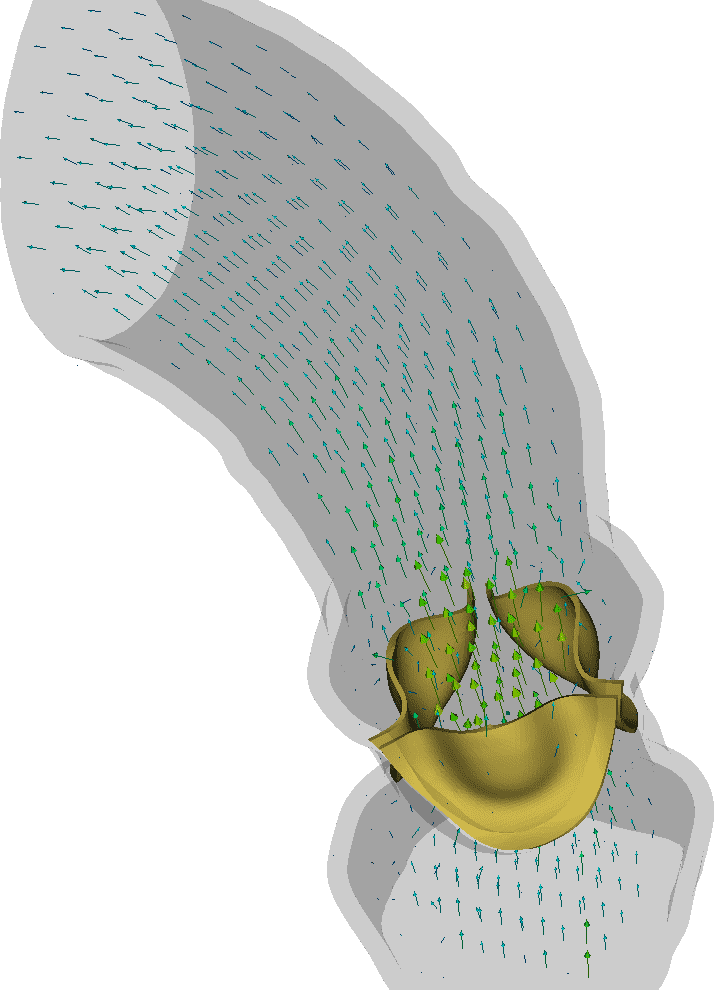} \ \
\includegraphics[width=0.235\textwidth]{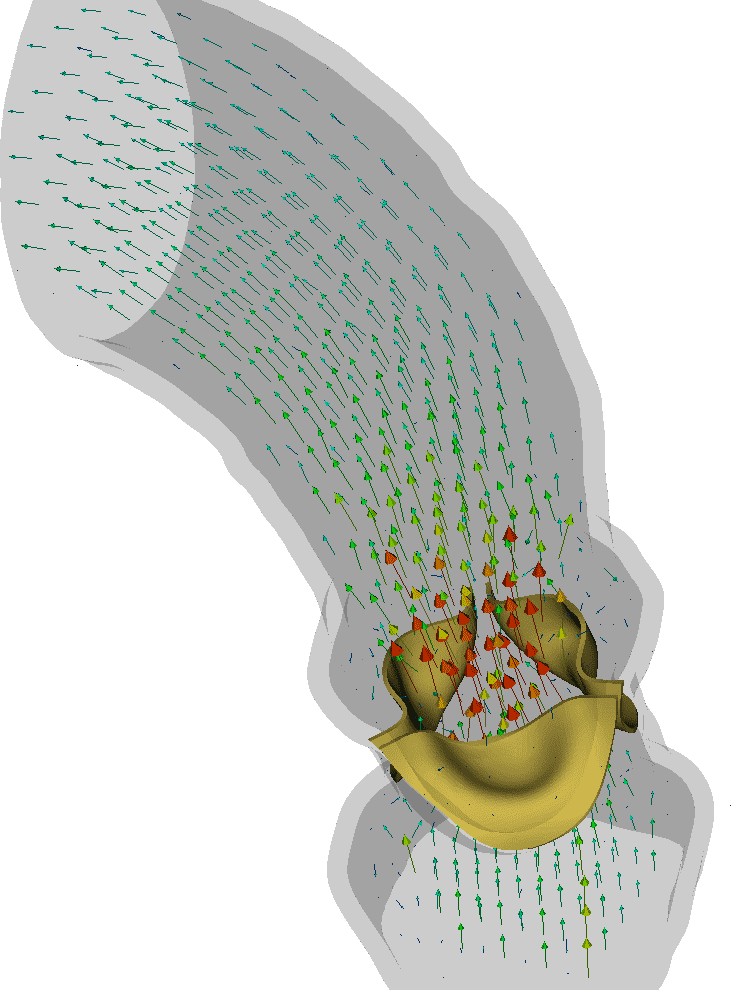} \ \
\includegraphics[width=0.235\textwidth]{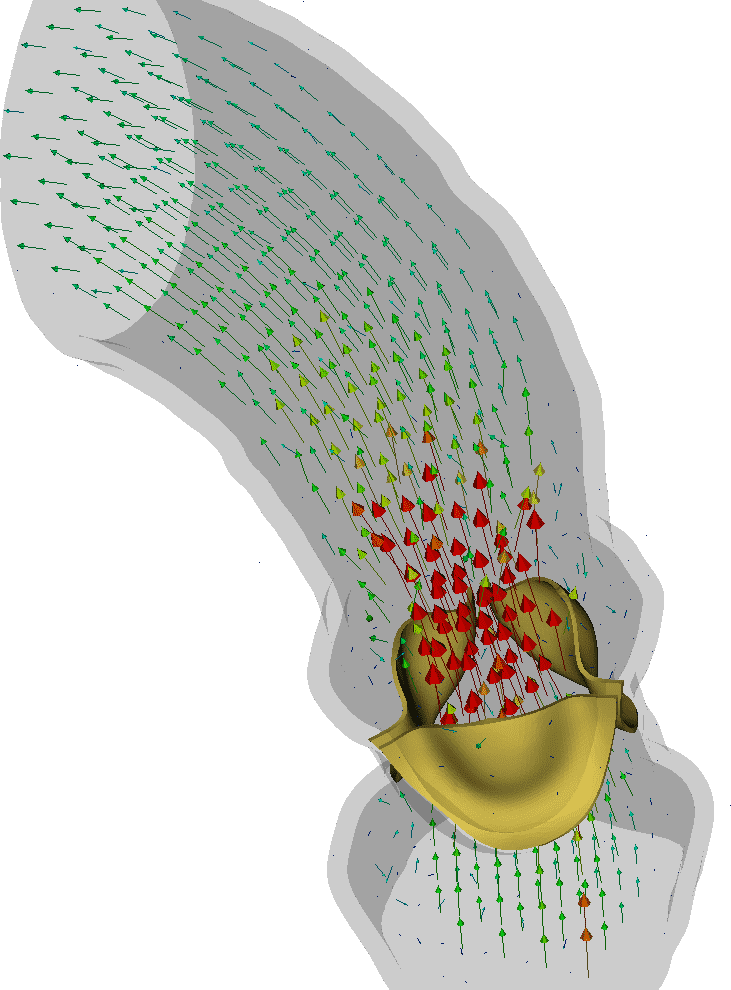}
\caption{Representative valve opening dynamics, showing the leaflet kinematics and fluid dynamics.  Peak flow velocities during early systole, highlighted in red, are approximately 2 m/s.}
\label{f:leaflet_flow}
\end{figure}

\begin{figure}
\centering
\sidesubfloat[][]{
	\includegraphics[width=0.195\textwidth]{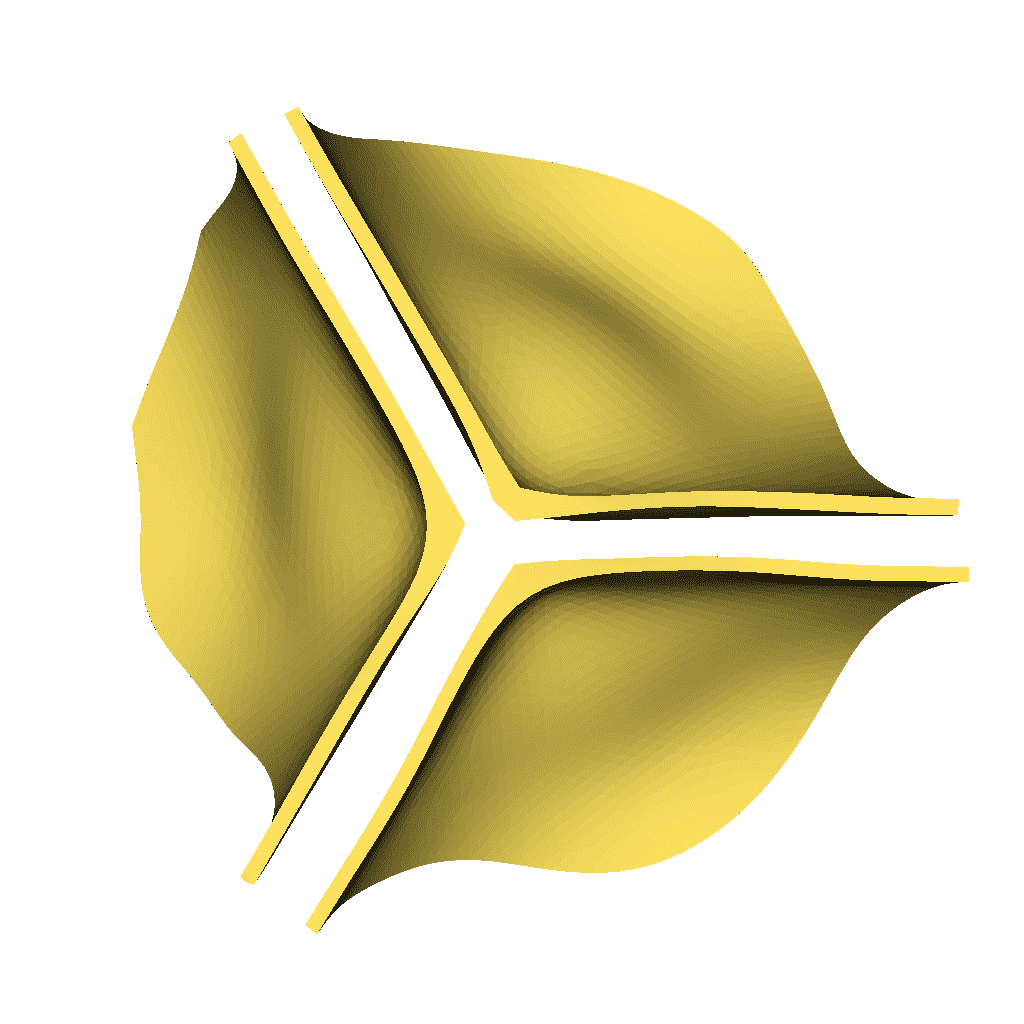}
	\includegraphics[width=0.195\textwidth]{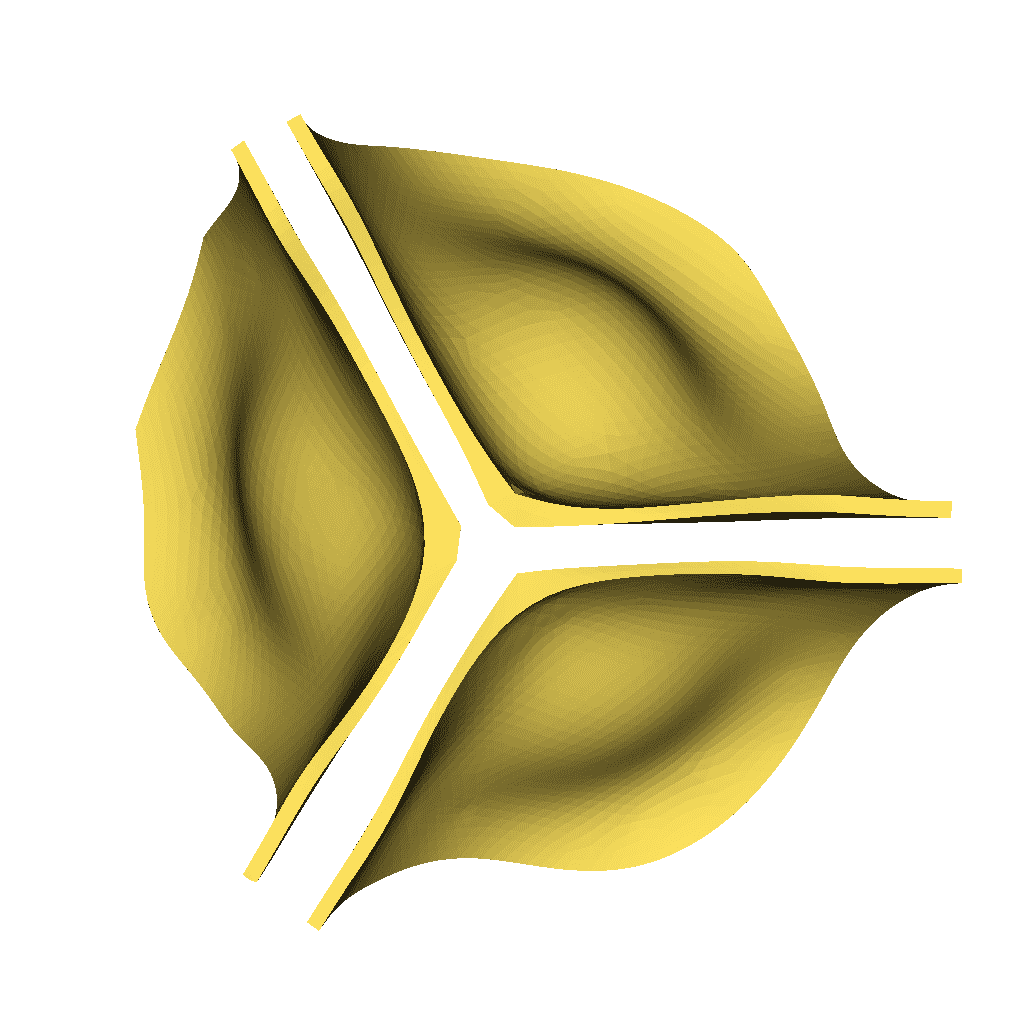}
	\includegraphics[width=0.195\textwidth]{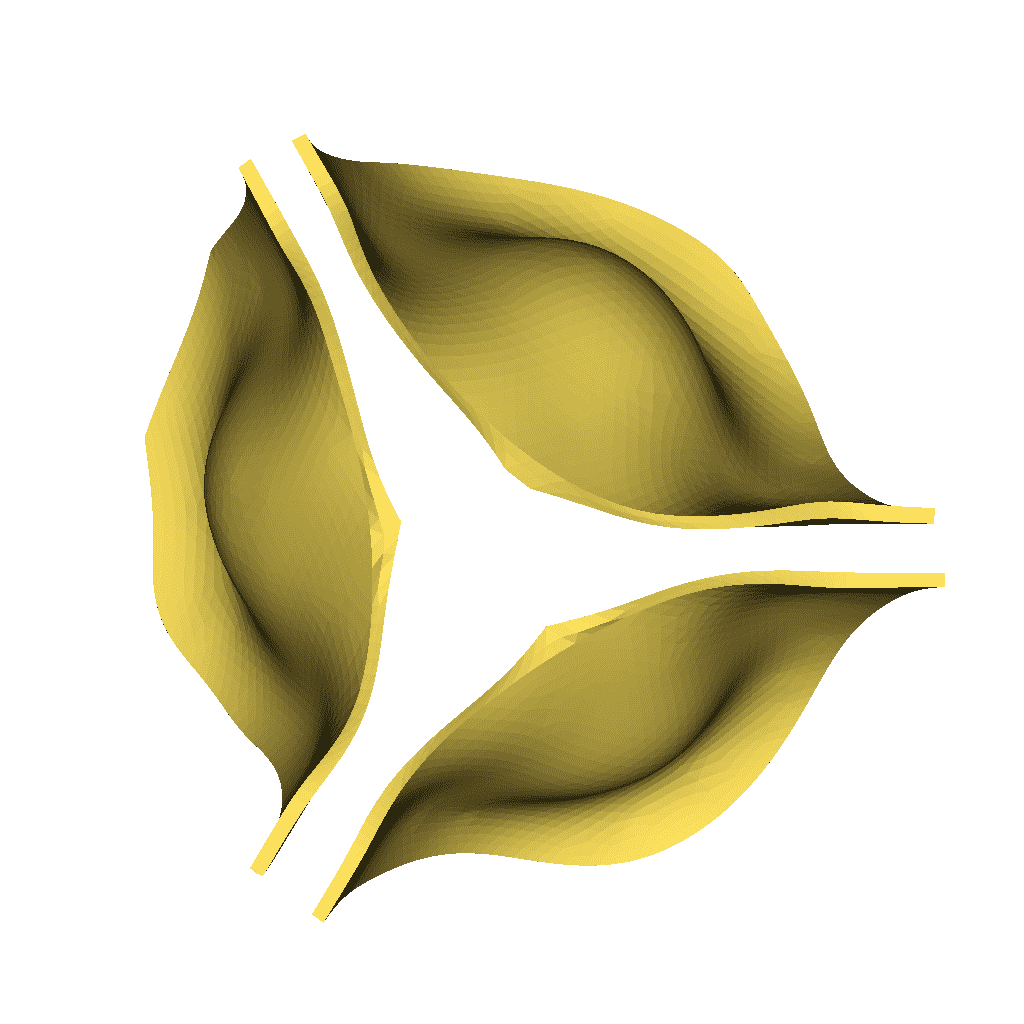}
	\includegraphics[width=0.195\textwidth]{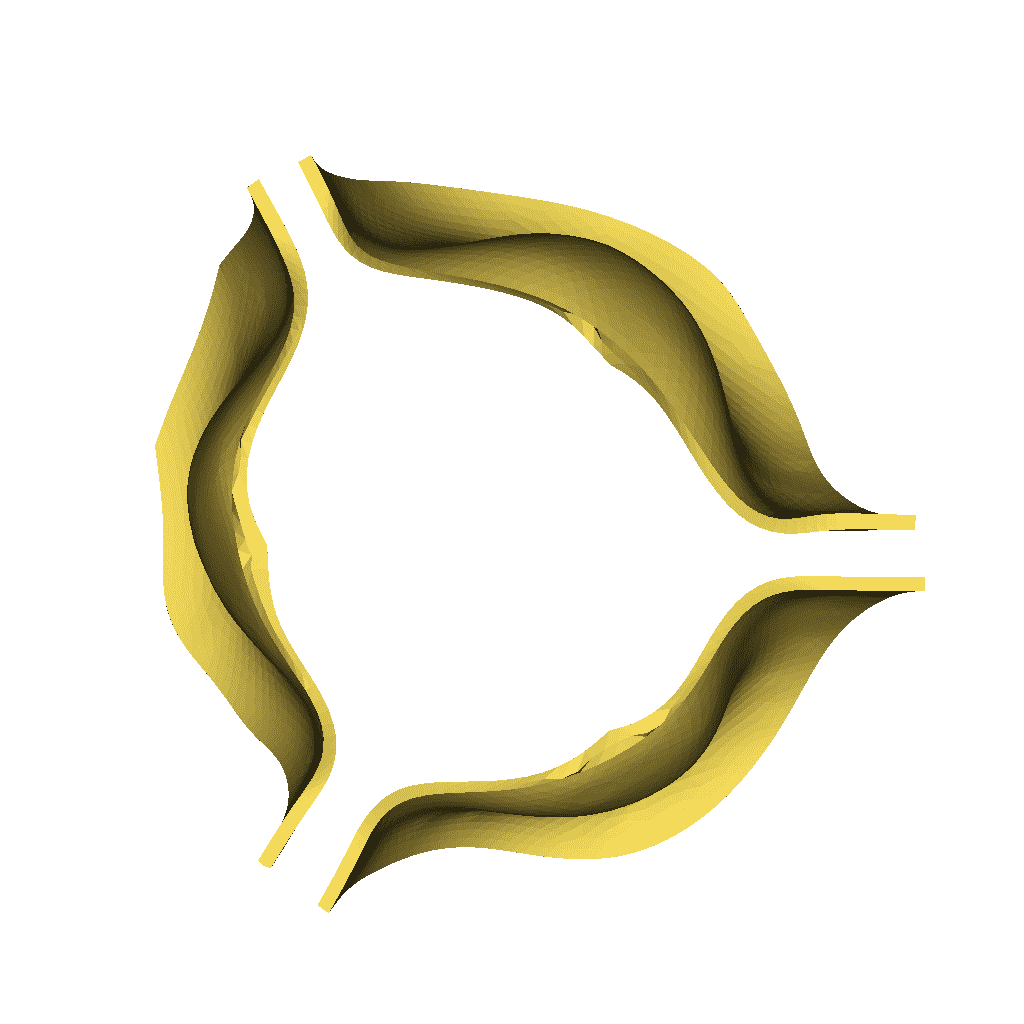}
	\includegraphics[width=0.195\textwidth]{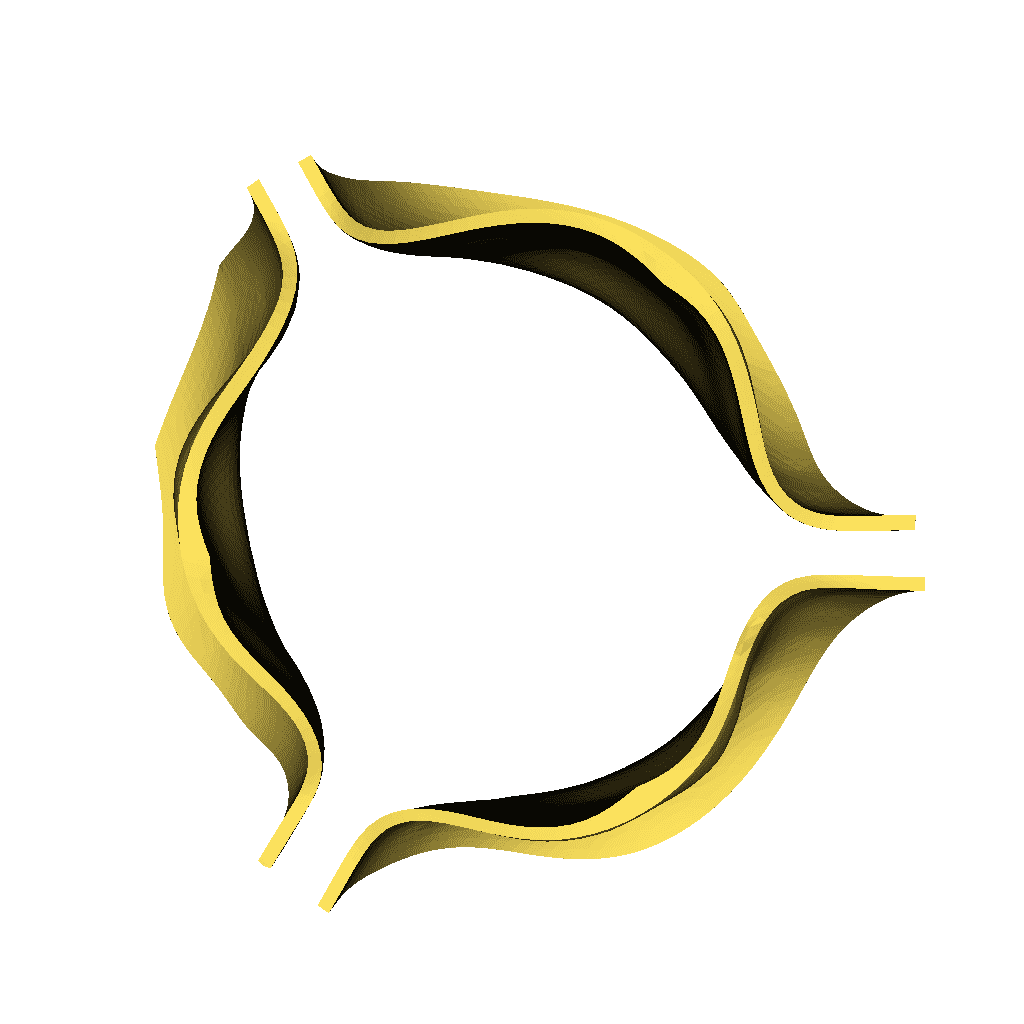}
}

\sidesubfloat[][]{
	\includegraphics[width=0.195\textwidth]{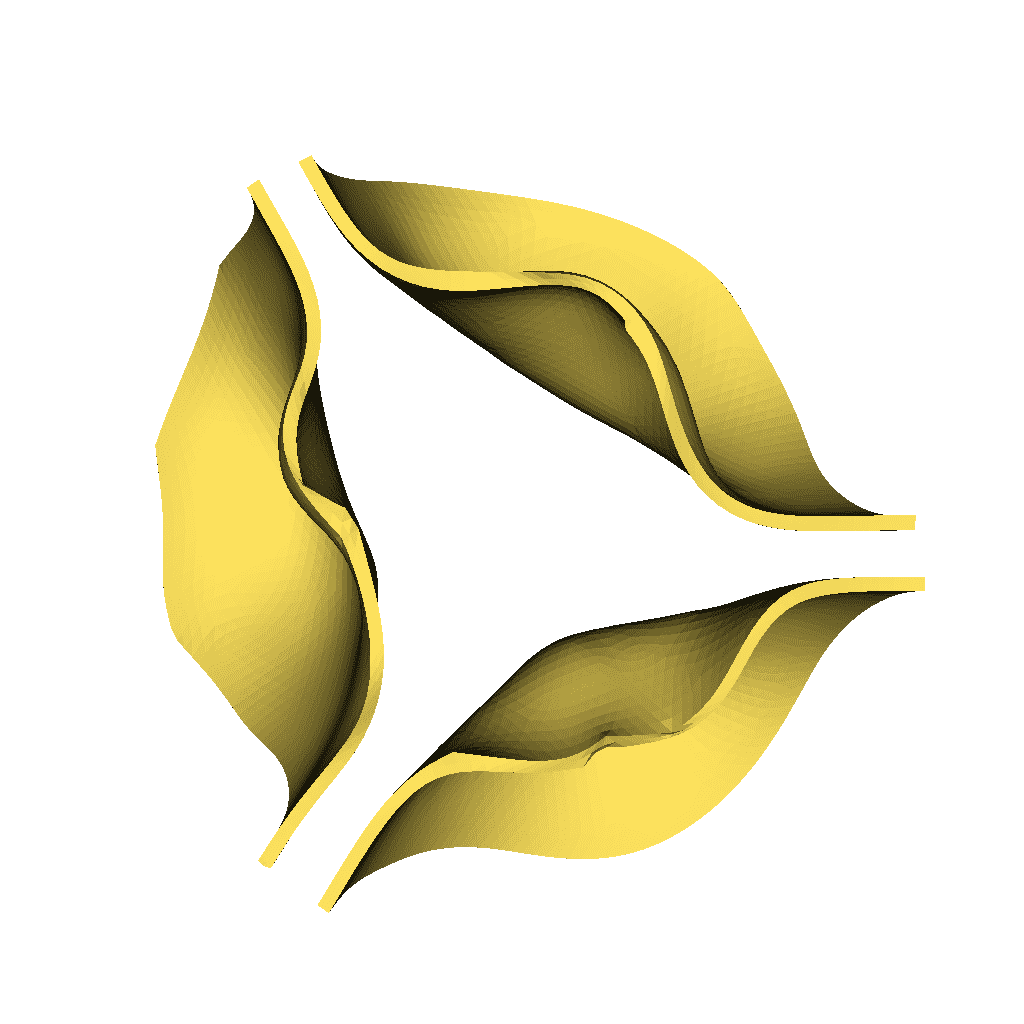}
	\includegraphics[width=0.195\textwidth]{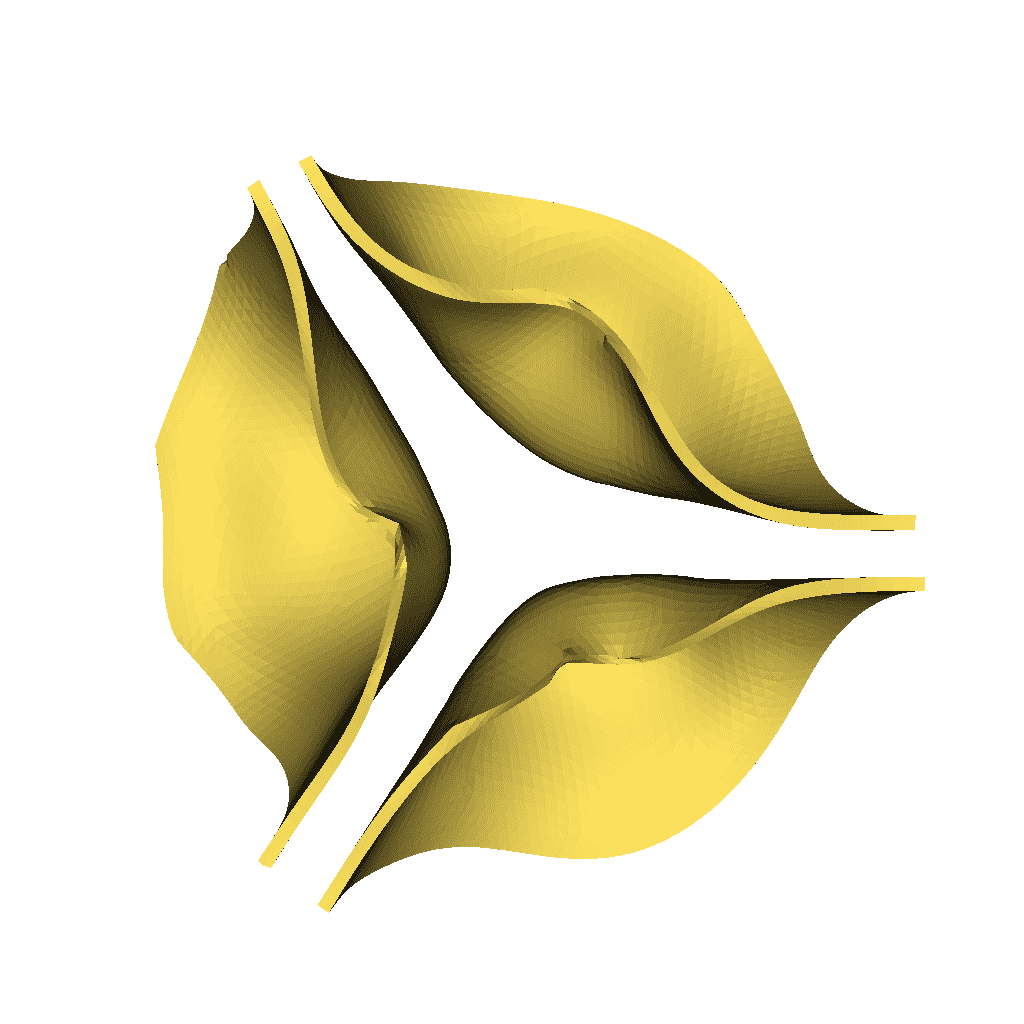}
	\includegraphics[width=0.195\textwidth]{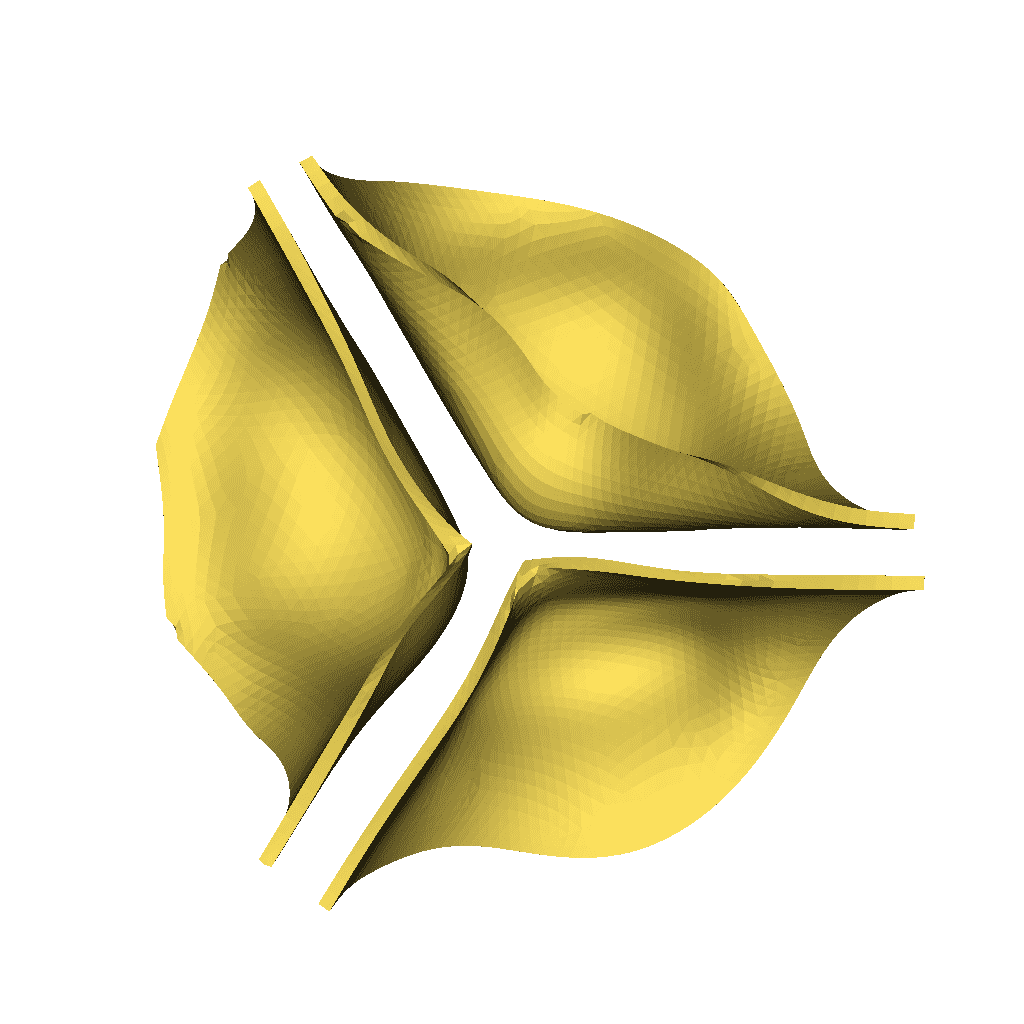}
	\includegraphics[width=0.195\textwidth]{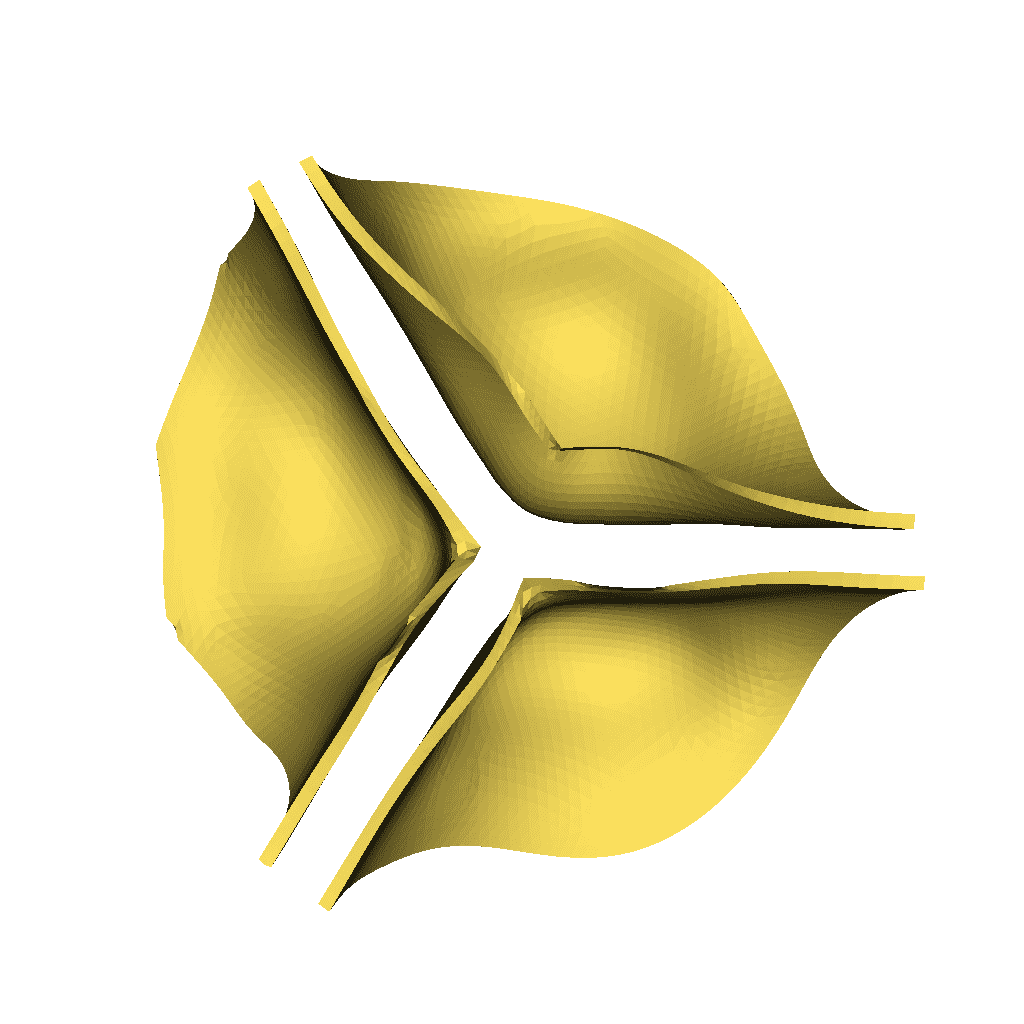}
	\includegraphics[width=0.195\textwidth]{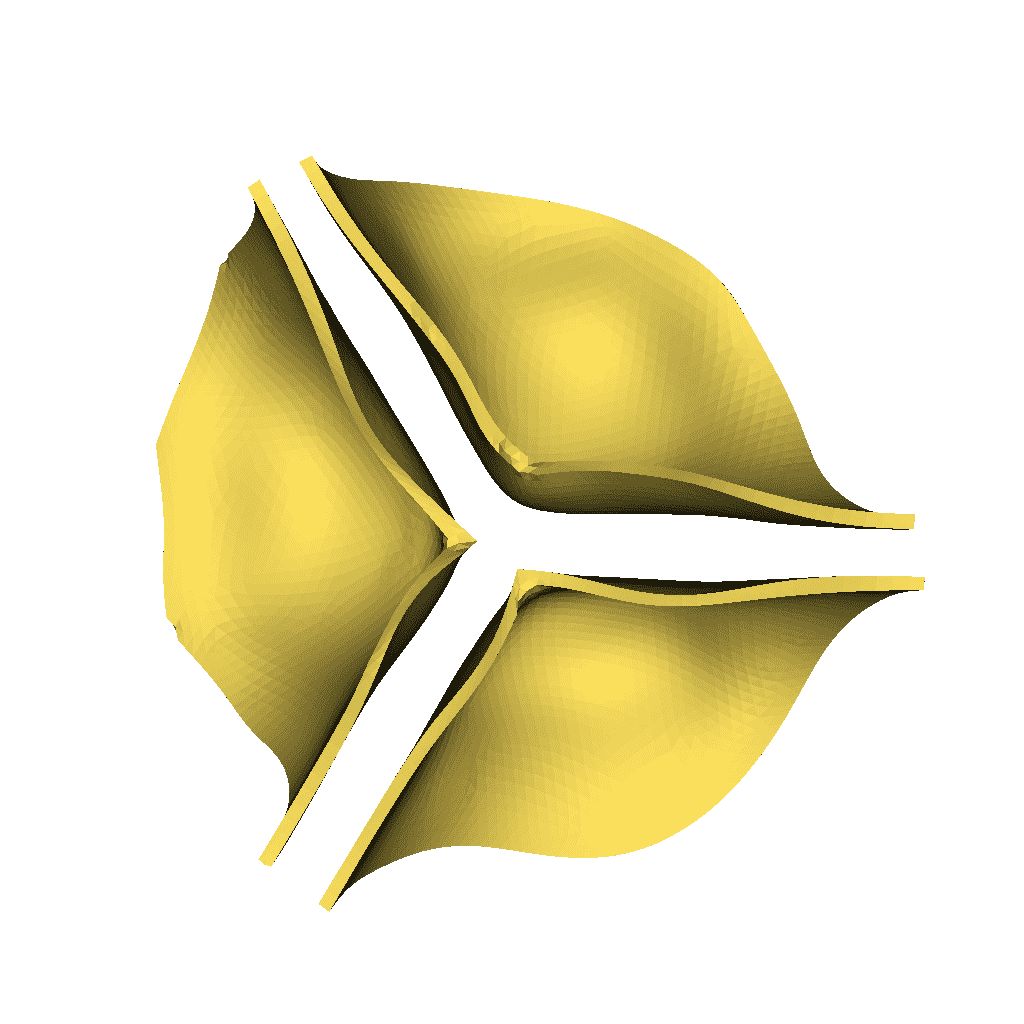}
}
\caption{
	Representative valve kinematics during (a)~opening and (b)~closure, shown at equally spaced intervals.
	The gap between the closed valve leaflets reflects the thickness of the regularized delta function.
	Despite this residual gap, the valve is sealed with respect to the fluid, as demonstrated in Fig.~\ref{f:comparison}, which shows that the closed valve supports a physiological pressure load without leak.
}
\label{f:leaflet_kinematics}	
\end{figure}

\begin{figure}
\centering
\includegraphics[width=0.475\textwidth]{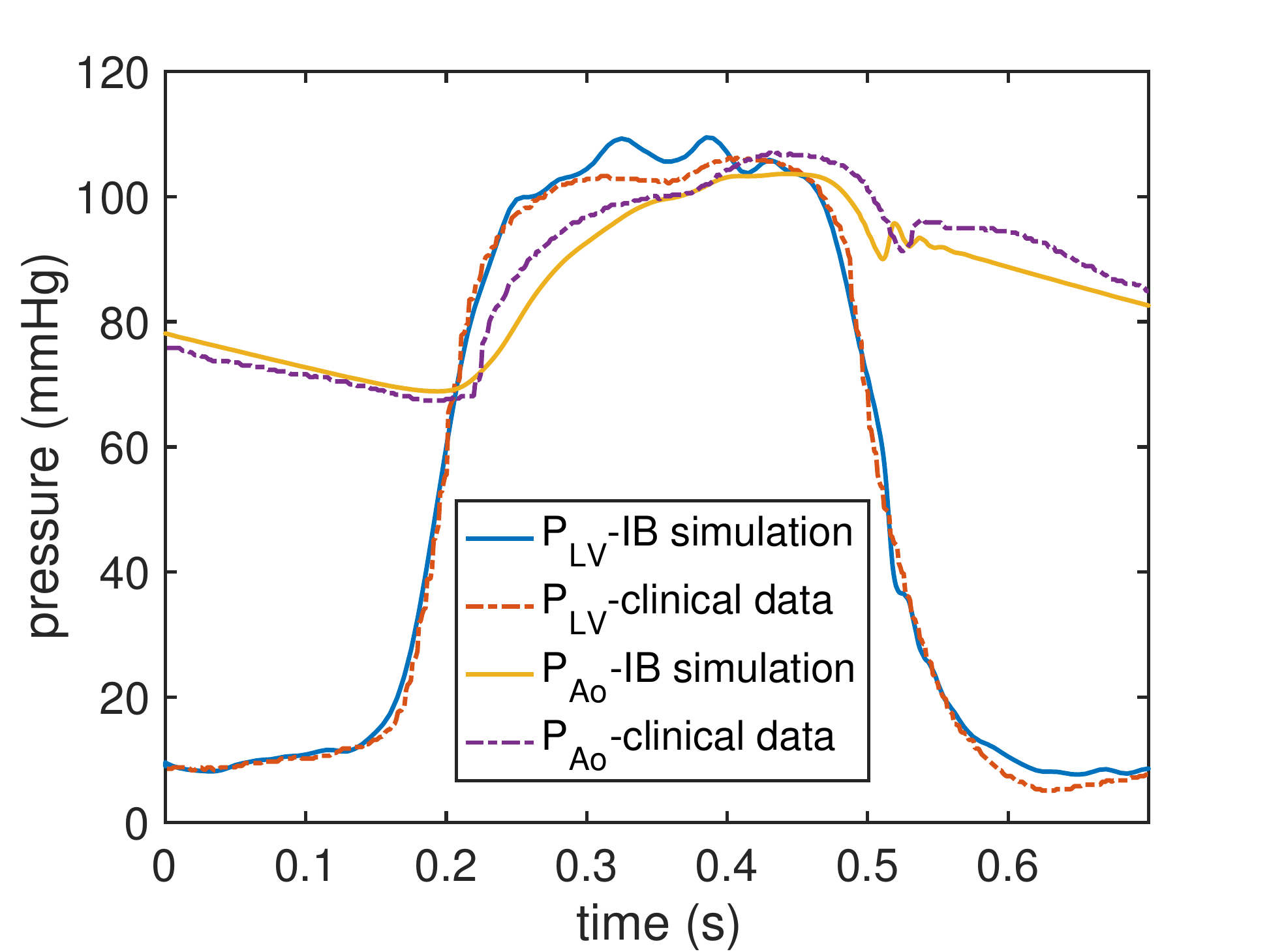}
\includegraphics[width=0.475\textwidth]{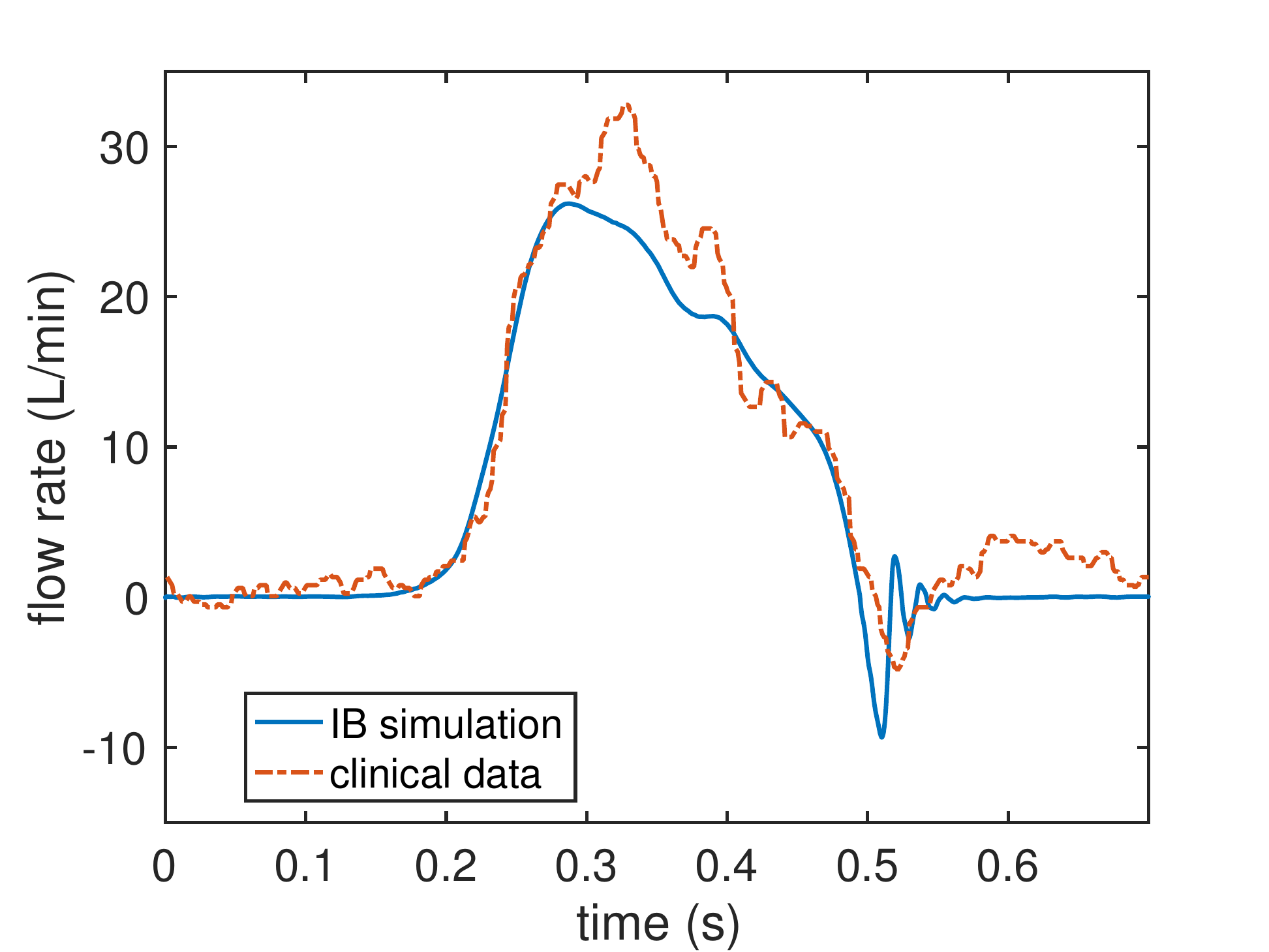}
\caption{Comparison of simulated and clinical left ventricular and aortic pressures (left) and flow rates (right).
Note that all computed pressures and flow rates shown here are determined by the model, and are not directly imposed as boundary conditions.
The computational and clinical results are in generally good agreement, although there are some discrepancies in the aortic pressures and flow rates; see text for further discussion.
}
\label{f:comparison}
\end{figure}

We first demonstrate representative results obtained using the model with fresh porcine valve leaflet parameters of Driessen et al.~\cite{NJBDriessen05}.
Specifically, we use $c_1 = 10~\text{kPa}$, $k_1 = 0.7~\text{kPa}$, and $k_2 = 9.9$ along with a fiber angle standard deviation of $10.7^\circ$.
In these computations, we use the coarser Cartesian grid spacing of $0.86~\text{mm}$ along with a corresponding discretization of the solid model components.
Fig.~\ref{f:leaflet_flow} shows leaflet kinematics and flow patterns in early systole, as the valve begins to open.
Fig.~\ref{f:leaflet_kinematics} provides additional details of the leaflet kinematics during valve opening and closure.
Fig.~\ref{f:comparison} compares the bulk hemodynamics generated by the model to the clinical data used to fit the Windkessel model.
Notice that the closed valve supports a realistic pressure load during diastole without leak.
There is some discrepancy between the simulated and measured aortic pressures.
The aortic pressure reported from the simulation is obtained as the mean pressure along the outflow boundary $\partial\Omega_\text{out}$, whereas the clinical pressure measurement was obtained via catheterization using a pressure probe.
Because of the high Reynolds number characteristic of flow in the ascending aorta, we would expect to see relatively large differences in the pressure waveform depending on the placement of the pressure sensor.
Similarly, the flow rate reported from the simulation is obtained as the net flow rate along the outflow boundary, whereas the clinical value was calculated from flow velocity measurements obtained via catheterization \cite{JPMurgo80}.
Because the upstream driving pressure essentially acts as a pressure source in our model, we do not expect to obtain identical flow rates.
(Notice, for instance, that fluctuations in the clinical data indicate nontrivial forward flow during diastole.)
Nonetheless, the pressures and flow rates generated by the model are in generally good agreement with the corresponding clinical values.
Stroke volume in the simulation is $79~\text{ml}$, which is somewhat lower than the clinical stroke volume of $90~\text{ml}$ but still in the physiological range.

\subsection{Effect of grid resolution on leaflet mechanics}

\begin{figure}
\centering
\sidesubfloat[][]{
	\includegraphics[height=0.195\textwidth]{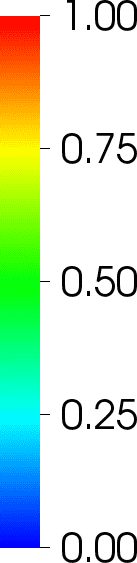} \
	\includegraphics[width=0.195\textwidth]{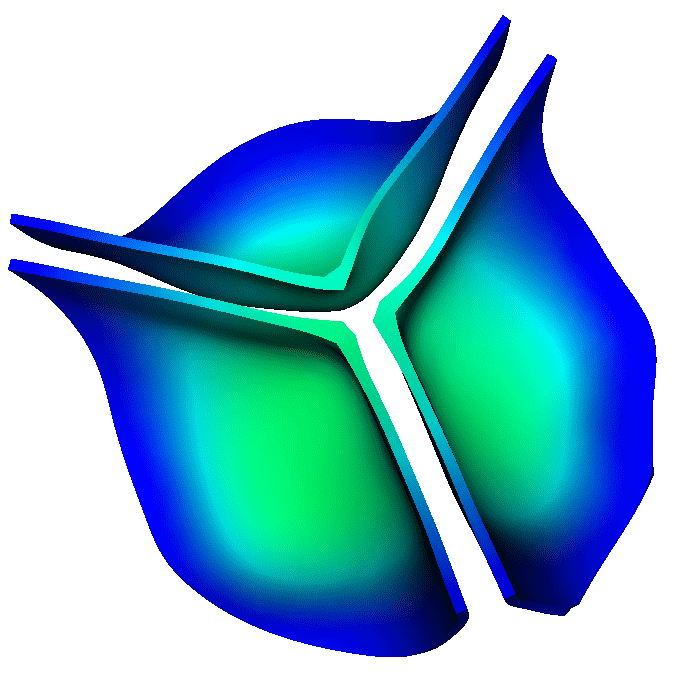}
	\includegraphics[width=0.195\textwidth]{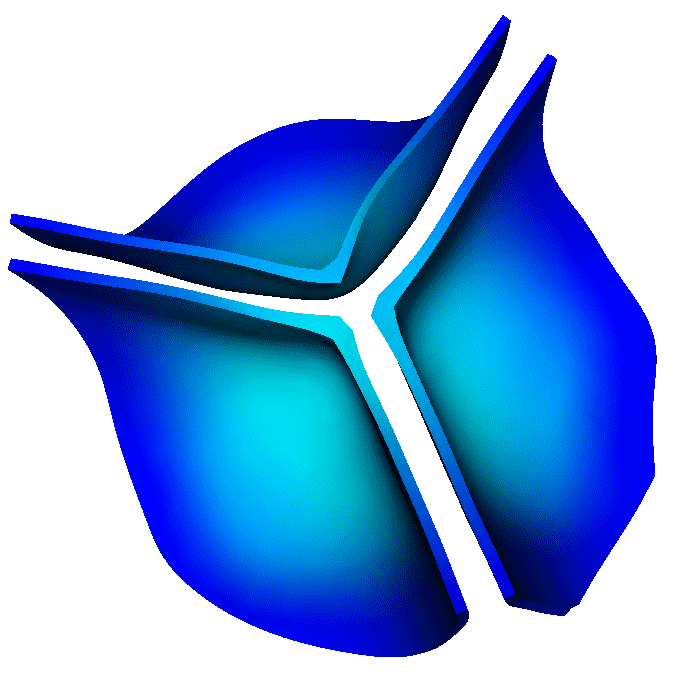}
	\includegraphics[width=0.195\textwidth]{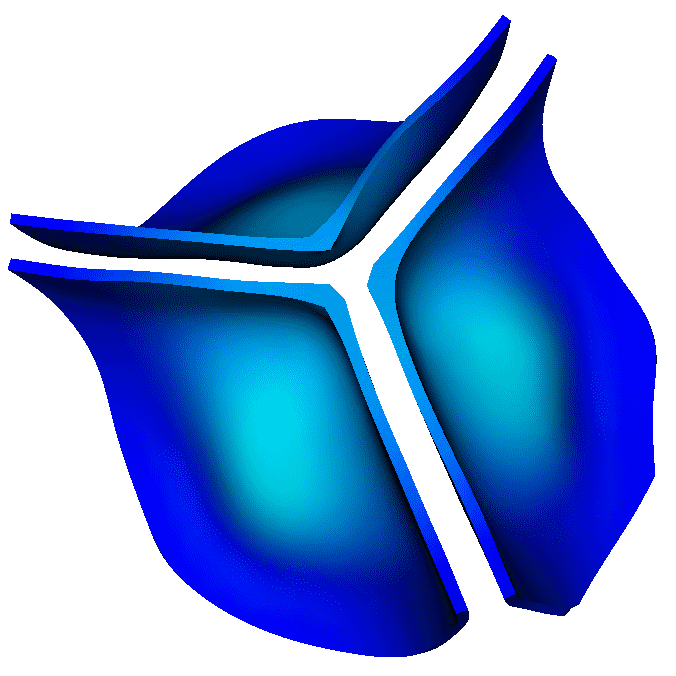}
	\includegraphics[width=0.195\textwidth]{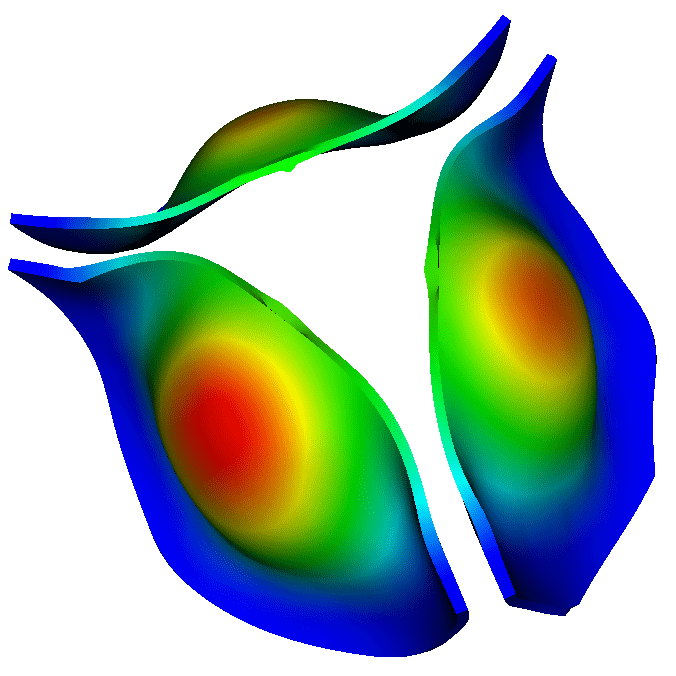}
	\includegraphics[width=0.195\textwidth]{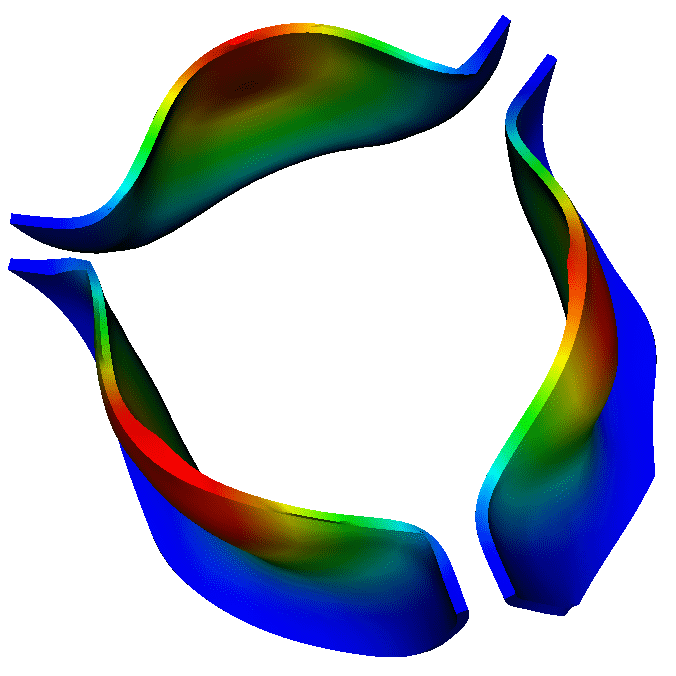}
}

\sidesubfloat[][]{
	\includegraphics[height=0.195\textwidth]{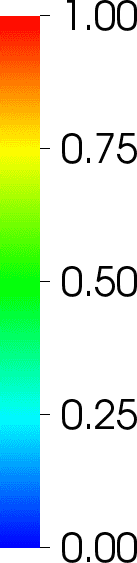} \
	\includegraphics[width=0.195\textwidth]{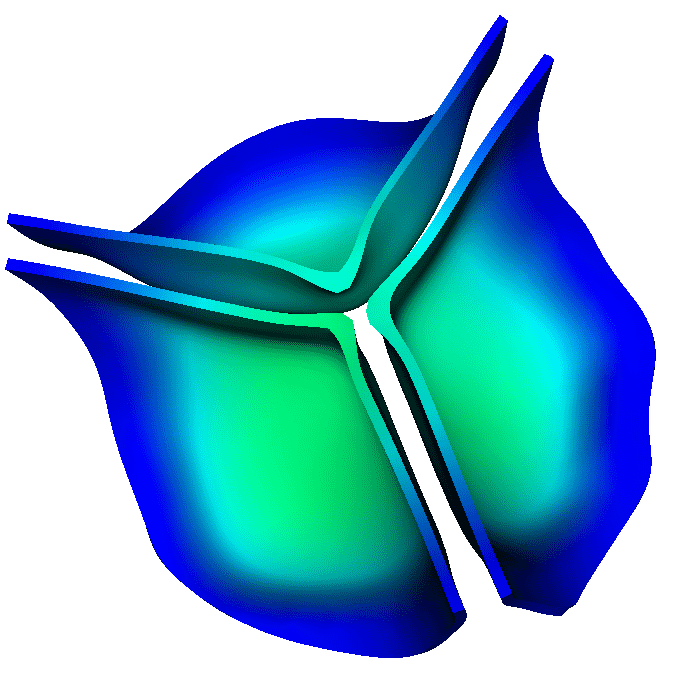}
	\includegraphics[width=0.195\textwidth]{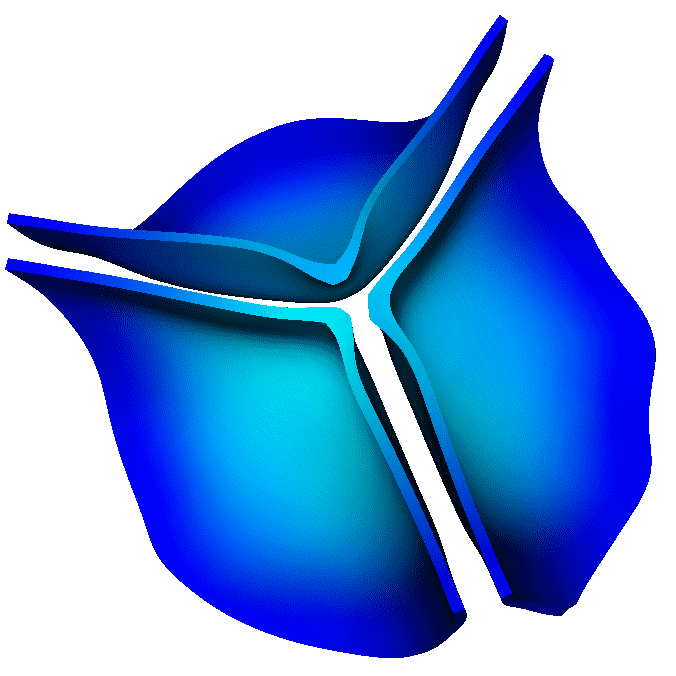}
	\includegraphics[width=0.195\textwidth]{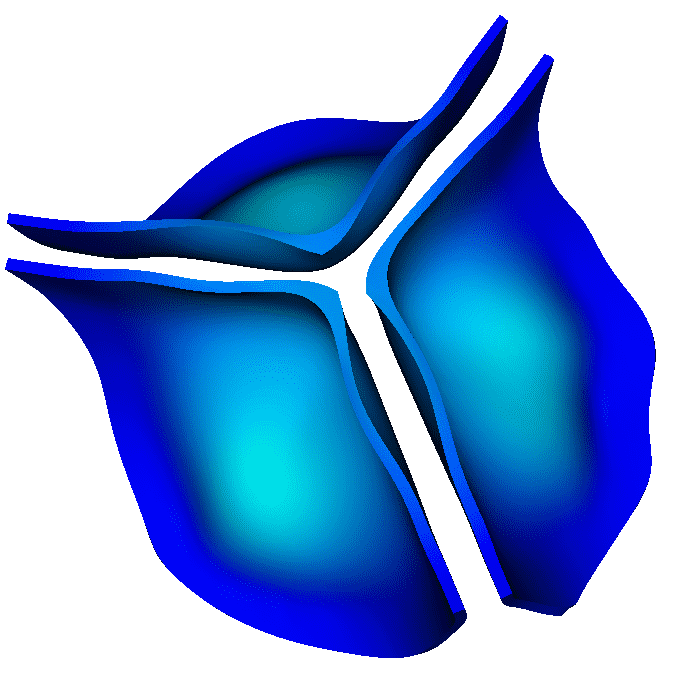}
	\includegraphics[width=0.195\textwidth]{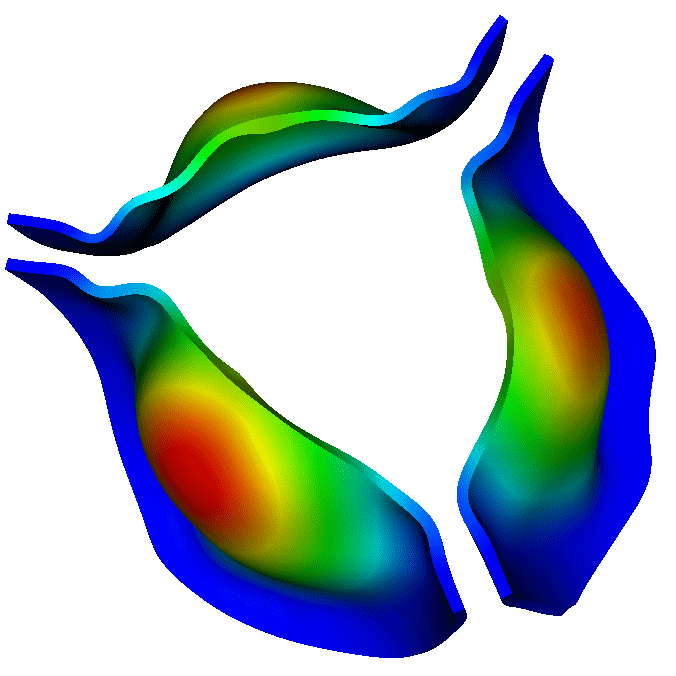}
	\includegraphics[width=0.195\textwidth]{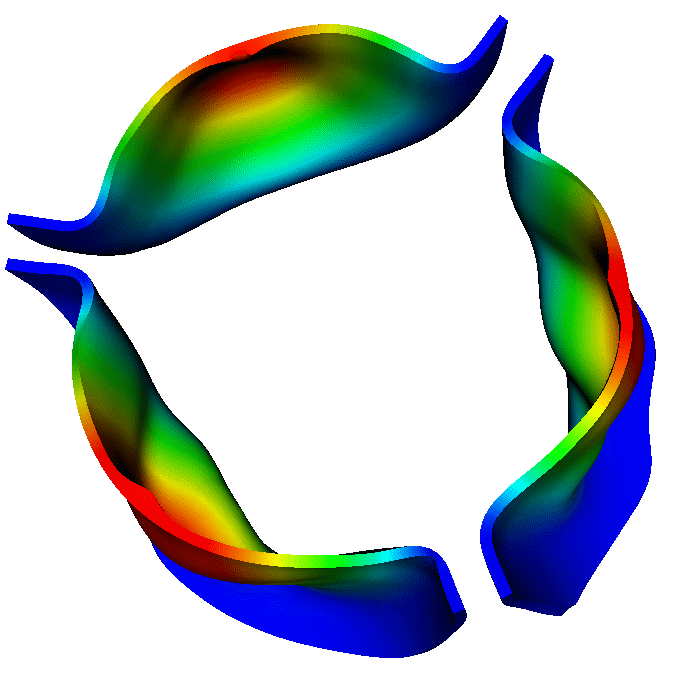}
}
\caption{Leaflet displacements (cm) obtained during diastole and early systole, using fresh porcine valve constitutive parameters.
Panel (a)~shows results obtained using a relatively coarse Cartesian grid spacing of $0.86~\text{mm}$, and panel (b) shows results obtained using a relatively fine spacing of $0.43~\text{mm}$.
The displacements are generally in good agreement.
See also Fig.~\ref{f:2D_slice_comparison} for details of the deformations and displacements of the center surface of the left coronary leaflet.
}
\label{f:dX_comparison}
\end{figure}

\begin{figure}
\centering
\sidesubfloat[][]{
	\includegraphics[height=0.195\textwidth]{aortic_root/native/N=128/dX/colorbar} \
	\includegraphics[width=0.195\textwidth]{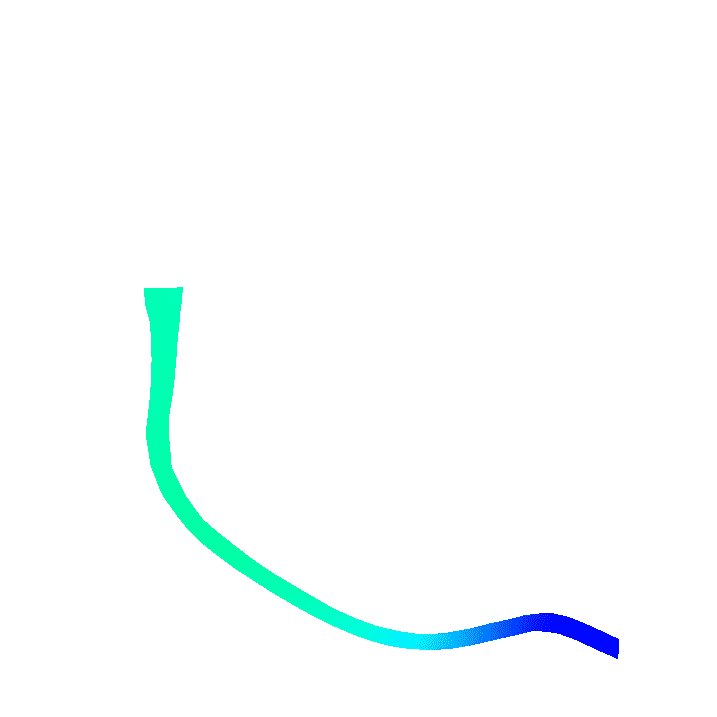}
	\includegraphics[width=0.195\textwidth]{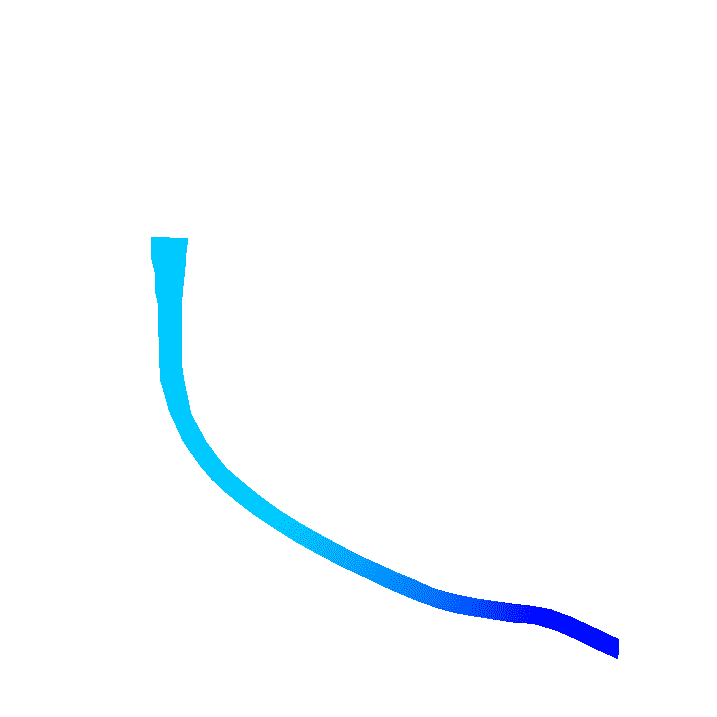}
	\includegraphics[width=0.195\textwidth]{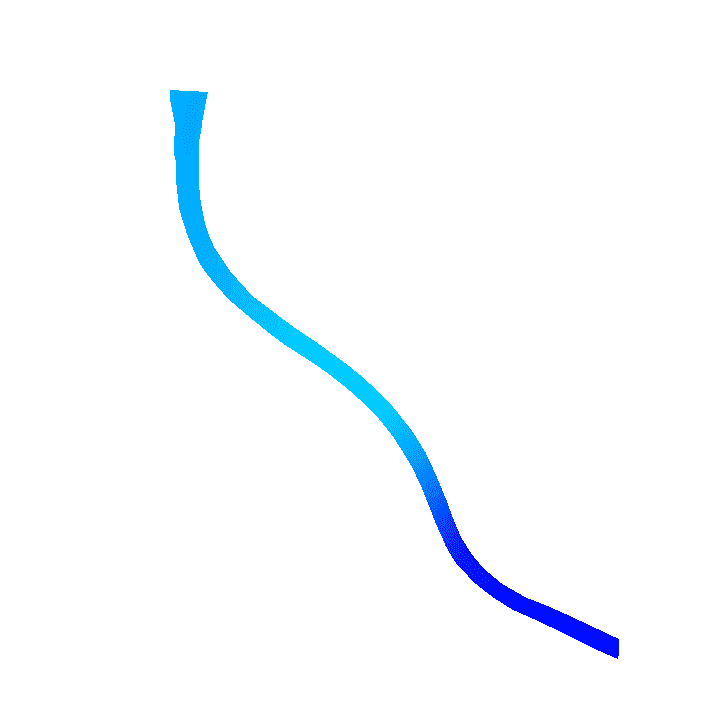}
	\includegraphics[width=0.195\textwidth]{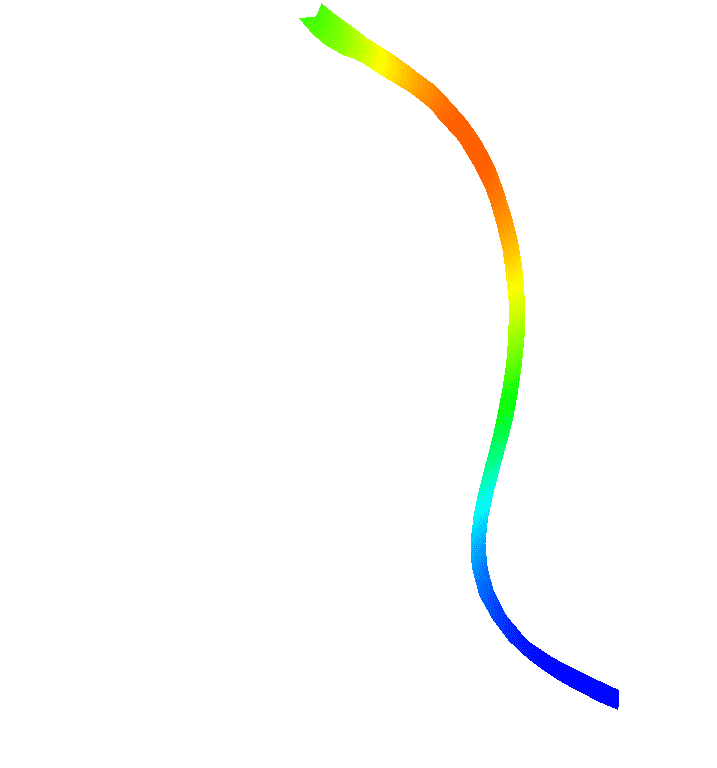}
	\includegraphics[width=0.195\textwidth]{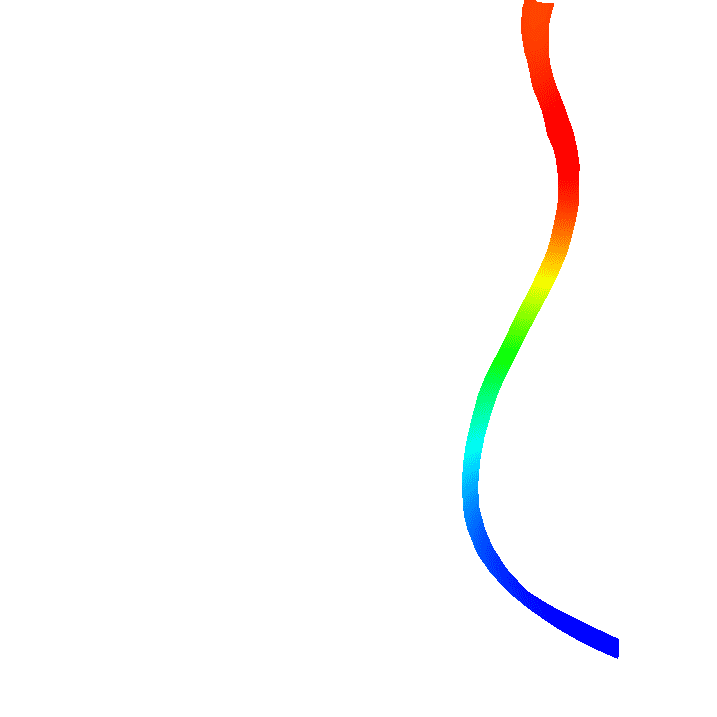}
}

\sidesubfloat[][]{
	\includegraphics[height=0.195\textwidth]{aortic_root/native/N=128/dX/colorbar} \
	\includegraphics[width=0.195\textwidth]{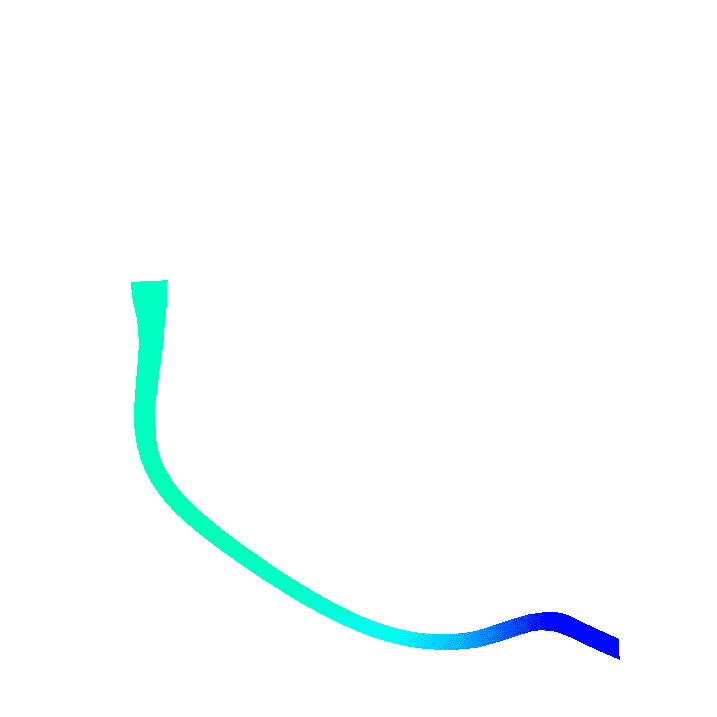}
	\includegraphics[width=0.195\textwidth]{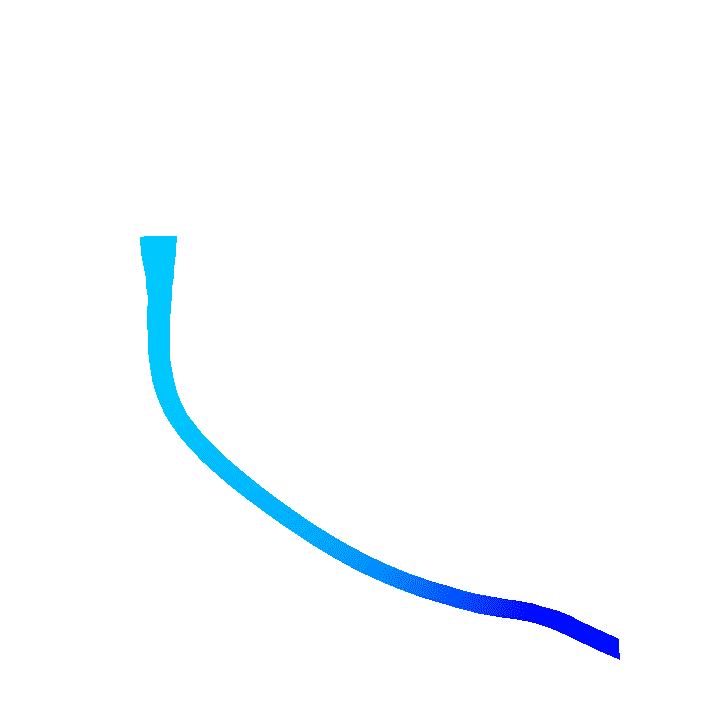}
	\includegraphics[width=0.195\textwidth]{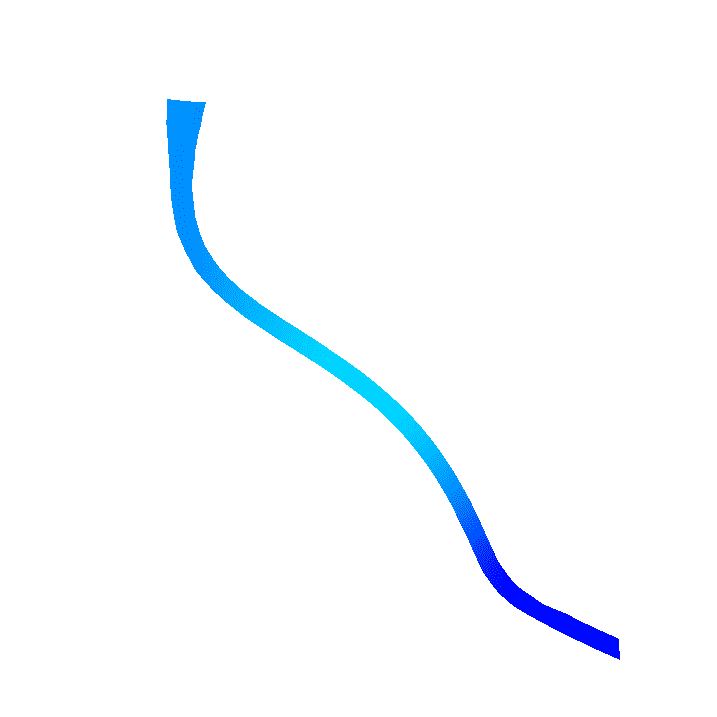}
	\includegraphics[width=0.195\textwidth]{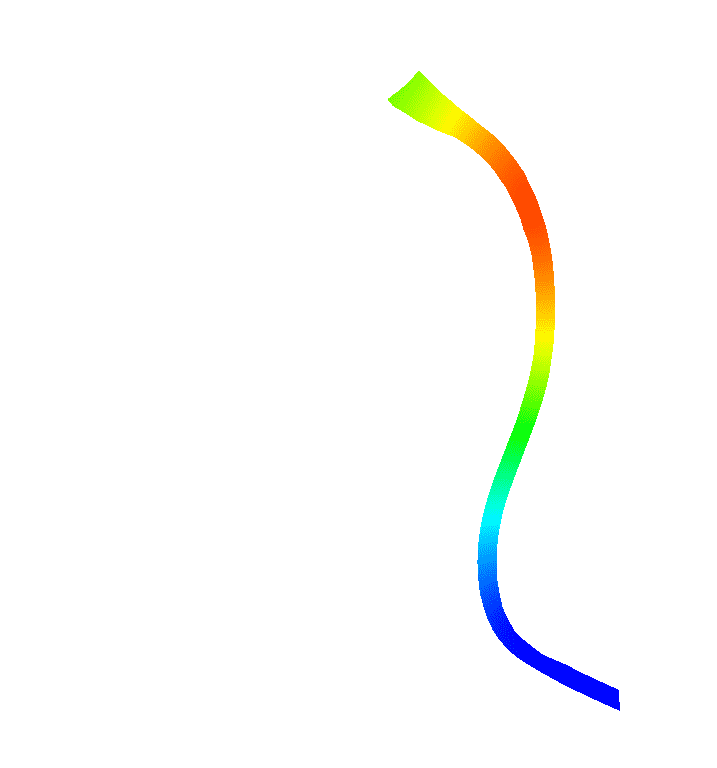}
	\includegraphics[width=0.195\textwidth]{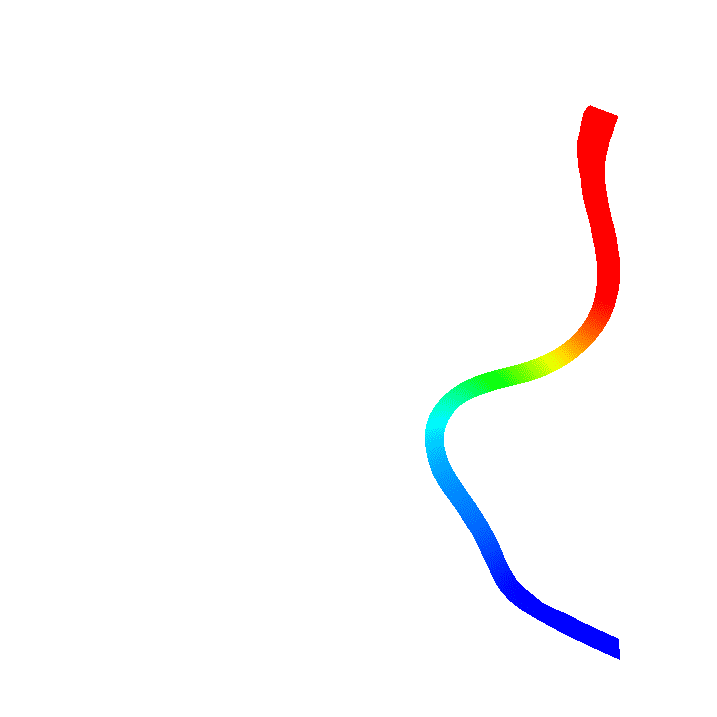}
}
\caption{Leaflet displacements (cm) along the center surface of the left coronary leaflet obtained during late diastole and early systole, using fresh porcine valve constitutive parameters.
Panel (a)~shows results obtained using a relatively coarse Cartesian grid spacing of $0.86~\text{mm}$, and panel (b) shows results obtained using a relatively fine spacing of $0.43~\text{mm}$.
The displacements are generally in good agreement, although some discrepancies are clearly observed once the valve is fully open.
We expect that these discrepancies arise from under-resolving the quasi-turbulent flow field.
See also Fig.~\ref{f:dX_comparison}.
}
\label{f:2D_slice_comparison}
\end{figure}

\begin{figure}
\centering
\sidesubfloat[][]{
	\includegraphics[height=0.195\textwidth]{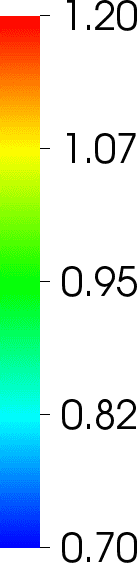} \
	\includegraphics[width=0.195\textwidth]{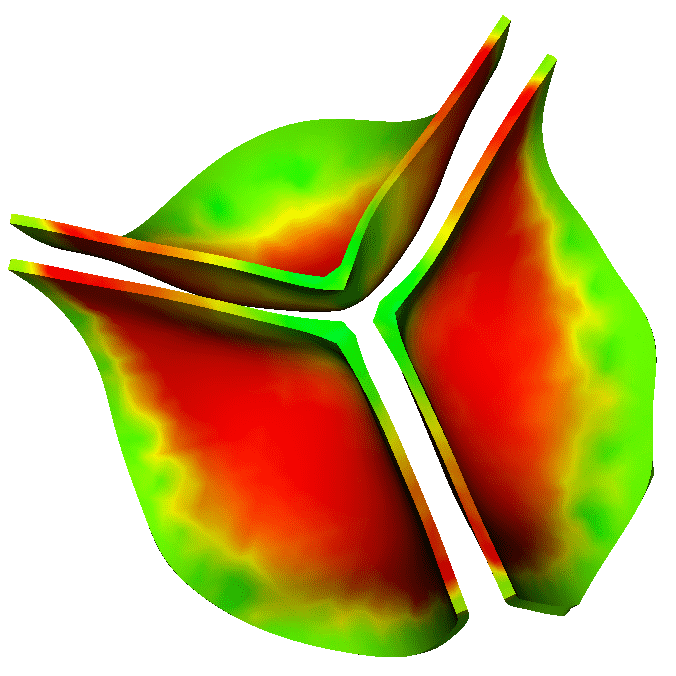}
	\includegraphics[width=0.195\textwidth]{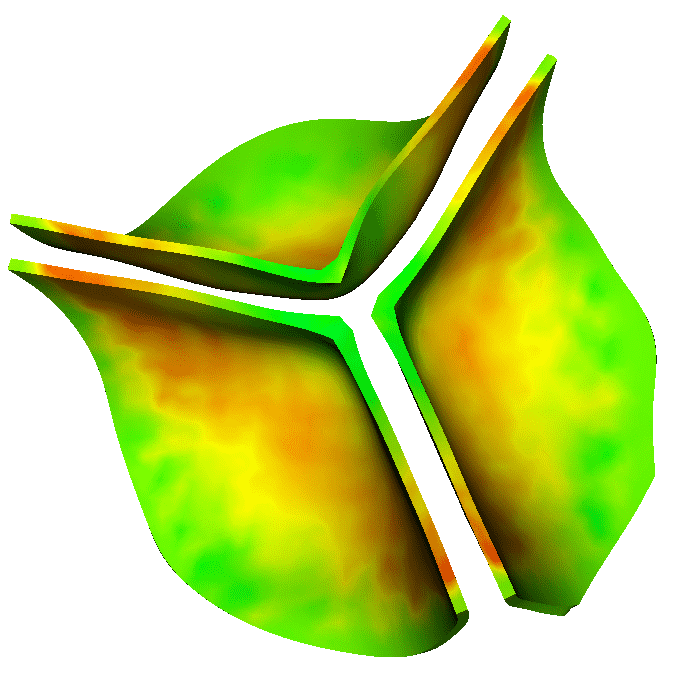}
	\includegraphics[width=0.195\textwidth]{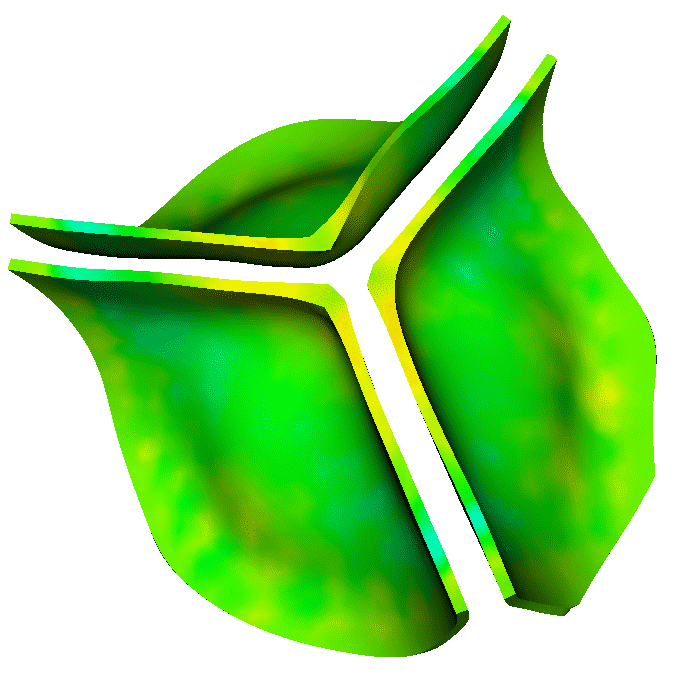}
	\includegraphics[width=0.195\textwidth]{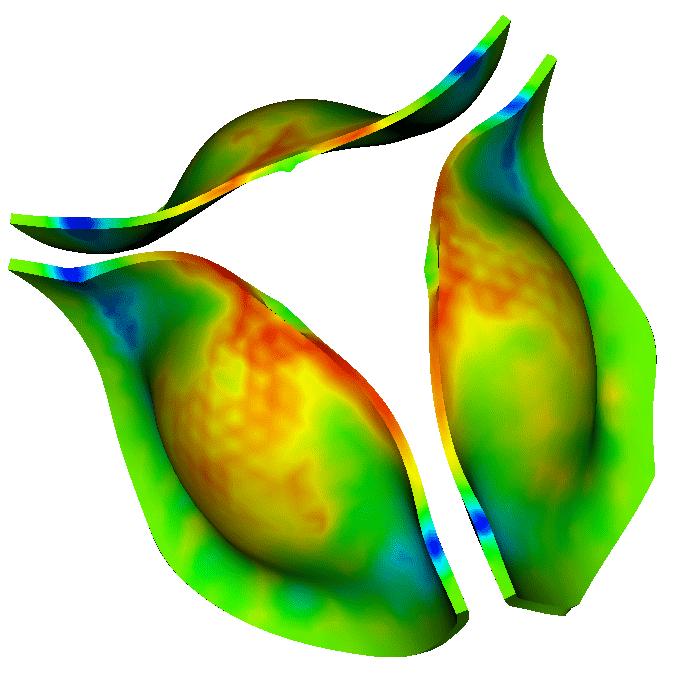}
	\includegraphics[width=0.195\textwidth]{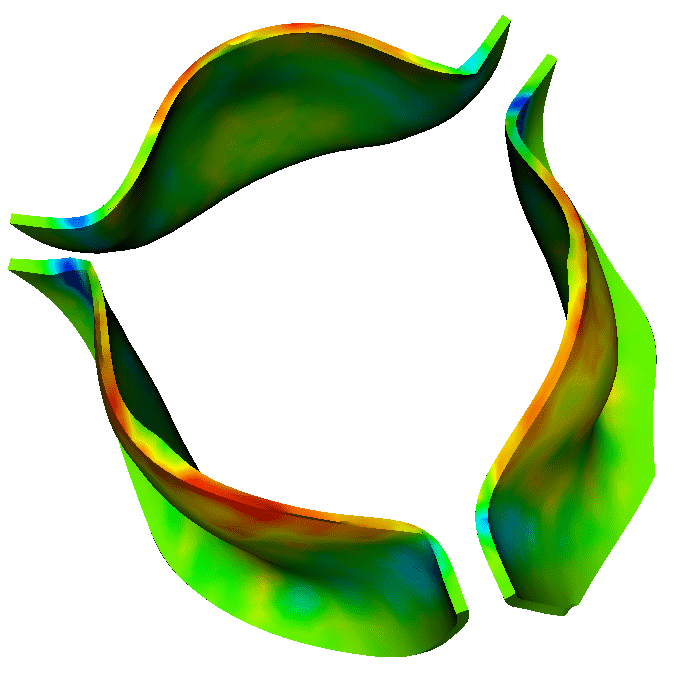}
}

\sidesubfloat[][]{
	\includegraphics[height=0.195\textwidth]{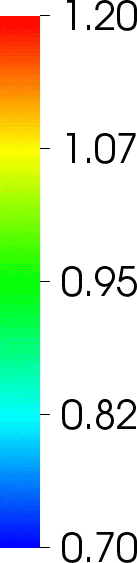} \
	\includegraphics[width=0.195\textwidth]{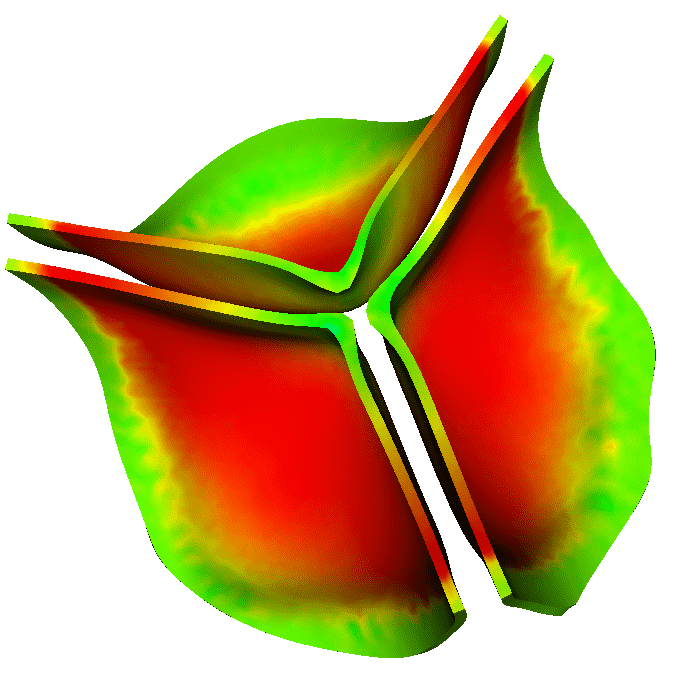}
	\includegraphics[width=0.195\textwidth]{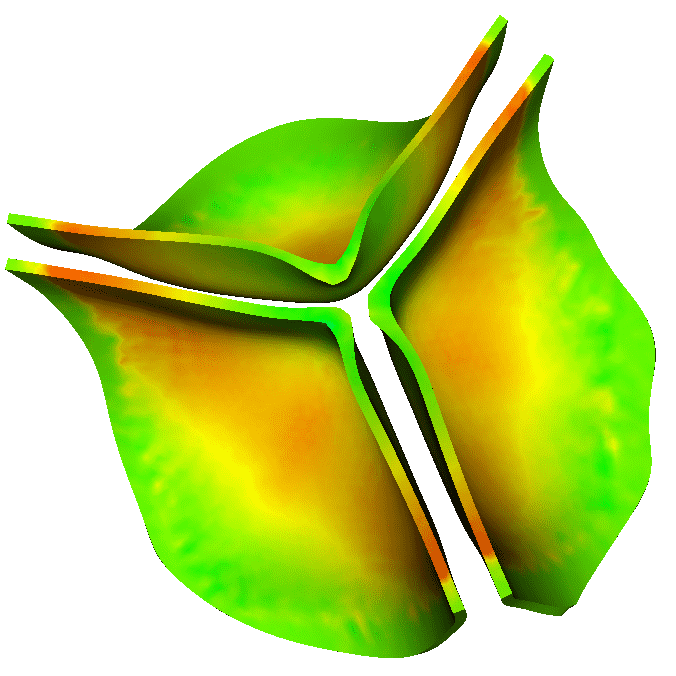}
	\includegraphics[width=0.195\textwidth]{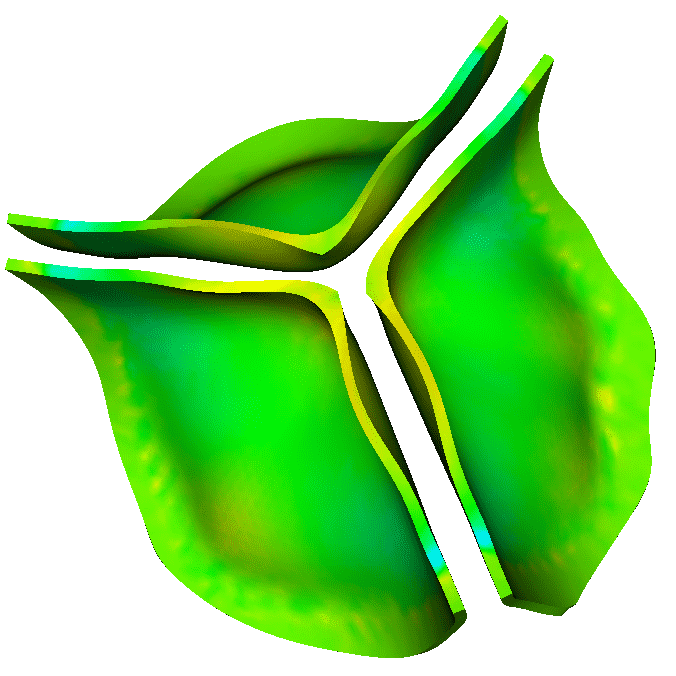}
	\includegraphics[width=0.195\textwidth]{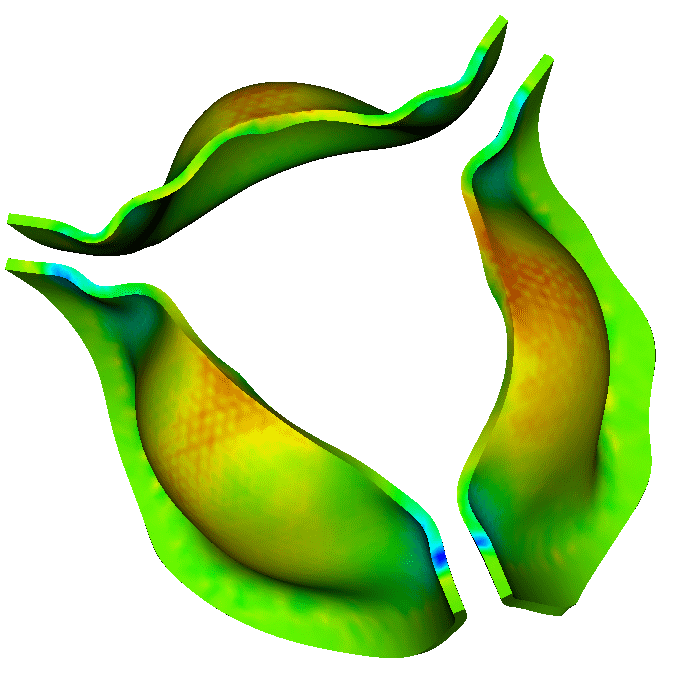}
	\includegraphics[width=0.195\textwidth]{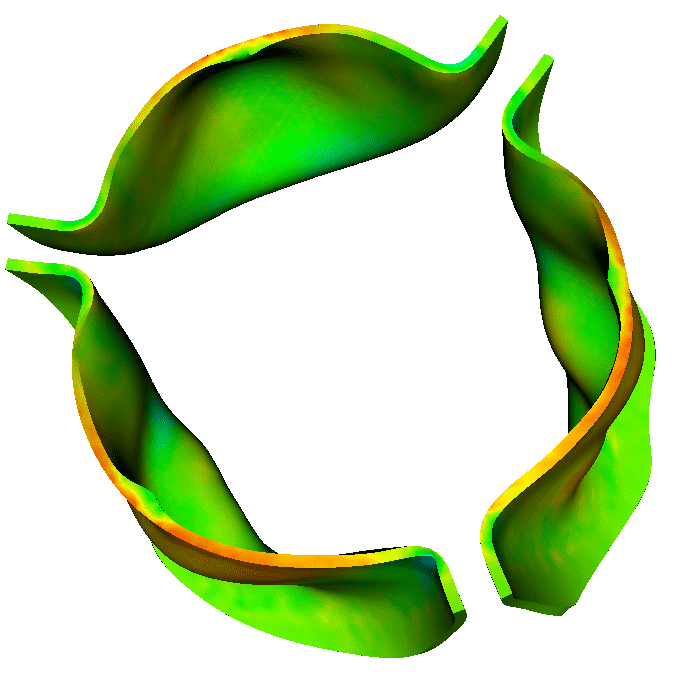}
}
\caption{Fiber stretch ratios obtained during late diastole and early systole, using fresh porcine valve constitutive parameters.
Panel (a)~shows results obtained using a relatively coarse Cartesian grid spacing of $0.86~\text{mm}$, and panel (b) shows results obtained using a relatively fine spacing of $0.43~\text{mm}$.
The strains are generally in good agreement.
Notice that the leaflets experience some compression near the commissures as the valve opens in early systole.
}
\label{f:lambda_f_comparison}
\end{figure}

\begin{figure}
\centering
\sidesubfloat[][]{
	\includegraphics[height=0.195\textwidth]{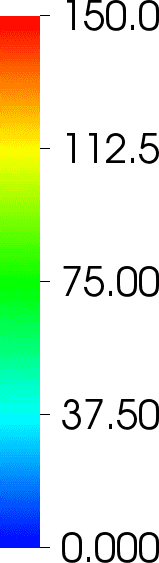} \
	\includegraphics[width=0.195\textwidth]{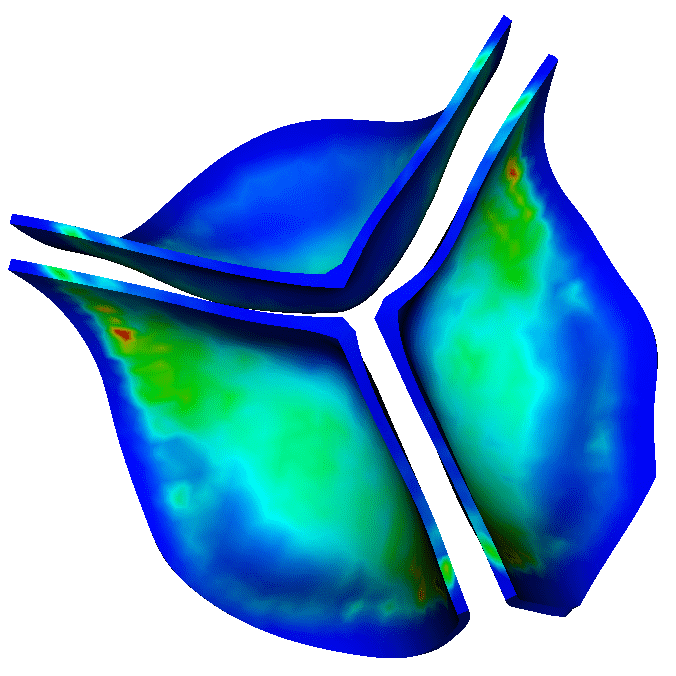}
	\includegraphics[width=0.195\textwidth]{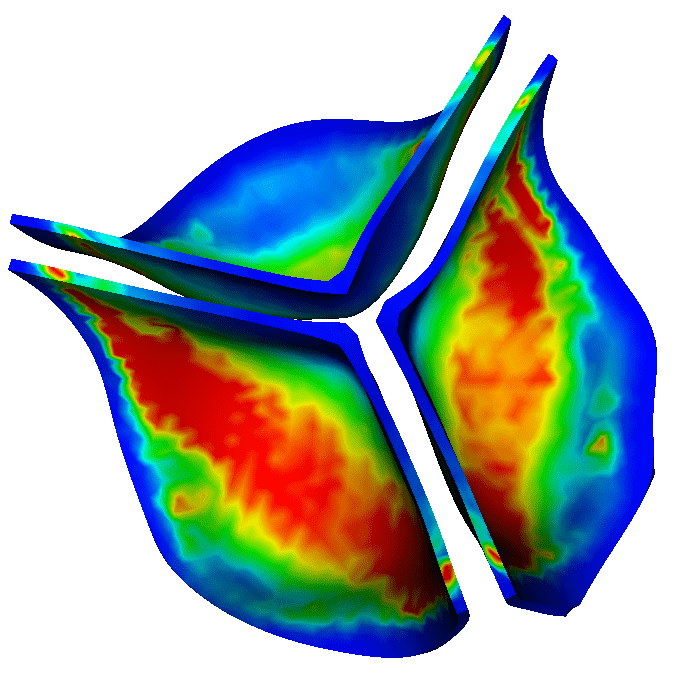}
	\includegraphics[width=0.195\textwidth]{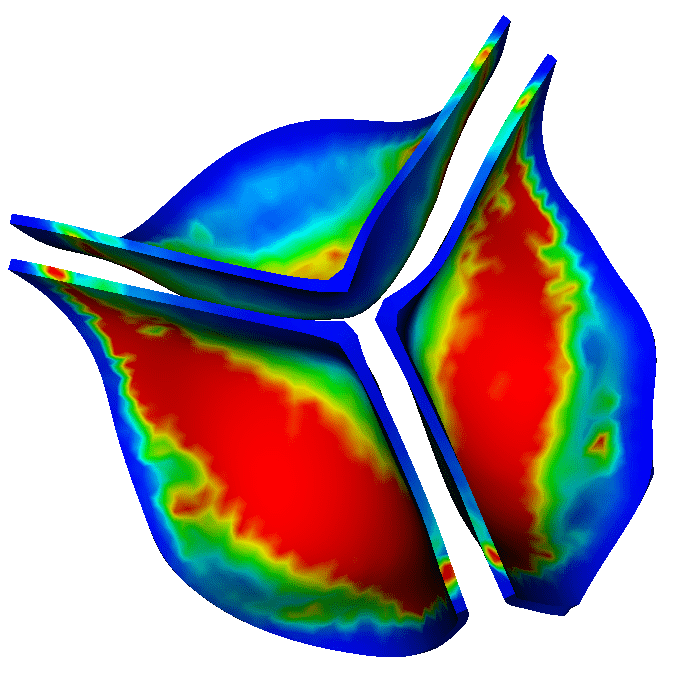}
	\includegraphics[width=0.195\textwidth]{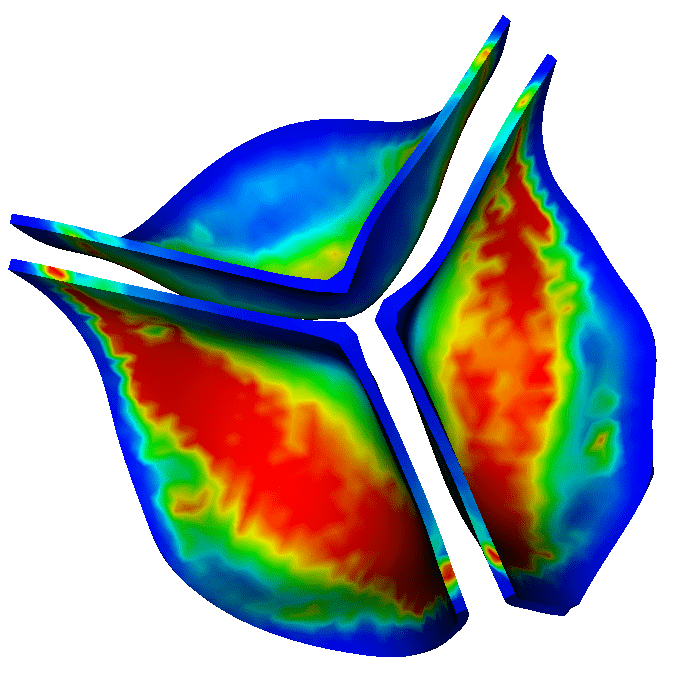}
	\includegraphics[width=0.195\textwidth]{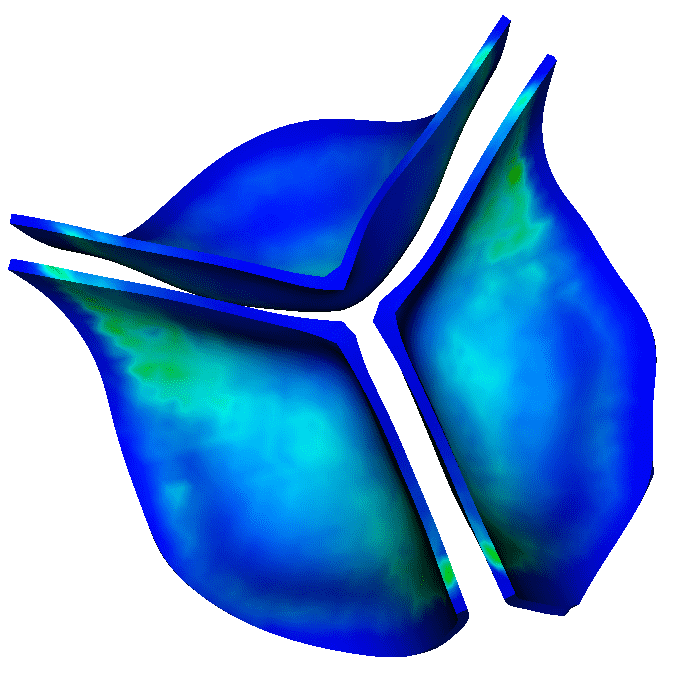}
}

\sidesubfloat[][]{
	\includegraphics[height=0.195\textwidth]{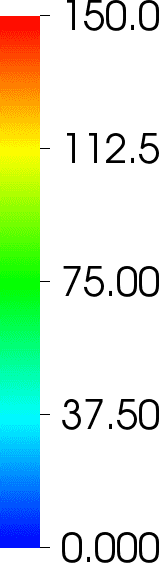} \
	\includegraphics[width=0.195\textwidth]{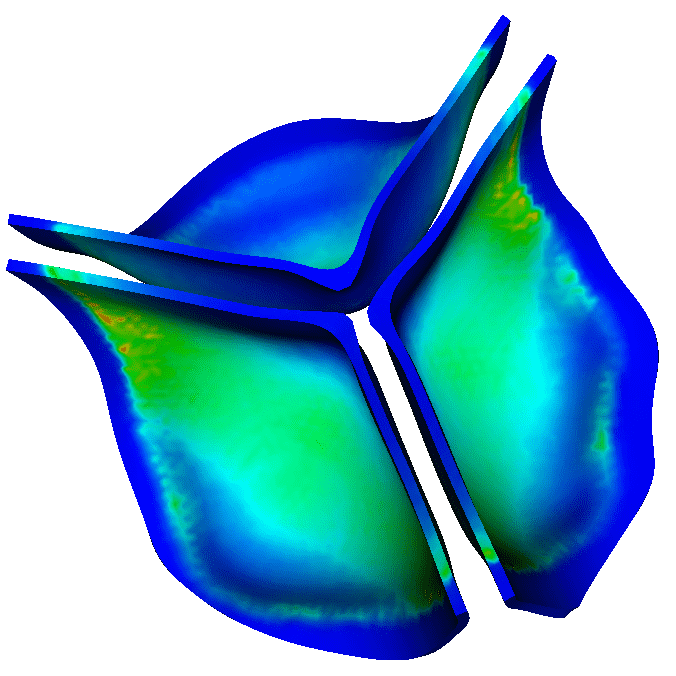}
	\includegraphics[width=0.195\textwidth]{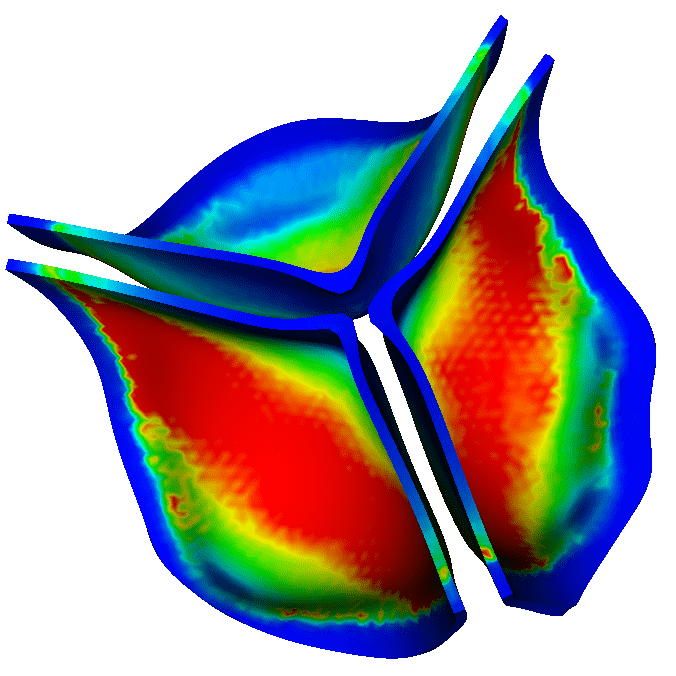}
	\includegraphics[width=0.195\textwidth]{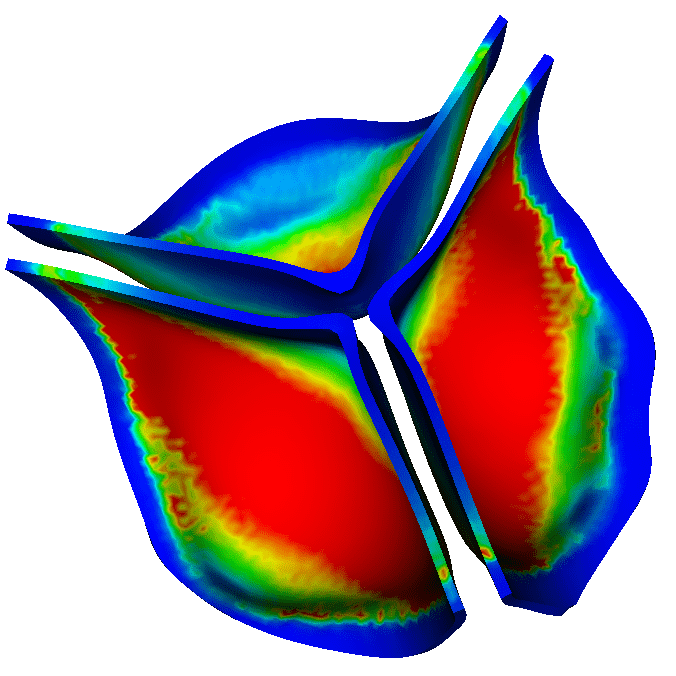}
	\includegraphics[width=0.195\textwidth]{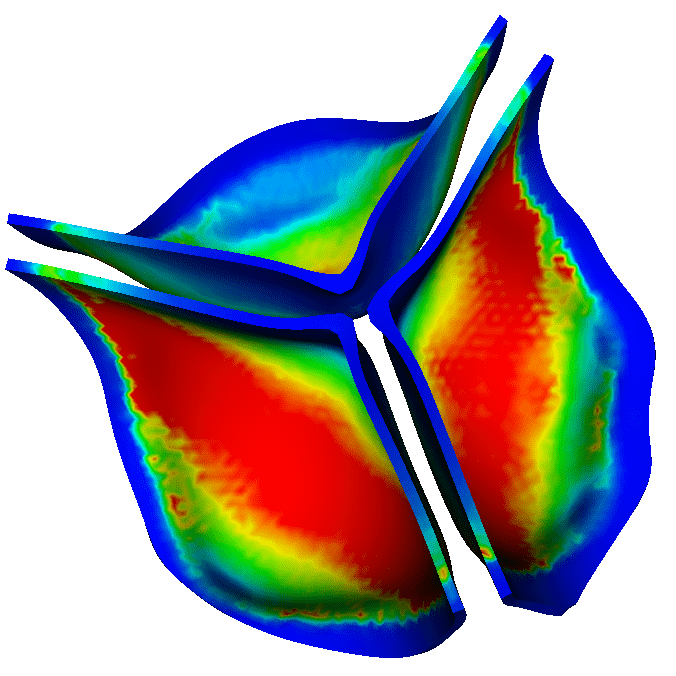}
	\includegraphics[width=0.195\textwidth]{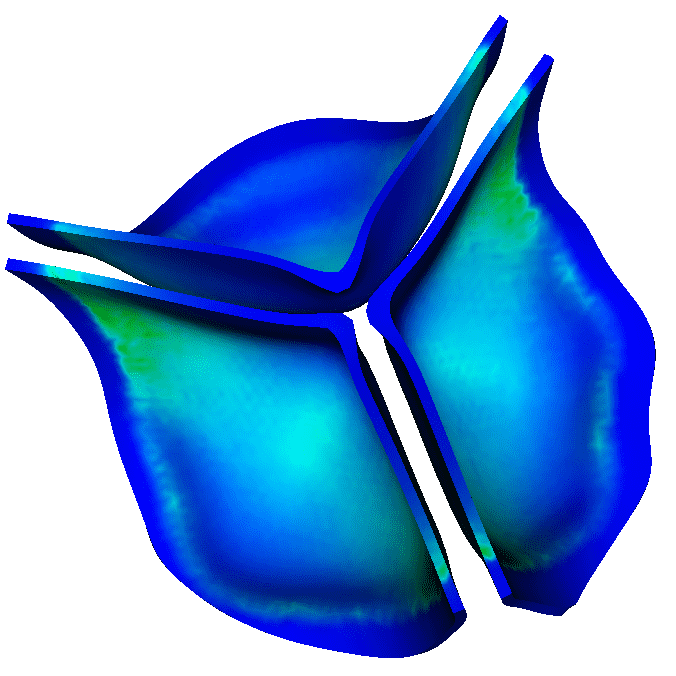}
}
\caption{von Mises stresses (kPa) obtained throughout diastole, up to early systole, using fresh porcine valve constitutive parameters.
Panel (a)~shows results obtained using a relatively coarse Cartesian grid spacing of $0.86~\text{mm}$, and panel (b) shows results obtained using a relatively fine spacing of $0.43~\text{mm}$.
The stresses are generally in good agreement.
The leaflets are supported by the collagen fibers that run from commissure to commissure, and the largest stresses occur during diastole, when the valve is fully loaded.
}
\label{f:sigma_v_comparison}
\end{figure}

We previously showed that grid resolutions similar to those used here are able to essentially resolve the bulk hemodynamics in IB models of the aortic root \cite{VFlamini16-aortic_root}, but that earlier study did not consider the leaflet mechanics in detail because it did not include a realistic model of the valve cusps.
Here we consider the effect of grid resolution on the leaflet mechanics.
Using the fresh valve parameters, we compare results obtained using a relatively coarse Cartesian grid spacing of $0.86~\text{mm}$ to those obtained using a relatively fine spacing of $0.43~\text{mm}$.
Figs.~\ref{f:dX_comparison} and \ref{f:2D_slice_comparison} compare the displacements during the same time points in early systole.
Some small discrepancies are observed in the leaflet deformations, especially after the valve opens fully.
The quasi-turbulent systolic flow generates complex flapping dynamics, and we expect that discrepancies in the leaflet kinematics arise from under-resolving this flow field.
Fig.~\ref{f:lambda_f_comparison} compares the fiber stretch ratios during early systole, as the valve opens.
The leaflets are supported by the collagen fibers that run from commissure to commissure, and the fibers in the ``belly'' of the leaflet see the largest strains.
Notice that the leaflets experience some compression near the commissures as the valve opens in early systole.
Fig.~\ref{f:sigma_v_comparison} compares the von Mises stresses during diastole and early systole.
The largest stresses occur during diastole, when the valve is fully loaded.
Notice that all of these quantities are all in good quantitative agreement between the two grid resolutions, suggesting that the model is able to resolve the details of the leaflet mechanics.

\subsection{Effect of material properties on leaflet mechanics}
\label{s:comparitive_leaflet_biomechanics}

\begin{figure}
\centering
\sidesubfloat[][]{
	\includegraphics[height=0.195\textwidth]{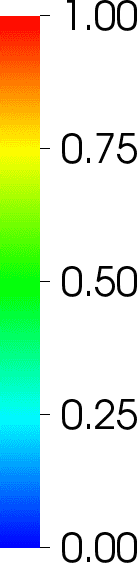} \
	\includegraphics[width=0.195\textwidth]{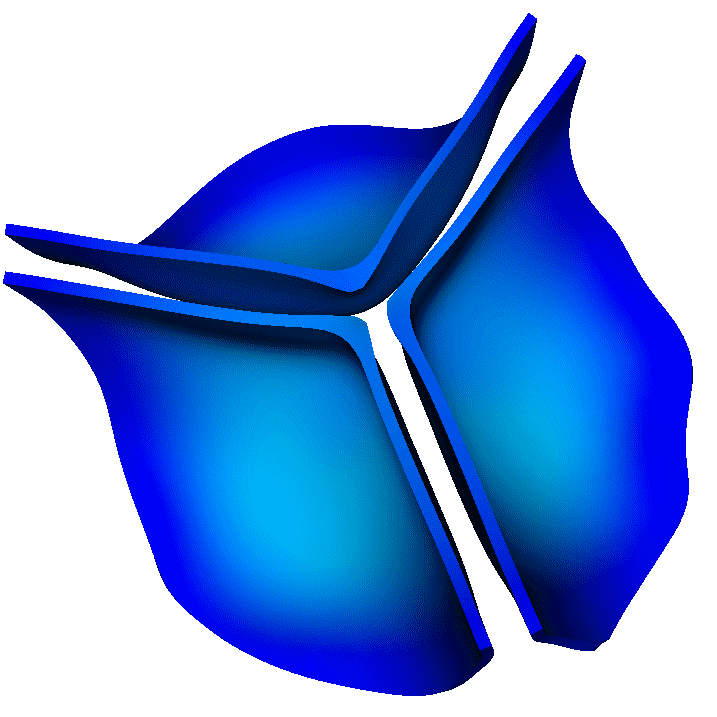}
	\includegraphics[width=0.195\textwidth]{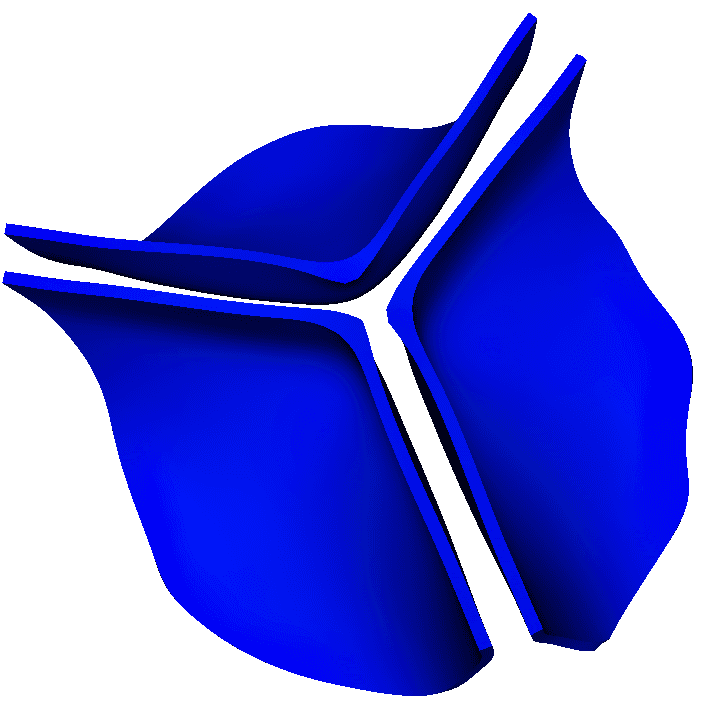}
	\includegraphics[width=0.195\textwidth]{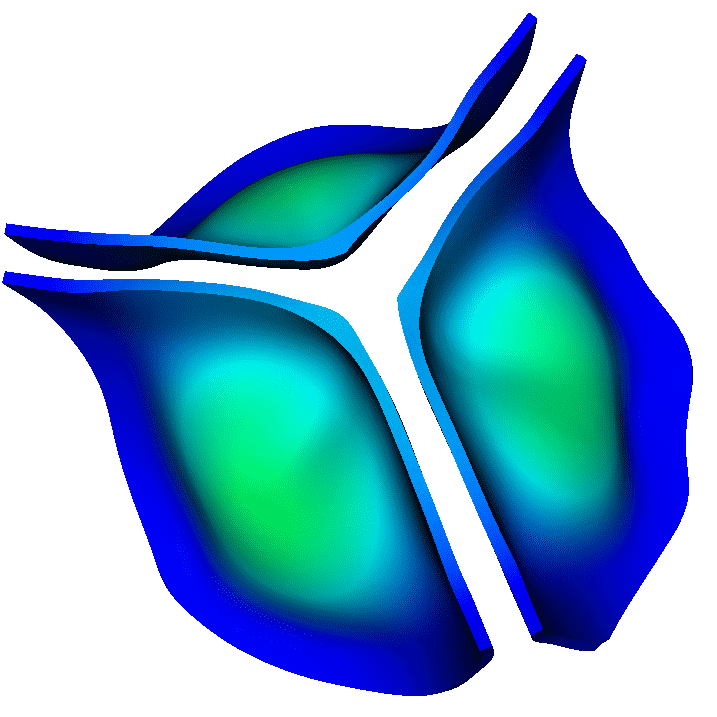}
	\includegraphics[width=0.195\textwidth]{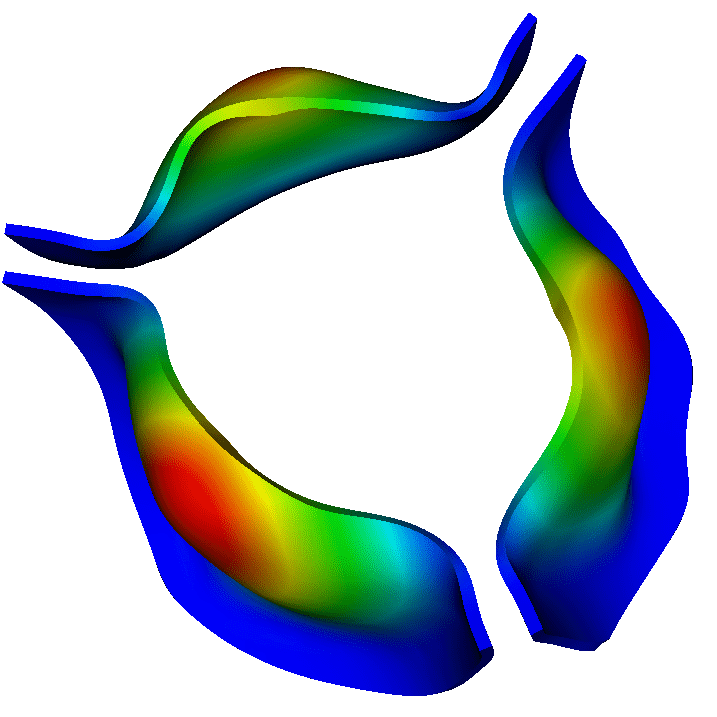}
	\includegraphics[width=0.195\textwidth]{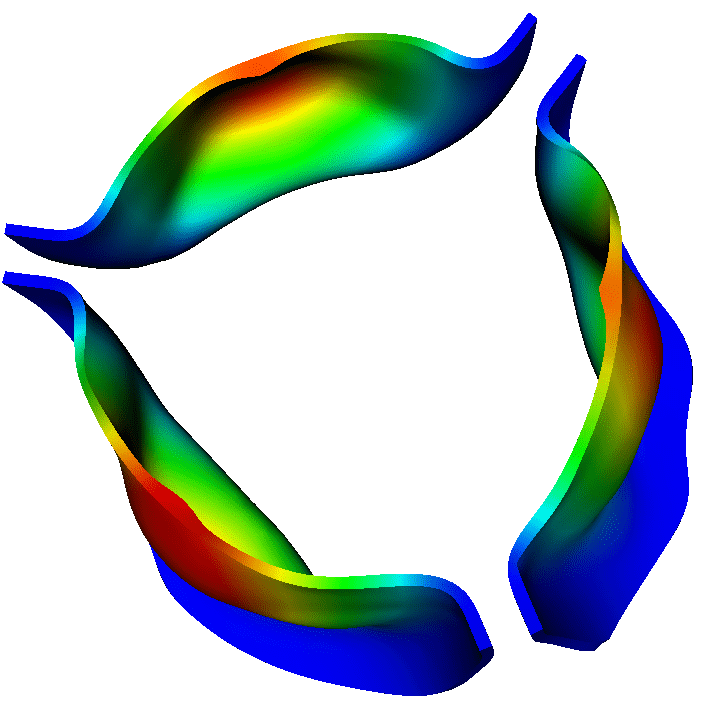}
}

\sidesubfloat[][]{
	\includegraphics[height=0.195\textwidth]{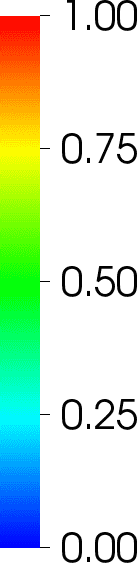} \
	\includegraphics[width=0.195\textwidth]{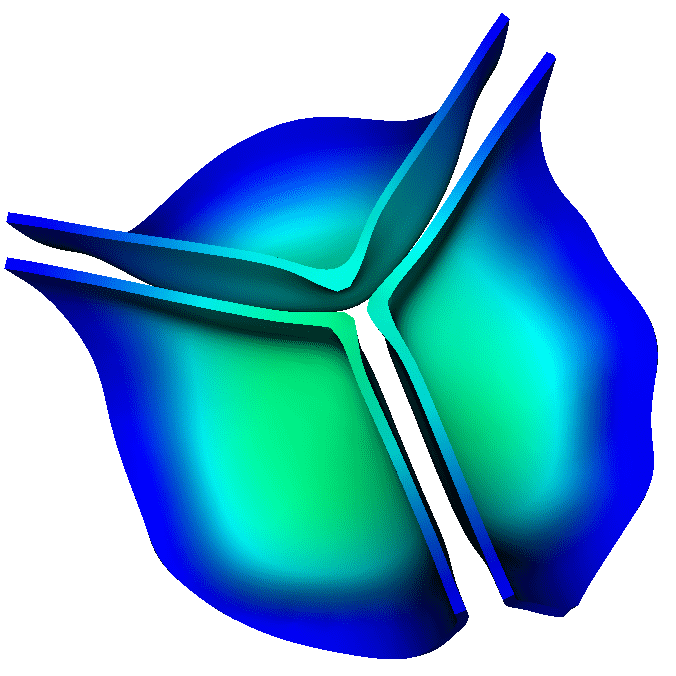}
	\includegraphics[width=0.195\textwidth]{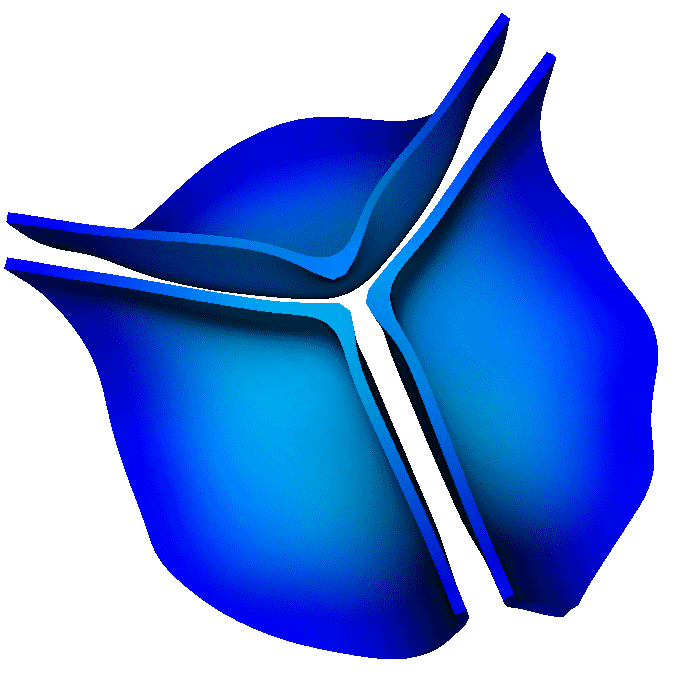}
	\includegraphics[width=0.195\textwidth]{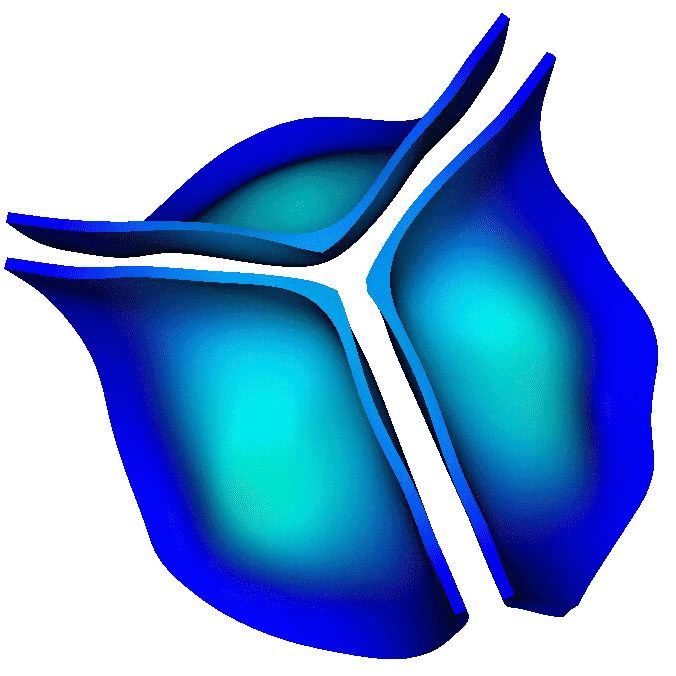}
	\includegraphics[width=0.195\textwidth]{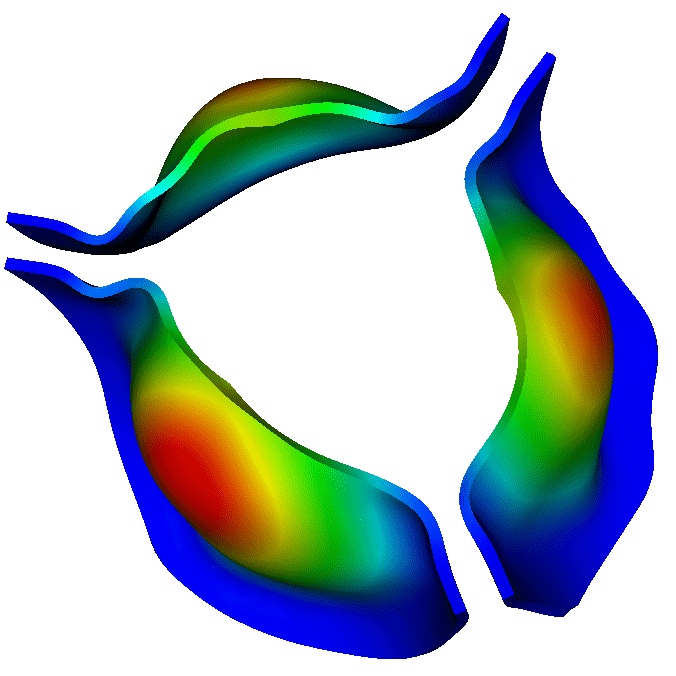}
	\includegraphics[width=0.195\textwidth]{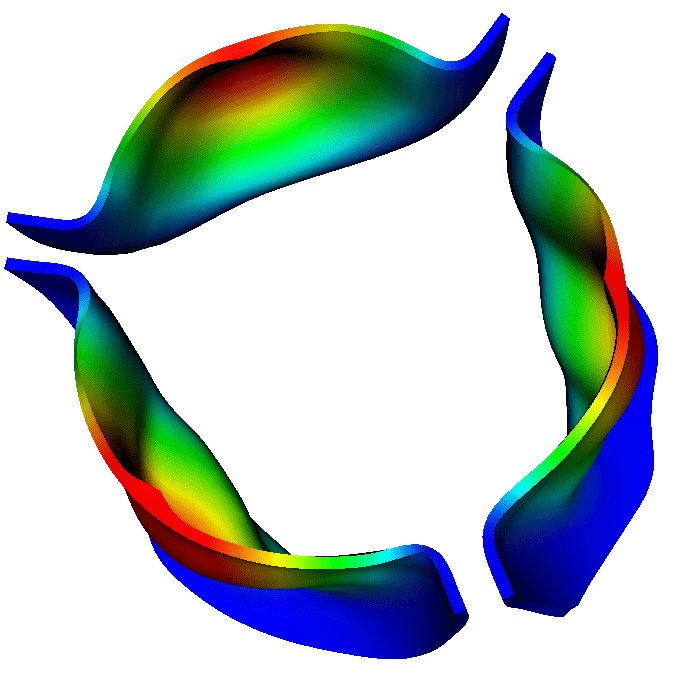}
}

\sidesubfloat[][]{
	\includegraphics[height=0.195\textwidth]{aortic_root/native/N=256/dX/colorbar} \
	\includegraphics[width=0.195\textwidth]{aortic_root/native/N=256/dX/dX0057}
	\includegraphics[width=0.195\textwidth]{aortic_root/native/N=256/dX/dX0059}
	\includegraphics[width=0.195\textwidth]{aortic_root/native/N=256/dX/dX0061}
	\includegraphics[width=0.195\textwidth]{aortic_root/native/N=256/dX/dX0063}
	\includegraphics[width=0.195\textwidth]{aortic_root/native/N=256/dX/dX0065}
}
\caption{Leaflet displacements (cm) obtained during late diastole and early systole, using a relatively fine spacing of $0.43~\text{mm}$ along with constitutive parameters for (a)~glutaraldehyde-fixed porcine leaflets with $4~\text{mmHg}$ fixation pressure, (b)~glutaraldehyde-fixed porcine leaflets with $0~\text{mmHg}$ fixation pressure, and (c)~fresh porcine leaflets.
Notice the glutaraldehyde-fixed porcine leaflets with $4~\text{mmHg}$ fixation pressure deform the least in diastole, and that the fresh leaflets experience the largest deformations.
See also Fig.~\ref{f:2D_slice_material_comparison} for details of the deformations and displacements of the center surface of the left coronary leaflet.
}
\label{f:dX_material_comparison}
\end{figure}

\begin{figure}
\centering
\sidesubfloat[][]{
	\includegraphics[height=0.195\textwidth]{aortic_root/BHV_p=4mmHg/N=256/dX/colorbar} \
	\includegraphics[width=0.195\textwidth]{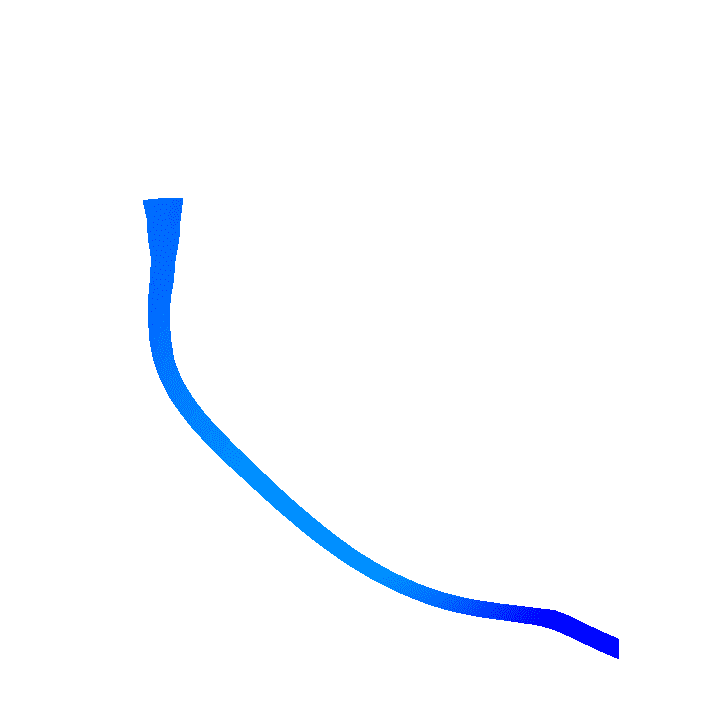}
	\includegraphics[width=0.195\textwidth]{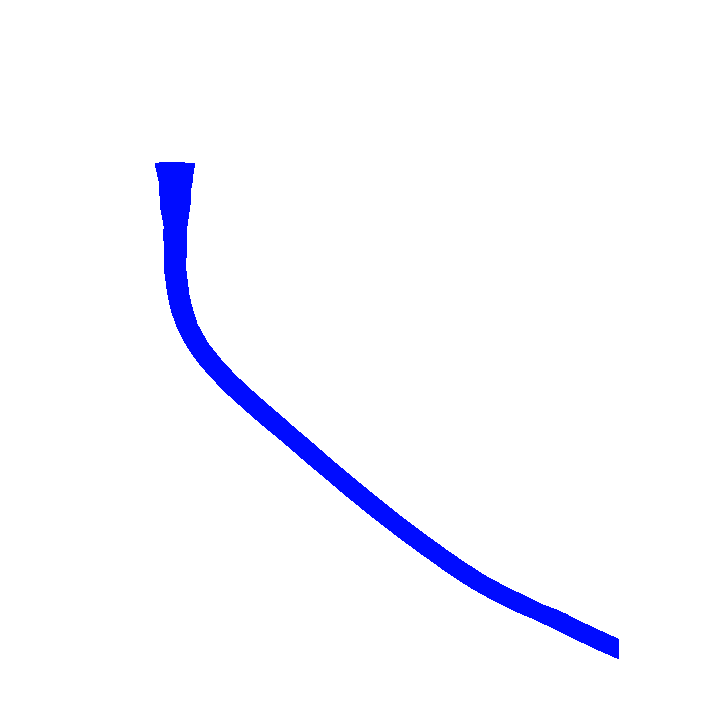}
	\includegraphics[width=0.195\textwidth]{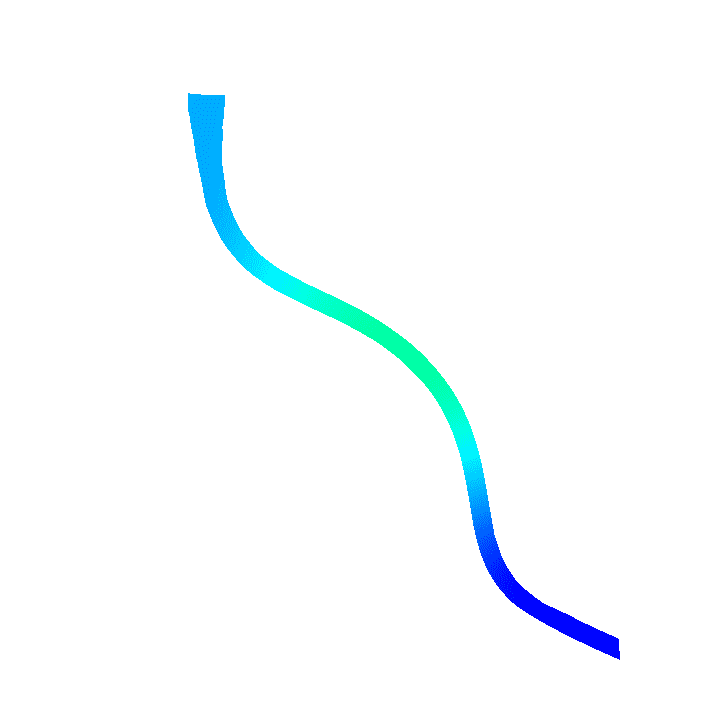}
	\includegraphics[width=0.195\textwidth]{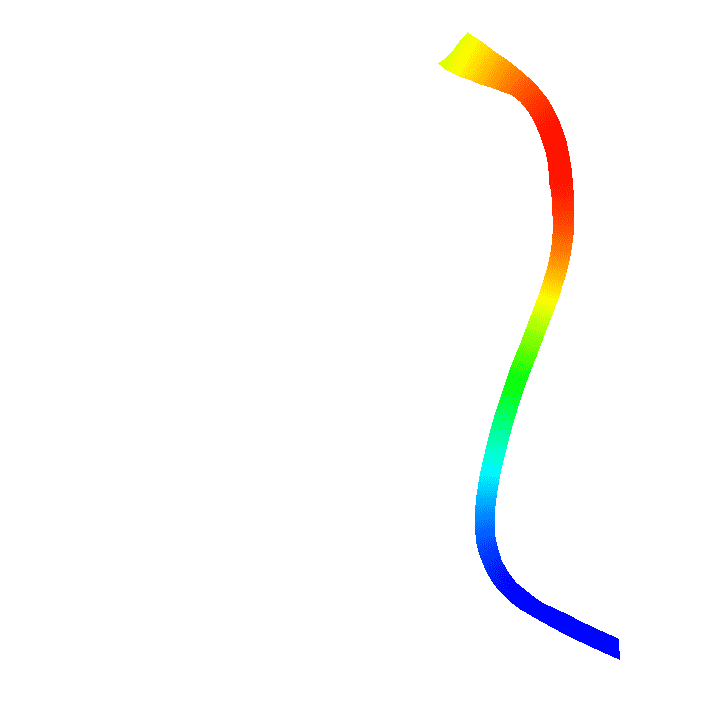}
	\includegraphics[width=0.195\textwidth]{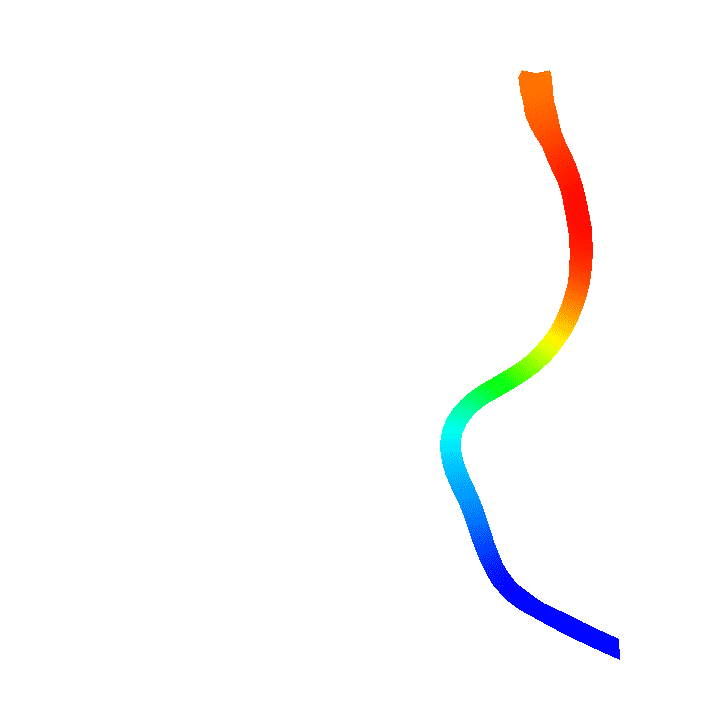}
}

\sidesubfloat[][]{
	\includegraphics[height=0.195\textwidth]{aortic_root/BHV_p=0mmHg/N=256/dX/colorbar} \
	\includegraphics[width=0.195\textwidth]{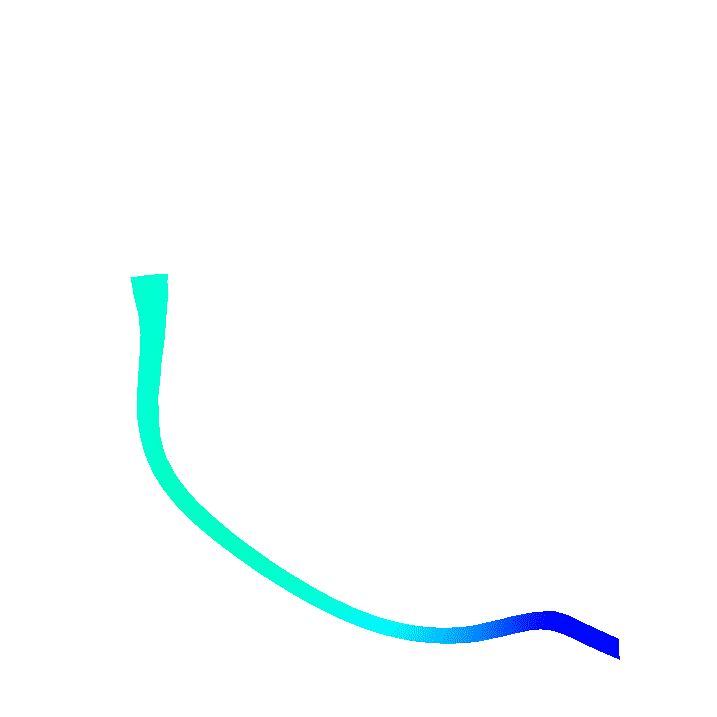}
	\includegraphics[width=0.195\textwidth]{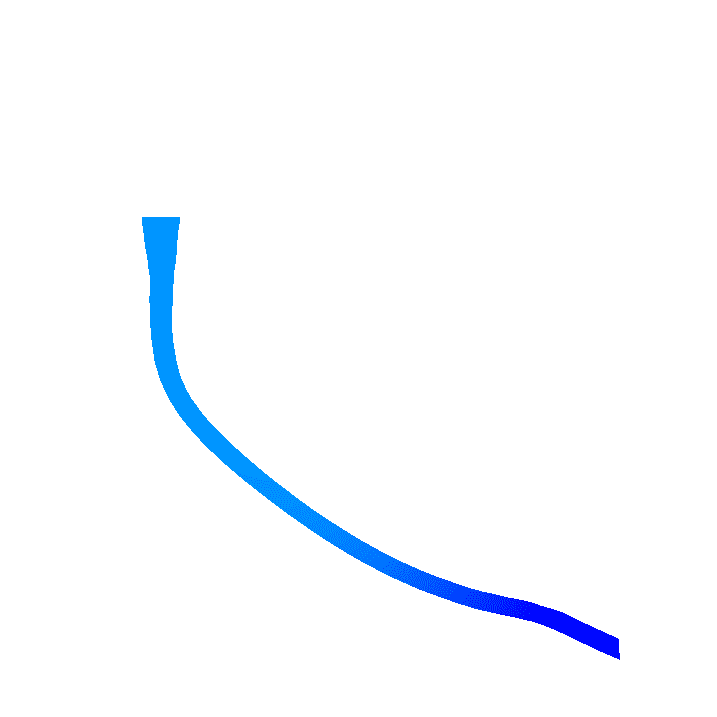}
	\includegraphics[width=0.195\textwidth]{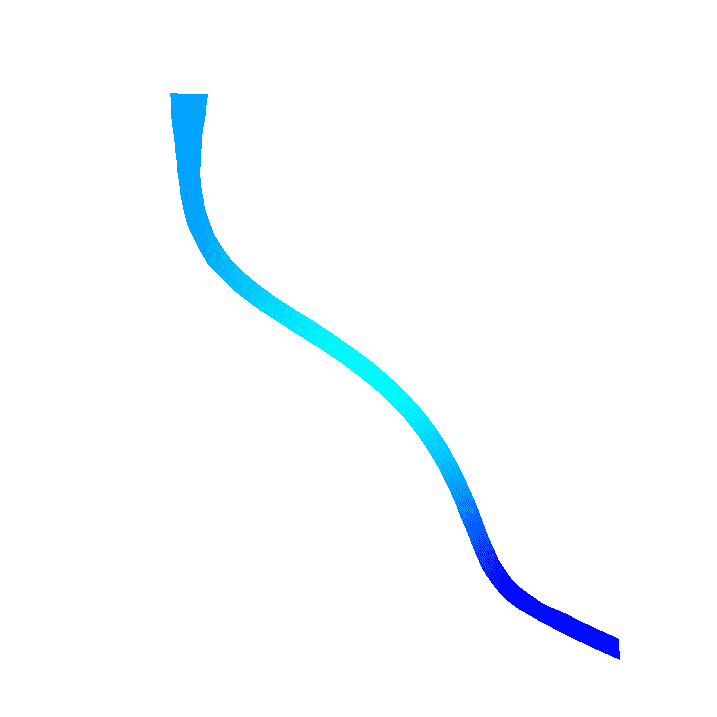}
	\includegraphics[width=0.195\textwidth]{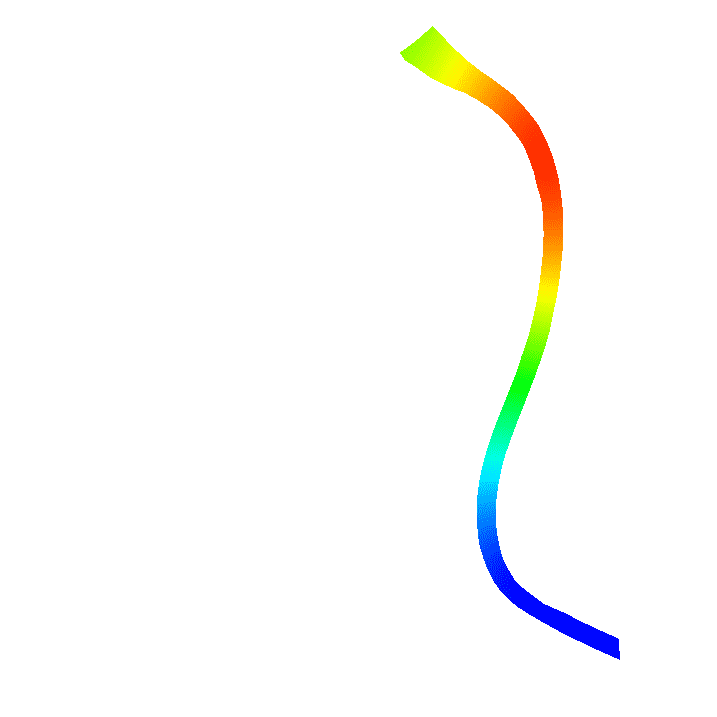}
	\includegraphics[width=0.195\textwidth]{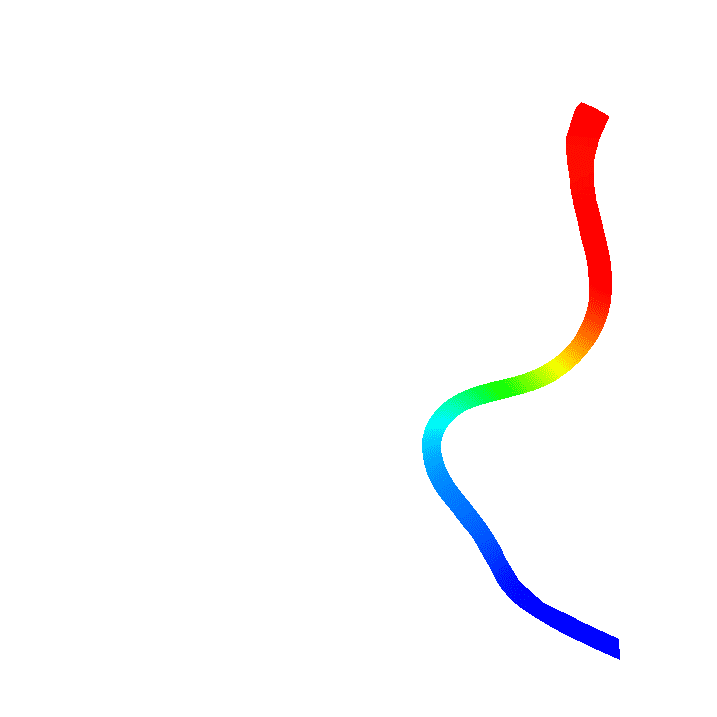}
}

\sidesubfloat[][]{
	\includegraphics[height=0.195\textwidth]{aortic_root/native/N=256/dX/colorbar} \
	\includegraphics[width=0.195\textwidth]{aortic_root/native/N=256/2D_slice/2D_slice0057}
	\includegraphics[width=0.195\textwidth]{aortic_root/native/N=256/2D_slice/2D_slice0059}
	\includegraphics[width=0.195\textwidth]{aortic_root/native/N=256/2D_slice/2D_slice0061}
	\includegraphics[width=0.195\textwidth]{aortic_root/native/N=256/2D_slice/2D_slice0063}
	\includegraphics[width=0.195\textwidth]{aortic_root/native/N=256/2D_slice/2D_slice0065}
}
\caption{
Similar to Fig.~\ref{f:dX_material_comparison}, but here showing leaflet displacements (cm) along the center surface of the left coronary leaflet obtained during late diastole and early systole.
Notice that the glutaraldehyde-fixed porcine leaflets with $4~\text{mmHg}$ fixation pressure (a) deform the least in diastole, and that the fresh leaflets (c) experience the largest deformations.
By contrast, the open configurations are similar.
}
\label{f:2D_slice_material_comparison}
\end{figure}

\begin{figure}
\centering
\sidesubfloat[][]{
	\includegraphics[height=0.195\textwidth]{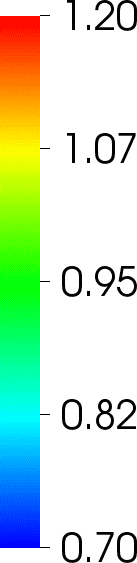} \
	\includegraphics[width=0.195\textwidth]{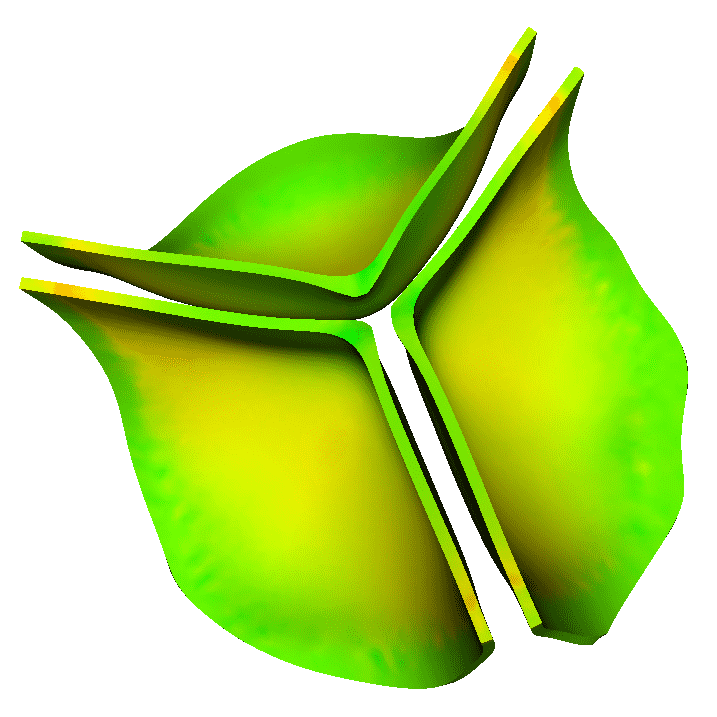}
	\includegraphics[width=0.195\textwidth]{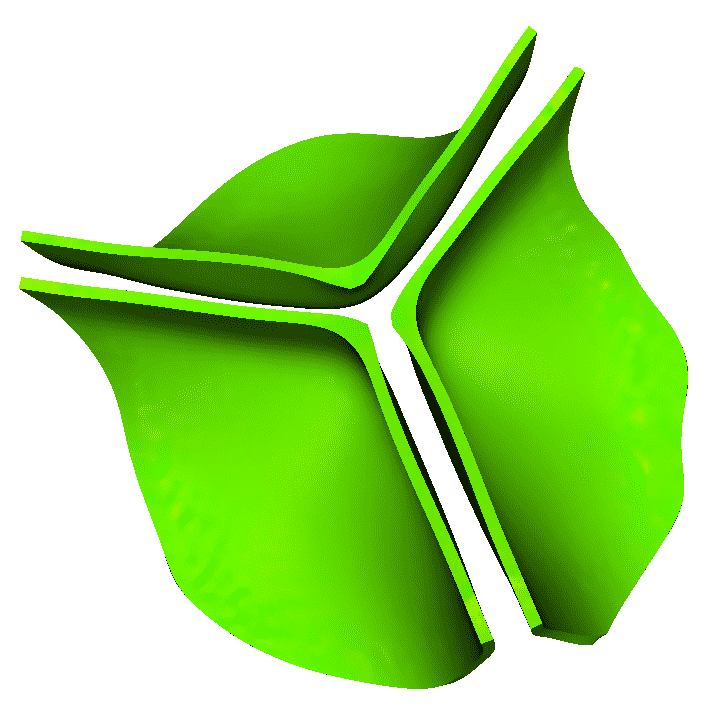}
	\includegraphics[width=0.195\textwidth]{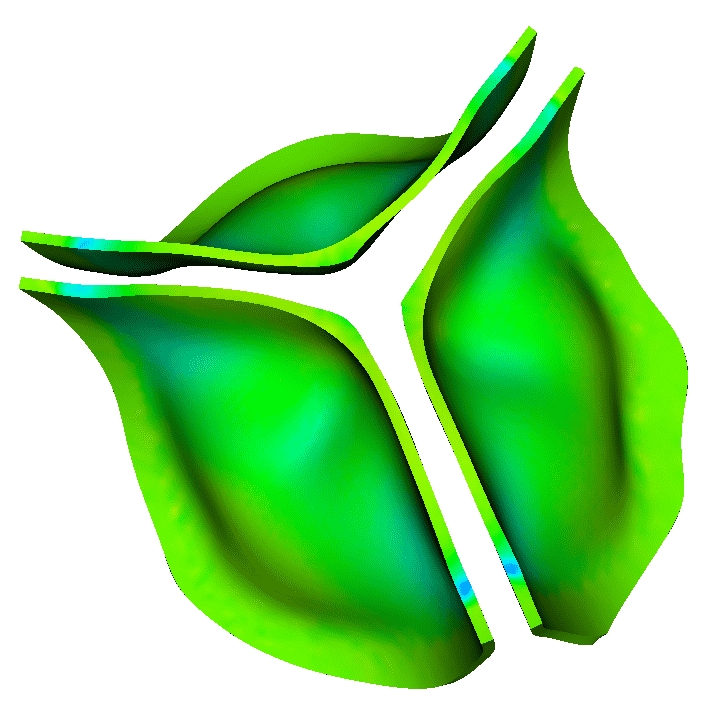}
	\includegraphics[width=0.195\textwidth]{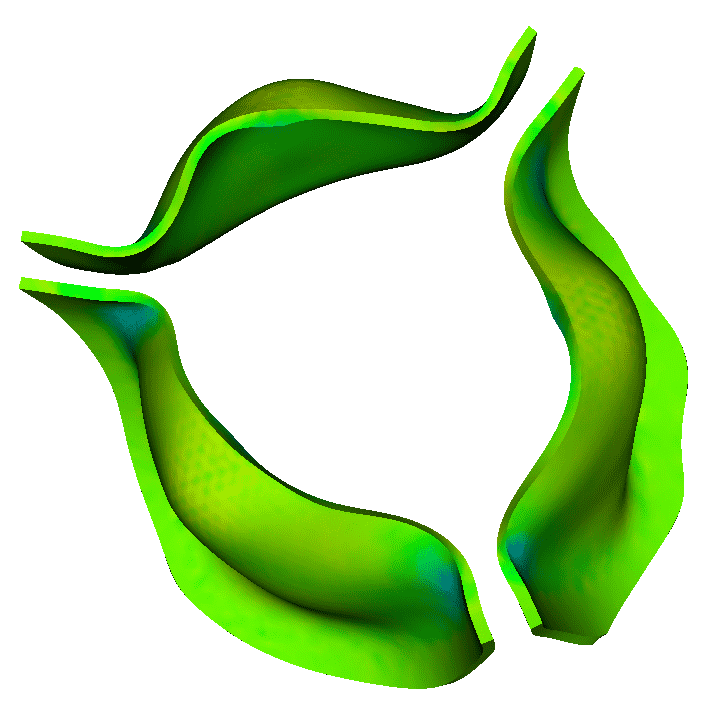}
	\includegraphics[width=0.195\textwidth]{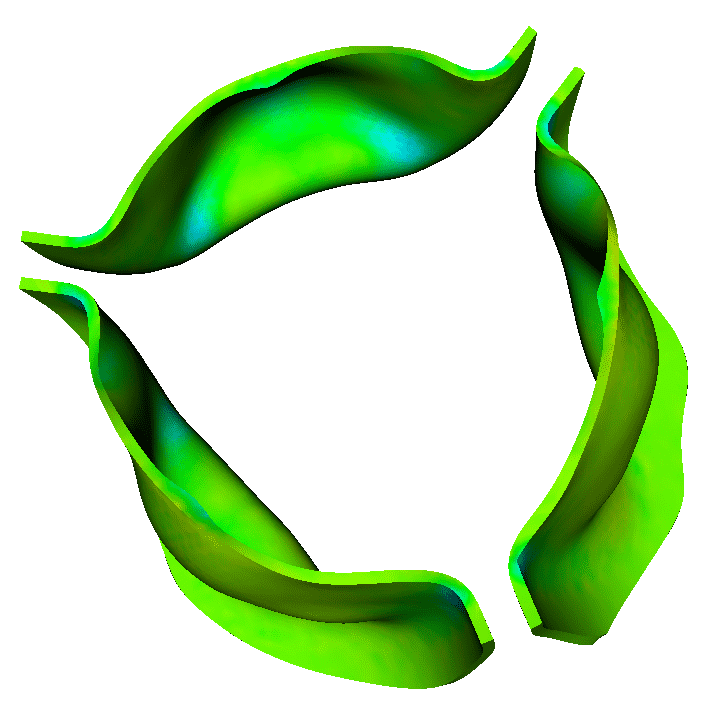}
}

\sidesubfloat[][]{
	\includegraphics[height=0.195\textwidth]{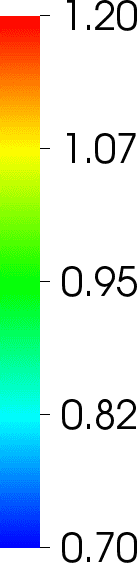} \
	\includegraphics[width=0.195\textwidth]{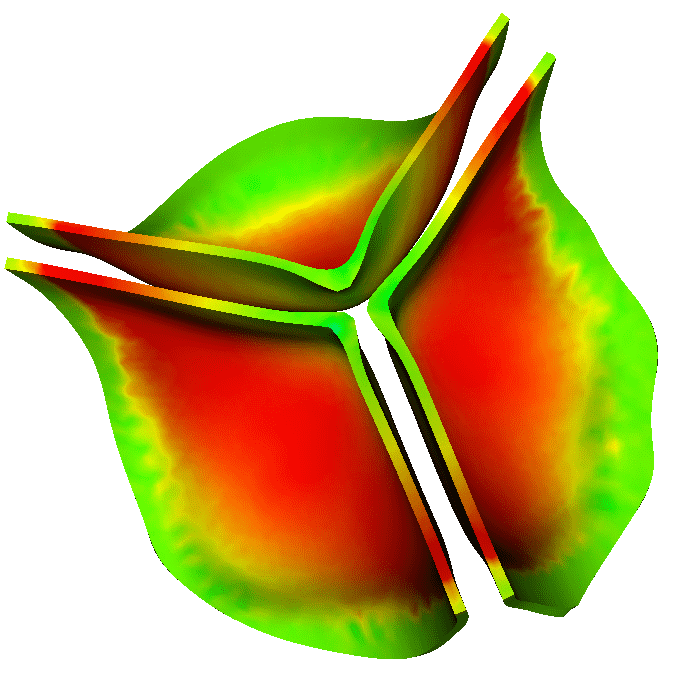}
	\includegraphics[width=0.195\textwidth]{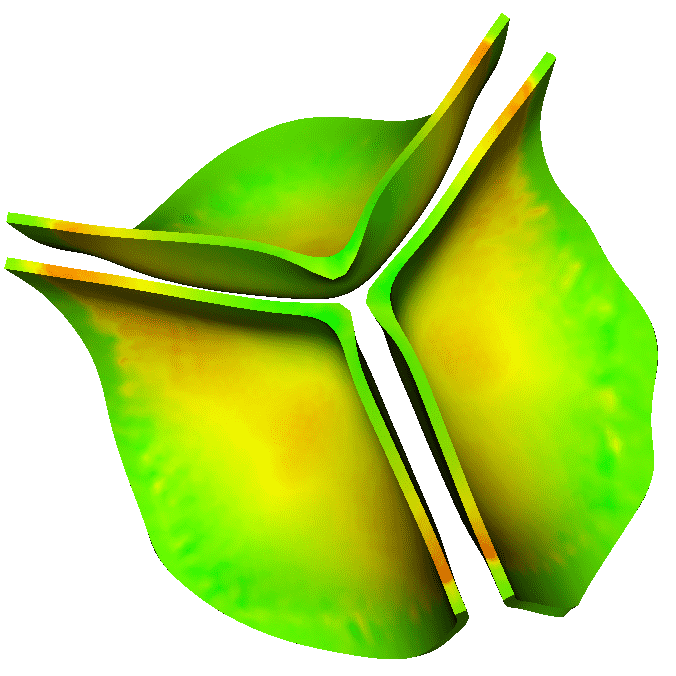}
	\includegraphics[width=0.195\textwidth]{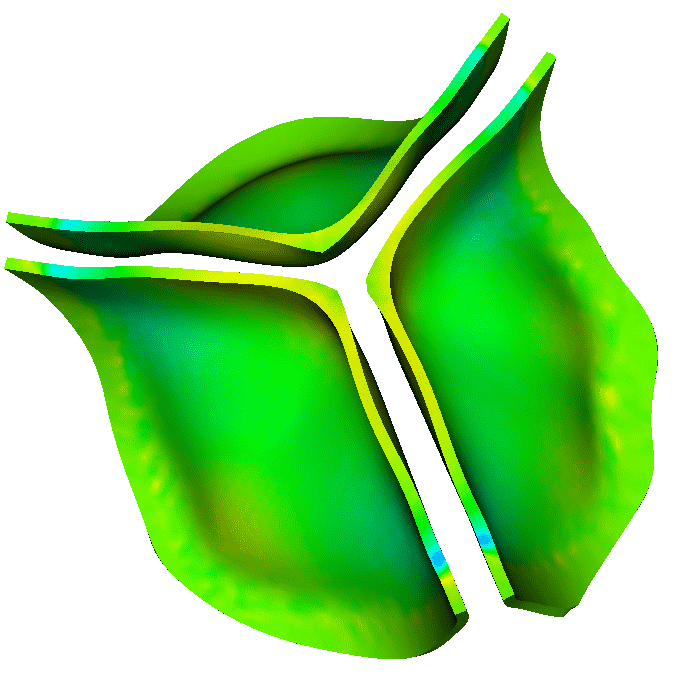}
	\includegraphics[width=0.195\textwidth]{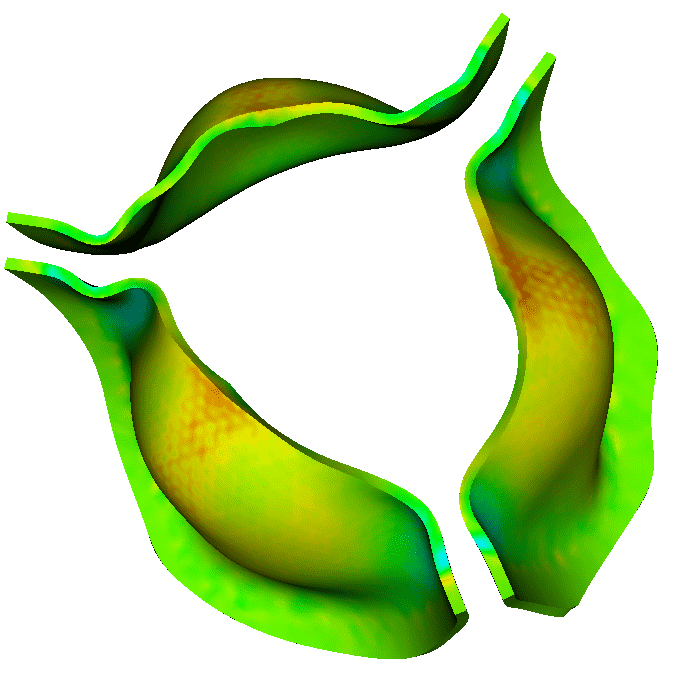}
	\includegraphics[width=0.195\textwidth]{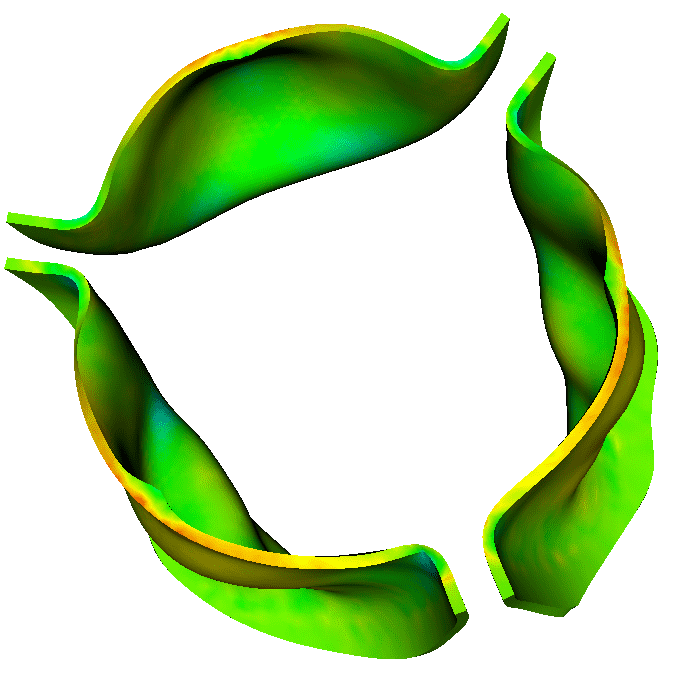}
}

\sidesubfloat[][]{
	\includegraphics[height=0.195\textwidth]{aortic_root/native/N=256/lambda_f/colorbar} \ 
	\includegraphics[width=0.195\textwidth]{aortic_root/native/N=256/lambda_f/lambda_f0057}
	\includegraphics[width=0.195\textwidth]{aortic_root/native/N=256/lambda_f/lambda_f0059}
	\includegraphics[width=0.195\textwidth]{aortic_root/native/N=256/lambda_f/lambda_f0061}
	\includegraphics[width=0.195\textwidth]{aortic_root/native/N=256/lambda_f/lambda_f0063}
	\includegraphics[width=0.195\textwidth]{aortic_root/native/N=256/lambda_f/lambda_f0065}
}
\caption{
Similar to Fig.~\ref{f:dX_material_comparison}, but here showing fiber stretch ratios obtained during late diastole and early systole.
Notice that the glutaraldehyde-fixed porcine leaflets with $4~\text{mmHg}$ fixation pressure (a) show the smallest fiber strains in diastole, and that the fresh leaflets (c) experience the largest fiber strains.
}
\label{f:lambda_f_material_comparison}
\end{figure}

\begin{figure}
\centering
\sidesubfloat[][]{
	\includegraphics[height=0.195\textwidth]{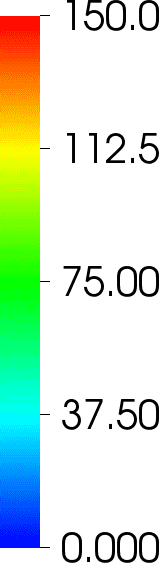} \
	\includegraphics[width=0.195\textwidth]{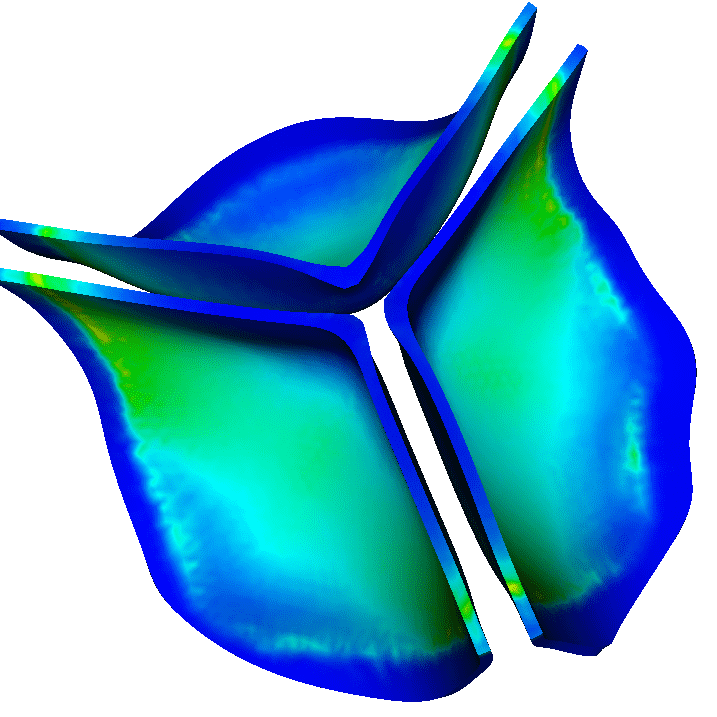}
	\includegraphics[width=0.195\textwidth]{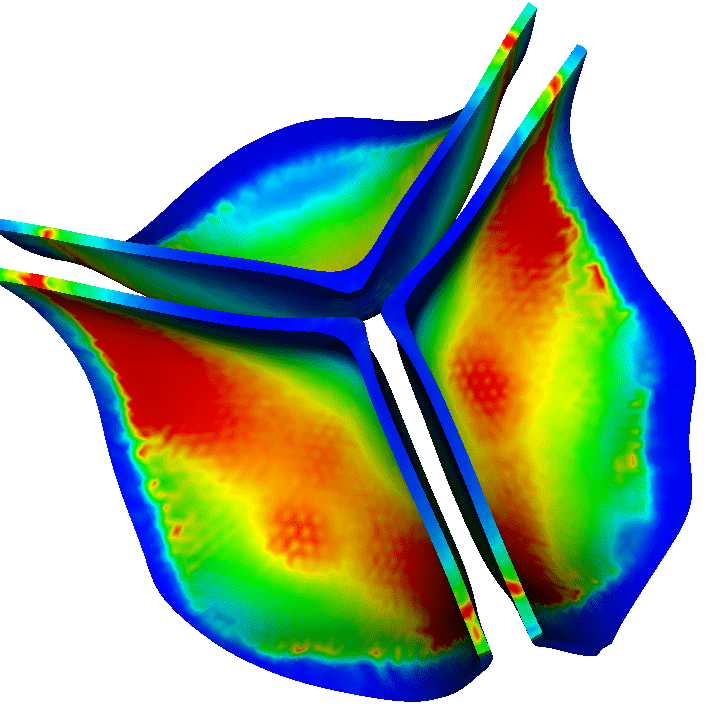}
	\includegraphics[width=0.195\textwidth]{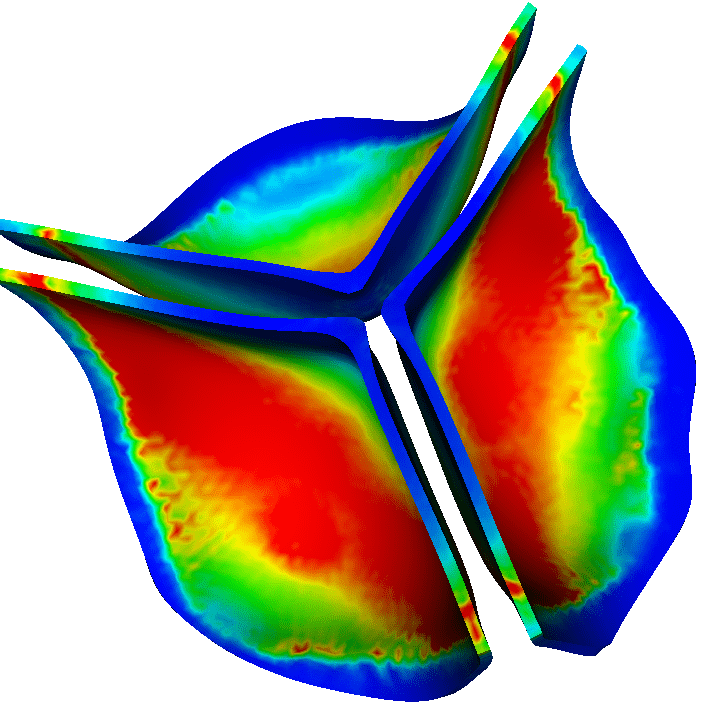}
	\includegraphics[width=0.195\textwidth]{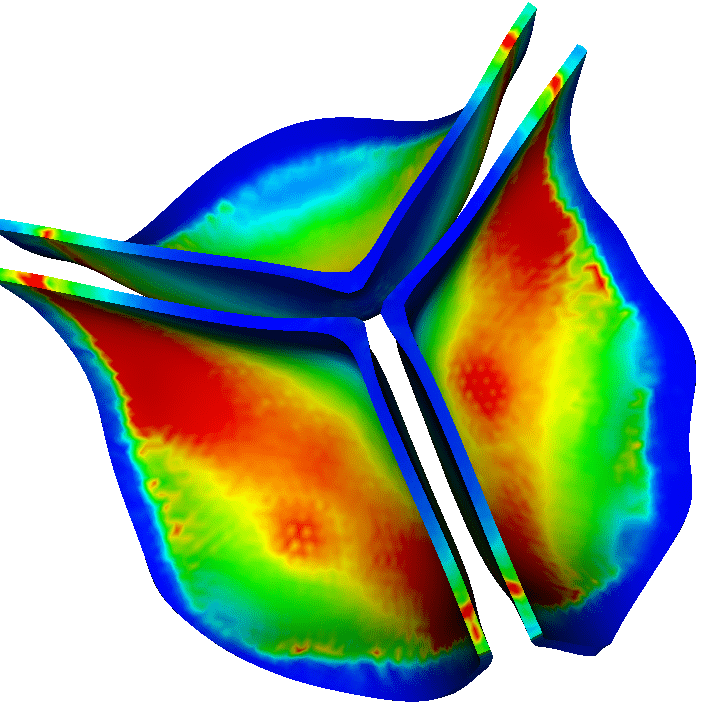}
	\includegraphics[width=0.195\textwidth]{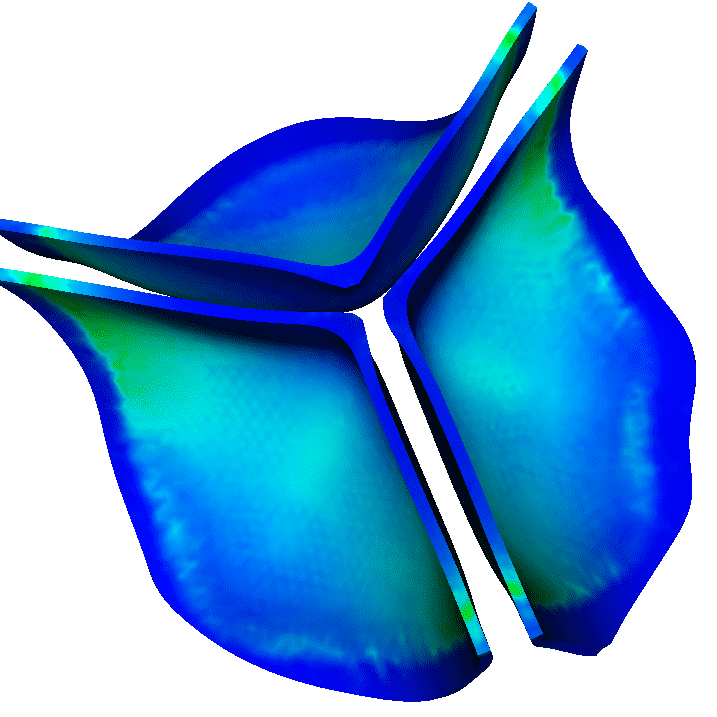}
}

\sidesubfloat[][]{
	\includegraphics[height=0.195\textwidth]{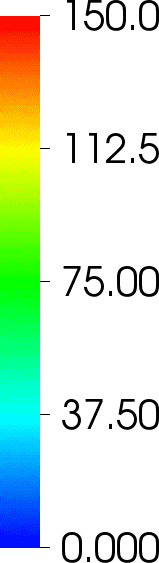} \
	\includegraphics[width=0.195\textwidth]{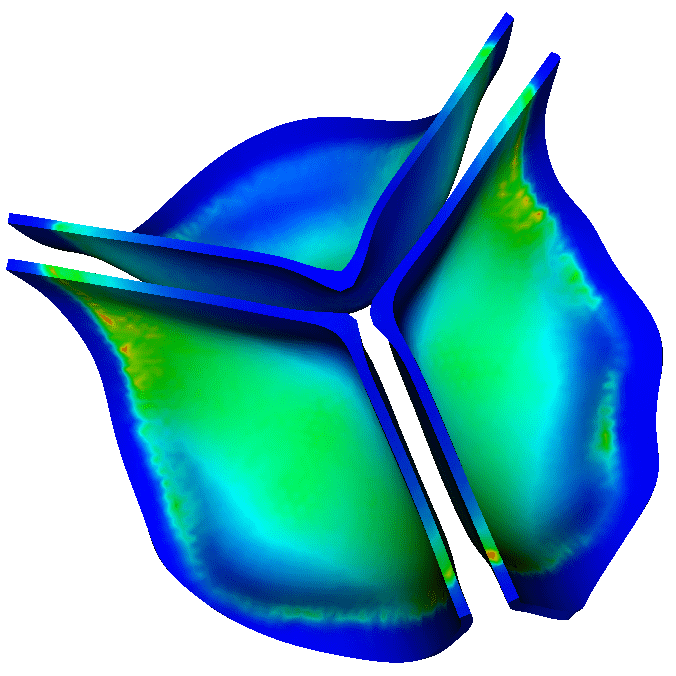}
	\includegraphics[width=0.195\textwidth]{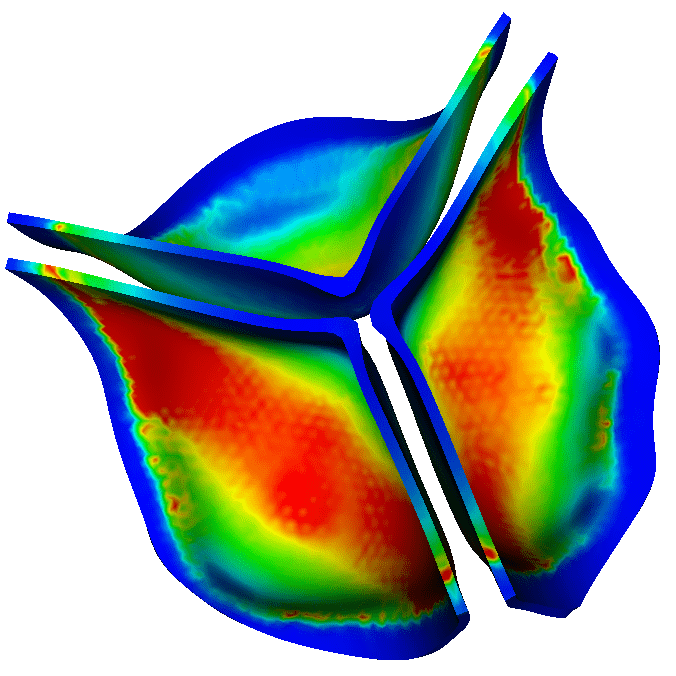}
	\includegraphics[width=0.195\textwidth]{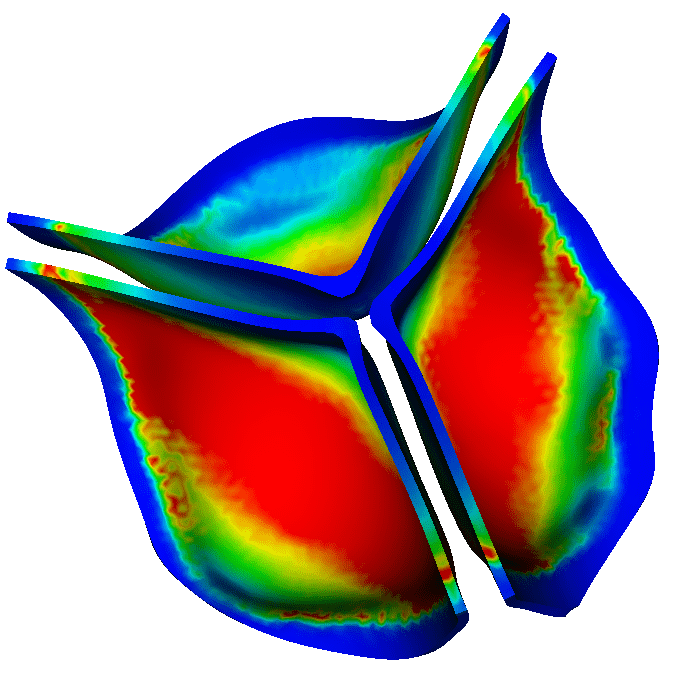}
	\includegraphics[width=0.195\textwidth]{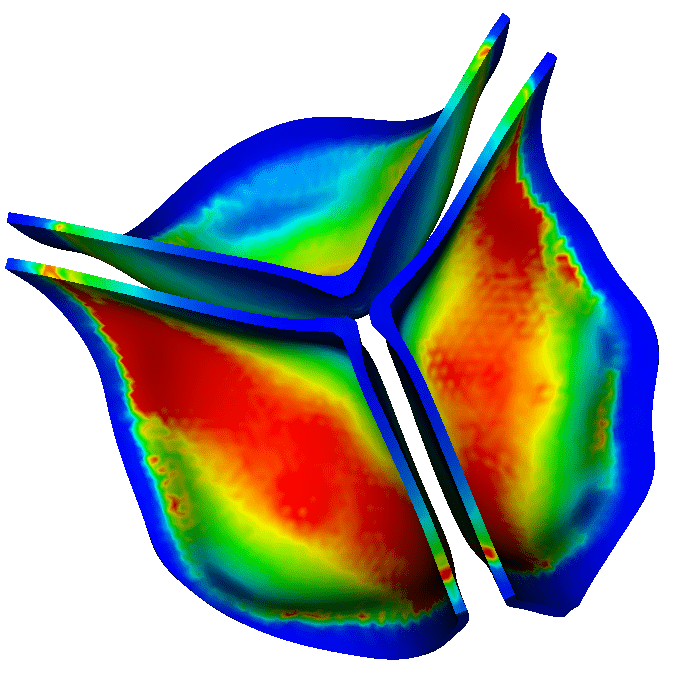}
	\includegraphics[width=0.195\textwidth]{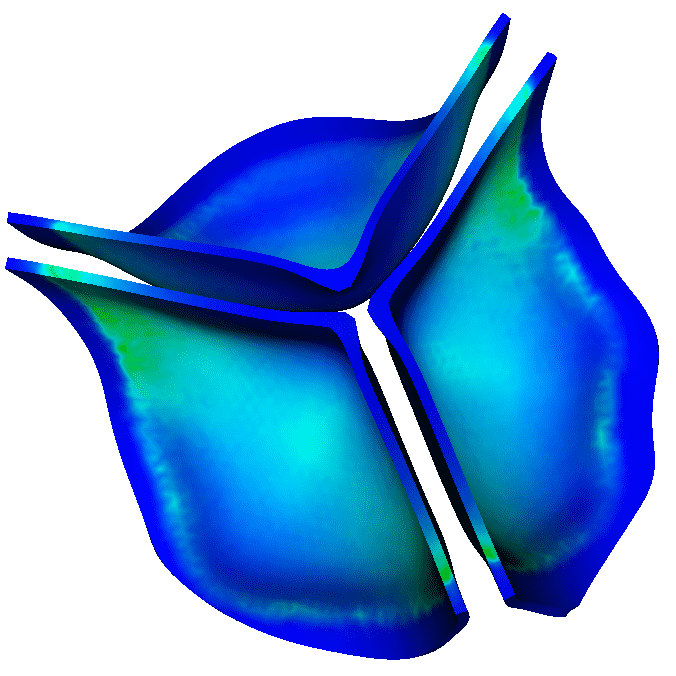}
}

\sidesubfloat[][]{
	\includegraphics[height=0.195\textwidth]{aortic_root/native/N=256/sigma_v/colorbar} \
	\includegraphics[width=0.195\textwidth]{aortic_root/native/N=256/sigma_v/sigma_v0009}
	\includegraphics[width=0.195\textwidth]{aortic_root/native/N=256/sigma_v/sigma_v0021}
	\includegraphics[width=0.195\textwidth]{aortic_root/native/N=256/sigma_v/sigma_v0033}
	\includegraphics[width=0.195\textwidth]{aortic_root/native/N=256/sigma_v/sigma_v0045}
	\includegraphics[width=0.195\textwidth]{aortic_root/native/N=256/sigma_v/sigma_v0057}
}
\caption{Similar to Fig.~\ref{f:dX_material_comparison}, but here showing on Mises stresses (kPa) obtained throughout diastole, up to early systole.
The stress distributions are similar in all three cases, although the glutaraldehyde-fixed porcine leaflets with $4~\text{mmHg}$ fixation pressure (a) experience slightly lower stresses than the glutaraldehyde-fixed leaflets with $0~\text{mmHg}$ fixation pressure (b), which are in turn slightly lower than the stresses experienced by the fresh leaflets (c).
}
\label{f:sigma_v_material_comparison}
\end{figure}

The aortic valve leaflet model of Driessen et al.~\cite{NJBDriessen05} was fit to tensile test data obtained by Billiar and Sacks \cite{KLBilliar00-I, KLBilliar00-II} from both fresh porcine aortic valve leaflets and also glutaraldehyde-fixed porcine leaflets at transvalvular fixation pressures of $0~\text{mmHg}$ and $4~\text{mmHg}$.
The glutaraldehyde-fixed leaflets are thereby similar to those used in porcine bioprosthetic heart valves.
In all cases, the model fits use $c_1 = 10~\text{kPa}$.
For $0~\text{mmHg}$ fixation pressure, the constitutive uses $k_1 = 5.35~\text{kPa}$, $k_2 = 5.85$, and a fiber angle standard deviation of $16.1^\circ$.
For $4~\text{mmHg}$ fixation pressure, the constitutive uses $k_1 = 55.3~\text{kPa}$, $k_2 = 5.75$, and a fiber angle standard deviation of $14.9^\circ$.
By comparison, for the fresh leaflets, the constitutive uses $k_1 = 0.7~\text{kPa}$, $k_2 = 9.9$, and a fiber angle standard deviation of $10.7^\circ$.
Thus, the fresh leaflets have a much softer initial response, but stiffen more rapidly under increasing fiber strain.
In addition, the collagen fibers are more highly aligned in the fresh valve leaflets than in the glutaraldehyde-fixed leaflets.

We compare the leaflet biomechanics in these three cases using a relatively fine Cartesian grid spacing of $0.43~\text{mm}$.
Figs.~\ref{f:dX_material_comparison} and \ref{f:2D_slice_material_comparison} compare the displacements during the same time points in early systole.
Notice the glutaraldehyde-fixed porcine leaflets with $4~\text{mmHg}$ fixation pressure deform the least in diastole, and that the fresh leaflets experience the largest deformations.
This is in good agreement with the results obtained by Driessen et al.~\cite{NJBDriessen05} using a solid mechanics model of the valve leaflets with an idealized leaflet geometry.
As the valve opens, however, the differences in the deformations obtained using the different valve models become relatively small.
Fig.~\ref{f:lambda_f_material_comparison} compares the fiber stretch ratios during early systole, as the valve opens.
The glutaraldehyde-fixed porcine leaflets with $4~\text{mmHg}$ fixation pressure show the smallest fiber strains in diastole, and the fresh leaflets experience the largest fiber strains.
This is in agreement with the results shown in Figs.~\ref{f:dX_material_comparison} and \ref{f:2D_slice_material_comparison}.
Finally, Fig.~\ref{f:sigma_v_material_comparison} compares the von Mises stresses during diastole and early systole.
The stress distributions are similar in all three cases, although the glutaraldehyde-fixed porcine leaflets with $4~\text{mmHg}$ fixation pressure experience slightly lower stresses than the glutaraldehyde-fixed leaflets with $0~\text{mmHg}$ fixation pressure, which are in turn slightly lower than the stresses experienced by the fresh leaflets.

\subsection{Effect of material properties on hemodynamics}

\begin{figure}
\centering
\includegraphics[width=0.475\textwidth]{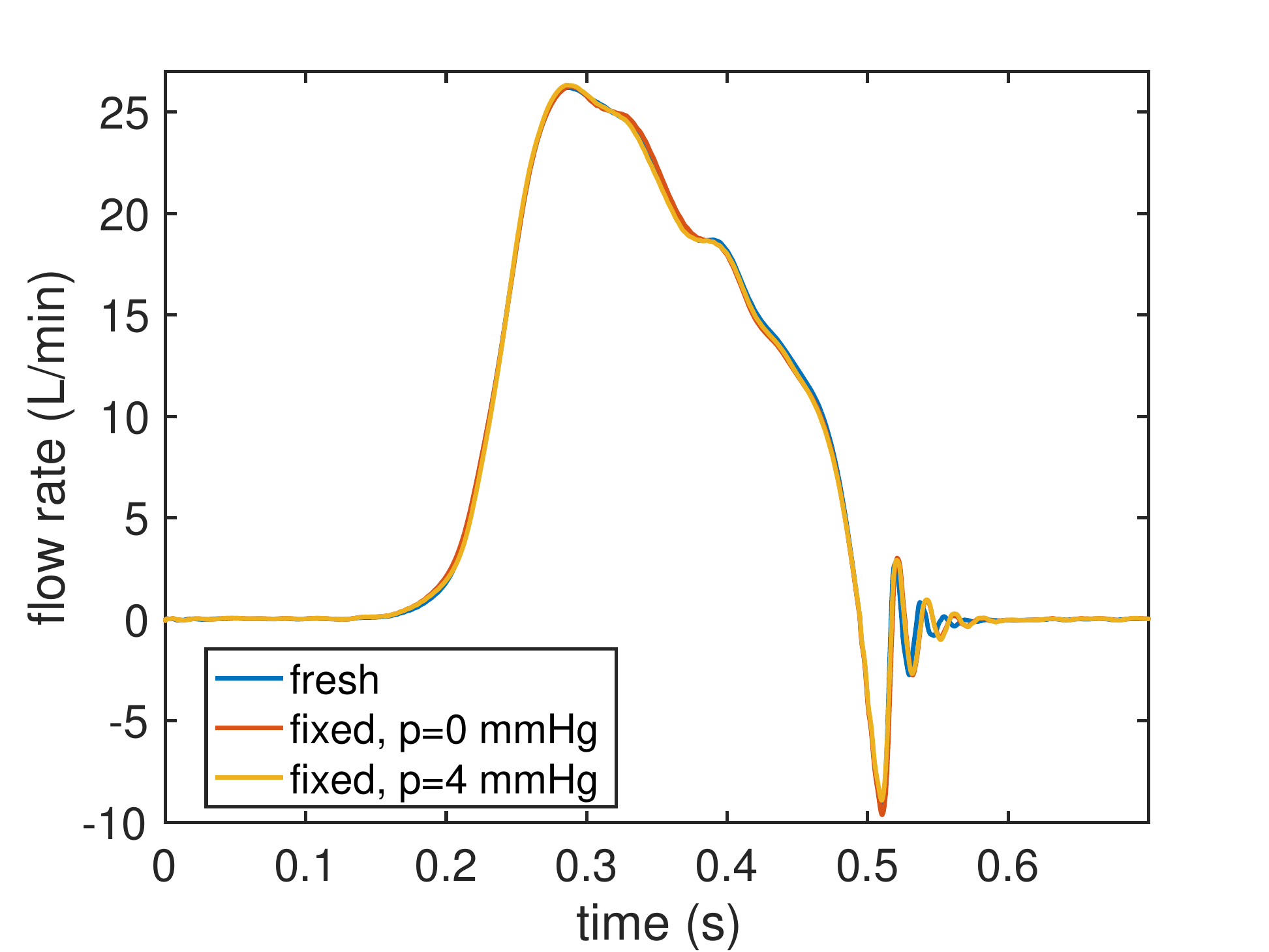}
\caption{Flow rates through the aortic root model using constitutive parameters for fresh and glutaraldehyde-fixed valve leaflets obtained using the relatively coarse Cartesian grid spacing of $0.86~\text{mm}$.
Notice that the flow rates are virtually identical.
}
\label{f:hemodynamics_comparison}	
\end{figure}

Finally, we consider the effect of leaflet material parameters on the hemodynamics of the aortic root and ascending aorta, here using the relatively coarse Cartesian grid spacing of $0.86~\text{mm}$.
The results of Sec.~\ref{s:comparitive_leaflet_biomechanics} suggest that there should be relatively little difference in the hemodynamics observed for the different material models, because the open configurations for the different cases are similar.
Fig.~\ref{f:hemodynamics_comparison} confirms that this is indeed the case; the volumetric flow rates are nearly identical for the three leaflet model parameter sets considered.
Likewise, the pressures are similar (data not shown).
Thus, although the deformations and stress distributions differ substantially between fresh and fixed leaflets, there is relatively little difference in the flow properties obtained using these different material models.

\section{Discussion and conclusions}

This paper presents initial results along with numerical and biomechanical tests using a dynamic fluid-structure interaction (FSI) model of the aortic root and ascending aorta based on the immersed boundary (IB) method.
This model employs a clinical image-based anatomical geometry along with a realistic fiber-reinforced hyperelastic model of the aortic valve leaflets.
Driving and loading conditions for the model are based on clinical measurements from normal humans \cite{JPMurgo80} and are provided by reduced-order models, including a Windkessel model fit to these clinical data \cite{Stergiopulos99}.
The model generates physiological pressures and flow rates that are in reasonable agreement with the clinical data, although there are some discrepancies that can be explained by differences in measurement techniques.
Numerical tests show that the model is able to resolve the leaflet biomechanics in diastole and early systole at the grid spacings considered here.
The model is also used to examine differences in the mechanics and fluid dynamics yielded by fresh valve leaflets as compared to glutaraldehyde-fixed leaflets that are similar to those used in bioprosthetic heart valves.
Relatively large differences are seen the biomechanics of the different valve models during diastole, but there are only relatively small differences during systole.
Consequently, similar bulk hemodynamics are obtained for the three material models considered in this study.

Limitations of the study include the assumption of a nearly rigid model of the aortic sinuses and ascending aorta, and the use of porcine material properties for the valve leaflets.
The real aorta has substantial compliance that is not accounted for by this model.
Employing a realistically compliant model of the aorta is an important extension of this model that is planned for future work.
Further, although previous results suggest that the grid spacings used in this work are adequate to resolve the bulk hemodynamics \cite{VFlamini16-aortic_root}, higher resolution simulations should be performed to resolve the fluid dynamics and leaflet kinematics more completely, which will allow us to obtain grid-resolved leaflet kinematics during systole.
We aim to perform higher resolution simulations in future work using higher-order accurate IB-like methods based on sharp-interface approaches \cite{LiLai01, LeeLeVeque03}.
Finally, the model has not yet seen substantial validation; in separate work, we are now pursuing an approach to model validation using in vitro experimental models.

If future work, we plan to deploy this model to simulate the dynamics of implanted cardiovascular medical devices, including both surgical and TAVR heart valve prostheses.
This same modeling approach could also be used to model the dynamics of bicuspid aortic valves and their effects on stress distributions in the ascending aorta.

\section*{Acknowledgments}

We gratefully acknowledge research support from the National Institutes of Health (NIH Award HL117063) and National Science Foundation (NSF Award ACI 1450327).

\section*{References}

\bibliography{short_bib}

\end{document}